\documentclass[10pt,a4paper,psfig,epsfig,graphicx,psfrag,bibtotoc]{scrbook}
\usepackage[bindingoffset=15mm,paperwidth=180mm,paperheight=260mm,scale=0.807,heightrounded,mag=1142,textwidth=147mm,textheight=204mm]{geometry}
\usepackage{epsfig}
\usepackage{graphicx}
\usepackage{amsfonts,amssymb,amsmath,bm,color}
\usepackage{psfrag}
\usepackage{hyperref}
\usepackage{geometry}
\usepackage[all]{xy}
\usepackage{feynmf}
\usepackage{sidecap}
\usepackage[spanish,english]{babel}
\usepackage[latin1]{inputenc}

\newcommand{\be}{\begin{equation}}
\newcommand{\ee}{\end{equation}}
\newcommand{\ba}{\begin{array}}
\newcommand{\ea}{\end{array}}
\newcommand{\bea}{\begin{eqnarray}}
\newcommand{\eea}{\end{eqnarray}}
\newcommand{\xbf}{\mathbf{x}}
\newcommand{\ybf}{\mathbf{y}}
\newcommand{\pbf}{\mathbf{p}}
\newcommand{\dH}{\dot{H}}
\newcommand{\Ham}{\mathcal{H}}
\newcommand{\Lag}{\mathcal{L}}
\newcommand{\Phimas}{\Phi^\dagger}
\newcommand{\Pimas}{\Pi^\dagger}
\newcommand{\vx}{\mathbf{x}}
\newcommand{\vp}{\mathbf{p}}
\newcommand{\deltat}{\delta^{(3)}}

\newcommand{\ena}{E_a}
\newcommand{\eb}{E_b}
\newcommand{\omp}{\omega_{\bm{p}}}
\newcommand{\intxt}{\int d^3\vx}
\newcommand{\intxc}{\int d^4 x}
\newcommand{\intpt}{\int \frac{d^3\vp}{(2\pi)^3}}
\newcommand{\intpc}{\int \frac{d^4 p}{(2\pi)^4}}
\newcommand{\udp}{U_{\vp}}
\newcommand{\vdp}{V_{\vp}}
\newcommand{\intauxx}{\int d^4 \tilde{x}}

\newcommand{\gaux}{\sqrt{\tilde{g}(\tilde{x})}}

\newcommand{\Lie}{\pounds_n}
\newcommand{\da}{\dot{a}}
\newcommand{\ddH}{\ddot{H}}
\newcommand{\dda}{\ddot{a}}
\newcommand{\al}{\alpha}
\newcommand{\mH}{\mathcal{H}}
\newcommand{\Ji}{\mathcal{\chi}}
\newcommand{\vk}{\mathbf{k}}
\newcommand{\Ro}{R_{(0)}}
\newcommand{\dRo}{\dot{R}_{(0)}}
\newcommand{\ddRo}{\ddot{R}_{(0)}}

\newcommand{\ddmu}{\overleftrightarrow{\nabla_\mu}}

\numberwithin{equation}{section}

\begin{document}

\frontmatter
\selectlanguage{english}

\begin{titlepage}
\vskip 2cm

\begin{center}
\textsf{Universidad de Zaragoza. Facultad de Ciencias.\\[.3cm]
Departamento de F\'isica Te\'orica.\\[3cm]}
{\LARGE {\bf Beyond Special and General Relativity. Neutrinos and Cosmology.}}\\[2cm]
{\Large Javier Indur\'ain Gaspar}\\[7cm]

\begin{figure}[htbp]
 \centerline{\includegraphics[height=3.5cm]{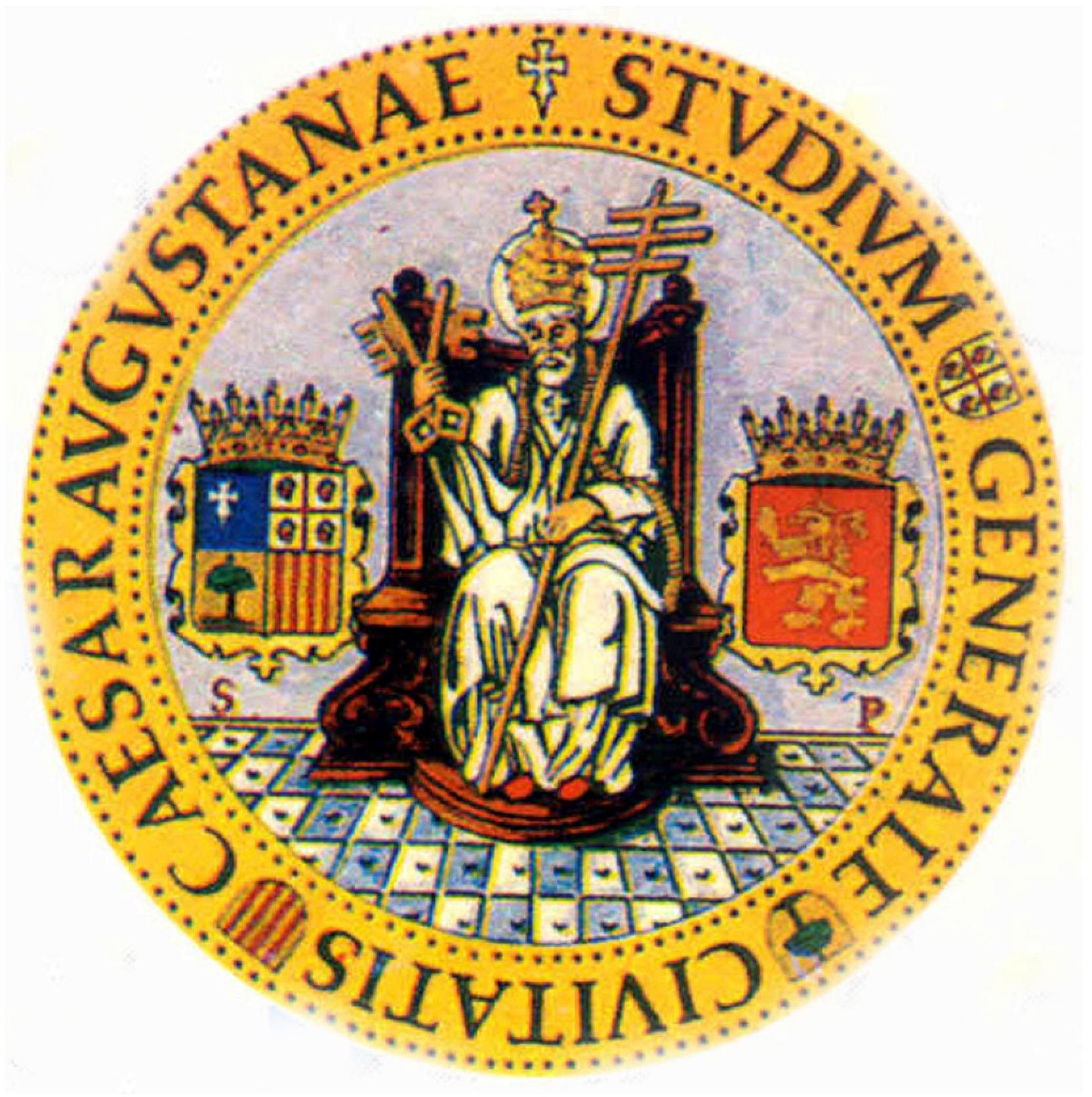}}
\end{figure}
{\Large Advisor: Prof. Jos\'e Luis Cort\'es Azcoiti\\[.2cm]
December, 2009}\\[1.5cm]

\end{center}

\end{titlepage}

\hspace{3cm}
\begin{flushright}
{\em A mis padres}.
\end{flushright}
\newpage

\thispagestyle{empty}

\chapter{Conventions and Abbreviations}

Throughout this thesis and unless it is otherwise stated, we will use units in which $\hbar = c = 1$. Moreover we will use the convention in which the signature of the metric is $+\, -\, -\, -$.

\bigskip

We will use the following abbreviations:

\bigskip

\textbf{ACM :}  Asymptotic Cosmological Model \bigskip

\textbf{ADM :}  Arnowitt-Deser-Misner formalism \bigskip

\textbf{BAO :}  Baryon Acoustic Oscillations \bigskip

\textbf{$\bm{\beta\beta 0\nu}$ :}  Neutrinoless Double Beta Decay \bigskip

\textbf{CMB :} Cosmic Microwave Background \bigskip

\textbf{C$\bm \nu$B :} Cosmic Neutrino Background \bigskip

\textbf{CSM :} Cosmological Standard Model \bigskip

\textbf{DE :} Dark Energy \bigskip

\textbf{DM :} Dark matter \bigskip

\textbf{DSR :} Doubly (or Deformed) Special Relativity \bigskip

\textbf{EFT :} Effective Field Theory \bigskip

\textbf{FRW :} Friedman-Robertson-Walker \bigskip

\textbf{GR :} General Theory of Relativity \bigskip

\textbf{GUT :} Grand Unified Theories \bigskip

\textbf{HDM :} Hot Dark Matter, for instance massive neutrinos \bigskip

\textbf{HST :} Hubble Space Telescope \bigskip

\textbf{ISW :} Integrated Sachs-Wolfe effect \bigskip

\textbf{IR :} Infrared \bigskip

\textbf{$\bm \Lambda$CDM :} the Concordance Model, in which the universe is dominated today by a mixture of Cold Dark Matter (CDM) and a Cosmological Constant ($\Lambda$) \bigskip

\textbf{LIV :}  Lorentz Invariance Violation \bigskip

\textbf{LQG :}  Loop Quantum Gravity \bigskip

\textbf{LSND :}  Liquid Scintillator Neutrino Detector \bigskip

\textbf{LSS :}  Large Scale Structure \bigskip

\textbf{MSW :}  Mikheev-Smirnov-Wolfenstein \bigskip

\textbf{NCQM :}  Noncommutative Quantum Mechanics \bigskip

\textbf{PMNS :}  Pontecorvo-Maki-Nakagawa-Sakata \bigskip

\textbf{QED :}  Quantum Electrodynamics \bigskip

\textbf{QFT :}  Quantum Field Theory \bigskip

\textbf{QNCFT :}  Quantum Theory of Noncanonical Fields \bigskip

\textbf{SM :}  Standard Model of particle physics \bigskip

\textbf{SME :}  Lorentz Violating Standard Model Extension \bigskip

\textbf{SNe :}  Type Ia Supernovae \bigskip

\textbf{SNO :}  Sudbury Neutrino Observatory \bigskip

\textbf{SR :}  Special Relativity \bigskip

\textbf{SSB :}  Spontaneous Symmetry Breaking \bigskip

\textbf{ST :}  String Theory (or Theories) \bigskip

\textbf{UV :}  Ultraviolet \bigskip

\textbf{vev :}  Vacuum expectation value \bigskip

\textbf{VSL :}  Varying Speed of Light \bigskip

\textbf{WMAP :}  Wilkinson Microwave Anisotropy Probe \bigskip

\textbf{WMAP5 :}  WMAP five year dataset \bigskip

\newpage

\tableofcontents

\newpage

\mainmatter
%%%%%%%%%%%%%%%%%%%%%%%%%%%%%%%
\chapter{Introduction}
%%%%%%%%%%%%%%%%%%%%%%%%%%%%%%%

By the beginning of the XXI century, the General Theory of Relativity (GR) and the Standard Model of particle physics (SM) seem to offer a very accurate description of all known physical phenomena... All of them? Not really, yet neutrino oscillations and cosmology still resist a full understanding.

\section{The Accelerating Universe}

Since the discovery of GR by Albert Einstein in 1915, there has been great interest in applying its equations to determine the behavior of the universe as a whole. Einstein was the first to open the Pandora's box in 1917, presenting a model of a static universe which required the presence of a `cosmological constant' in the field equations. Friedman found in 1922 that the equations admitted expanding universe solutions without the presence of this term. The expansion of the universe was also studied by Lema\^itre, Robertson and Walker, and confirmed seven years later by Hubble. Thus Einstein rejected the cosmological constant, calling it his `greatest blunder'.

Consider the Einstein's equations of GR

\be
R_{\mu\nu} -\frac 12 g_{\mu\nu}R = 8\pi G T_{\mu\nu}\, ,
\ee
where $R_{\mu\nu}$ is the Ricci curvature tensor, $g_{\mu\nu}$ is the metric of spacetime, $R = R_{\mu\nu}g^{\mu\nu}$ is the curvature scalar and $T_{\mu\nu}$ is the stress-energy tensor of the fluid filling spacetime. An isotropic and homogeneous universe is described by the Friedman-Robertson-Walker (FRW) metric

\be
ds^2 = dt^2 - a(t)^2 \left[ \frac{dr^2}{1-kr^2} + r^2 d\Omega_3^2 \right] \, ,
\ee
where $\{ t, r, \theta, \phi \}$ are the comoving coordinates, $a(t)$ is the scale factor, $k$ is the spatial curvature, and $\Omega_3$ is the solid angle of three-dimensional space $d\Omega_3^2 = d\theta^2 + sin^2 \theta d\phi^2$. The present value of the scale factor $a_0 = a(t_0)$ is usually normalized to one.

According to GR, this metric evolves following the relation (known as Friedman Equation)

\be
\frac{8\pi G}{3}\rho = H^2 + \frac{k}{a^2}\, ,
\ee
where $\rho$ is the total energy density and $H$ is the Hubble parameter, the rate of expansion of the universe ($H = \frac{da}{adt}$). Moreover the energy density of massive (matter) or massless (radiation) particles decreases with the scale factor as $\rho_m \sim a^{-3}$ or $\rho_r \sim a^{-4}$, respectively. This scenario, together with the conditions of an initially radiation-dominated universe which undergoes an adiabatic expansion, makes up the `Cosmological Standard Model' (CSM, as named by Guth \cite{Guth:1980zm}, by analogy with the Standard Model of particle physics). The solutions of the resulting differential equation for the scale factor are always decelerating ($\ddot a < 0$).

The CSM has achieved a number of successes, including the correct prediction of the relative densities of nuclei heavier than hydrogen (Nucleosynthesis) or the prediction of the Cosmic Microwave Background (CMB), the most perfect black body radiation ever measured. Since the first observations by Hubble it has however been very difficult to directly measure the precise effect of the cosmic expansion on the motion of galaxies.

This has been connected to the difficulty of finding 'standard candles', astronomical objects with known luminosity $L$ (energy released per unit time). Given the luminosity of an object and its spectrum, it is possible to directly measure its redshift ($z = \frac{a_0}a -1$) and bolometric flux $F$ (the energy detected per unit of area and time), where the relation between luminosity and flux is encoded in the luminosity distance $d_l$ to the astronomical object, defined as $d_l\equiv \left(\frac{L}{4\pi F } \right)^{1/2}$. The luminosity distance vs. redshift (which is in one to one correspondence with time) relation enables us to determine the evolution of the scale factor and the Hubble parameter with time\footnote{This relation is computed in the framework of metric spacetimes with a FRW metric. For objects of sufficient size, it is possible to compare this quantity to their angular size distance $d_A\equiv l/\Delta\theta$, where $l$ is the transverse size of the object and $\Delta\theta$ is its angular size (the angle that it subtends in the sky). In a metric spacetime, $d_l = d_A (1+z)^2$. This relation can be used to test the validity of metric theories as a good description of spacetime \cite{Bassett:2003vu,Sahni:2006pa}. However experimental data seem to support this relation \cite{DeBernardis:2006ii}. Thus we will not leave the framework of generally covariant metric spacetimes. },

\be
d_l (z) = c(1+z) \int^z_0 \frac {dz'}{H(z')} \, ,
\ee
and therefore gives us information about the relative importance of radiation, matter or spatial curvature in the evolution of the universe. As we told above, for any universe driven by a combination of those (the massless and massive particles contained in the Standard Model of particle physics (SM) or not, and the spatial curvature allowed in the FRW metric) will undergo a decelerated expansion, as gravity acts as an attractive force.

However, measurements of distant Type Ia Supernovae (SNe) (widely accepted standard candles) distance vs. redshift relation reveal that the universe is presently undergoing an accelerated expansion \cite{Riess:1998cb,Perlmutter:1998np}, in contradiction with the decelerated expansion predicted by GR!

\subsection{A conventional explanation: bring back the cosmological constant!}

Now it is not a secret that GR admits a cosmological constant $\Lambda$, a constant term in the Einstein-Hilbert lagrangian

\be
S = \intxc \sqrt{-g} [\frac{R -2 \Lambda}{16 \pi G} + \Lag_m ]\, ,
\ee
where $g$ is the determinant of the metric and $\Lag_m$ is the Lagrangian of the matter content of spacetime from which the stress-energy tensor is derived. The cosmological constant term $\Lambda$ is able to produce the desired accelerated expansion, as we will explain below. It modifies the Friedman equation giving

\be
\frac{8 \pi G}{3}\rho = H^2 + \frac{k}{a^2} - \frac \Lambda 3 \, . \label{11fried}
\ee

Given that the universe is expanding, at late times the effect of the energy density $\rho$ and curvature $k$ will be negligible and the universe will expand almost exponentially with

\be
H = \sqrt{\frac \Lambda 3} \, .
\ee

We arrive then to the Concordance Model ($\Lambda$CDM), in which the universe is homogeneous and isotropic in cosmological scales, with a spatial curvature $k$ and a cosmological constant $\Lambda$, and is filled by a mixture of radiation (mainly photons and neutrinos), massive particles of the standard model (mainly baryons, as electrons have a much smaller mass) and new massive particles which are non-baryonic and electrically neutral which we call Dark Matter (DM). If we divide \eqref{11fried} by $H^2$ and regroup terms we arrive to the cosmic balance equation,

\be
\Omega_r + \Omega_b + \Omega_{DM} + \Omega_k + \Omega_\Lambda = 1 \, , \label{11balan}
\ee
where $\Omega_i = \rho_i/\rho_c$ for $i=r,b,DM$, $\rho_c = \frac{3H^2}{8\pi G}$ is the critical density, $\Omega_k = -\frac k{a^2 H^2}$ and $\Omega_\Lambda = \frac \Lambda{3 H^2}$. The $\Omega$'s represent the relative importance of each of the components of the universe at a given time, and this relative importance vary.

This simple explanation fits comfortably the data collected by the SNe searchers, from the first to the last datasets \cite{Riess:1998cb,Perlmutter:1998np,Astier:2005qq,Riess:2006fw,Kowalski:2008ez}, which comprehends a few hundreds of SNe. It also provides a good explanation of the measured spectrum of CMB anisotropy \cite{Jaffe:2000tx,Spergel:2003cb,Spergel:2006hy,Dunkley:2008ie} and of the spectrum of matter anisotropy \cite{Peacock:2001gs,Tegmark:2003ud,Eisenstein:2005su,Cole:2005sx} if the universe is taken to be spatially flat, $k=0$ \cite{de Bernardis:2000gy}. In particular, the joint fit of the model to the whole set of experimental data gives, in the present time:

$$
\ba{cc}
H_0  = 72 \pm 3 \, km/sMpc\, , &
\Omega_r = (4.8 \pm 0.5)\times 10^{-5} \, , \\
\Omega_b = 0.044 \pm 0.004 \, , &
\Omega_{DM} = 0.22 \pm 0.2 \, , \\
-0.018 < \Omega_k  < 0.008 \, , &
\Omega_\Lambda = 0.726 \pm 0.015 \, .
\ea
$$

Most of the errors in the data are dominated by the error in the determination of $H_0$. If this is the end of the story, what is the point of the cosmic acceleration that we are not understanding?

\subsection{The Problems of $\Lambda$CDM}

\paragraph{The Cosmological Constant Problem}
Although a low value of the cosmological constant is compatible with the observations, theory predicts that it has a high value. It would not be any problem if the theory were classical, but the theory describing the source of the gravitational interaction $\langle T_{\mu\nu}\rangle$ is QFT in curved spacetime. Thus, the semiclassical Einstein equations valid for cosmology at curvatures well below the Planck scale ($E_P = 1.2209 \,\times\, 10^{19}\, GeV$ / $l_P = 1.6165 \,\times\,  10^{-35}\, m$)\footnote{At this scales gravity cannot be treated as a classical interaction any more, as we will explain later.} are

\be
R_{\mu\nu} -\frac 12 g_{\mu\nu}(R + \Lambda) = 8\pi G \langle\hat T_{\mu\nu}\rangle\, ,
\ee
where the curvature terms are treated classically and $\langle\hat T_{\mu\nu}\rangle$ is the expectation value of the stress-energy tensor computed for a QFT in curved spacetime \cite{BD}. In order to renormalize the contributions of the matter Lagrangian to the total action of matter and gravitation, it is necessary to add counterterms also in the gravitational Lagrangian that cancel some of the infinities arising in the matter Lagrangian. Some of these counterterms are of the same form of the ones appearing in the classical Lagrangian ($R$, $\Lambda$) and some present more than two derivatives of the metric.

The term of interest is the divergent term which has the same form as the cosmological constant $\Lambda$. If we compute the value of this contribution in the frame of Effective Field Theory (EFT) \cite{Burgess:2007pt}, we assume that the SM is valid up to a high energy scale (which we will call ultraviolet (UV) scale), which we assume to be of the order of the Planck scale. Then this contribution is $\sim 10^{71} GeV^4 $, that is, $120$ orders of magnitude greater than the observed value, $\sim 10^{-47} GeV^4 $ \cite{Weinberg:1988cp}. In order to explain the data, the quantum corrections $ 8\pi G \langle\hat T^{vac}_{\mu\nu}\rangle$ should cancel the first $120$ significant figures of the classical contribution $\Lambda$, but not the next one! Such an unnatural cancelation is called the Cosmological Constant Problem.

The solutions to the Cosmological Constant Problem among the quantum gravity community are not very convincing. In String Theory (ST) the natural value of the cosmological constant is negative and of order $E_p^4$. This has led the ST community to resort to anthropic arguments in order to save the face (see Ref.~\cite{Carroll:2000fy} for arguments and references). In Loop Quantum Gravity (LQG) it has been claimed that a positive small cosmological constant is possible in a solution called Kodama state \cite{Smolin:2002sz}. However the dispersion relation of particles living in this spacetime have at energies well below the Planck scale a modified dispersion relation of the form

\be
E^2 = \vp^2 + m^2 + \alpha \frac{E^3}{E_p} \, ,
\ee
which has been already almost ruled out by observation \cite{Jacobson:2002ye,Mattingly:2005re,Collaborations:2009zq}.

The problem is that we have quantum corrections coming from the UV behavior of a QFT to an otherwise infrared (IR) scale, i.e. of lower energy than the typical energies in particle physics, of GR. Is there any possibility that the quantum corrections coming from QFT do not stop the cosmological constant being IR? EFT relies in some basic principles, and the one responsible of the quantum corrections to the vacuum energy being of the order of the UV scale is the decoupling theorem \cite{Appelquist:1974tg}. Any mechanism able to avoid UV quantum corrections to the stress-energy tensor must violate this property. It would then be possible to have a QFT which is valid in a window of energies bounded by an IR and an UV scale, in which the main quantum corrections to the Einstein's equations come from the IR.

\paragraph{The Horizon Problem} The present homogeneous, isotropic domain of the universe is at least as large as the present horizon scale $c t_0 \sim c/H_0$, where $H_0$ is the present value of the Hubble parameter. This domain has grown up with the scale factor. As inhomogeneity cannot be erased with the decelerated expansion, we can conclude that the size of the patch of the universe from which this domain originated was at $t = t_i$ at least as big as $l_i \sim c t_0 a_i/ a_0$. If we compare this with the causal horizon at the same time, $l_c = c t_i$, then the ratio between the two scales is,

\be
\frac{l_i}{l_c}\sim \frac {t_0}{t_i}\frac{a_i}{a_0}\, .
\ee

Thus, if we assume some random initial conditions at $t_i$, which we will assume to be of Planckian size because we consider that GR is valid up to that scale, and $a\sim 1/T$ under radiation domination, we get

\be
\frac{l_i}{l_c} \sim \frac{10^{17}s}{10^{-43}s}\frac{2.73 K}{10^{32} K} \sim 10^{28}\, ,
\ee
and therefore that patch was not causally connected, so it did not need to be homogeneous at all! This is a big problem, as the initial conditions should be very fine-tuned to reproduce our current universe. This is the Horizon Problem.

Unless there is something we are missing. Let us assume that the scale factor grows with a power of time, and therefore $\dot a \sim a/t$. Thus we can rewrite

\be
\frac{l_i}{l_c} \sim \frac {\dot a_i}{\dot a_0}  \, .
\ee

From the last expression we can read that the size of our universe was initially larger than that of a causal patch by the ratio of the corresponding expansion speeds. Only if these are of the same order can the initial conditions at $t_i$ thermalize and give rise to an homogeneous universe. And as gravity has been decelerating this rate for a wide period of time, there must have been a fast accelerated expansion before the decelerated one that has correctly described so much physics. This points out to new particle or gravitational physics in the UV, but below the Planck scale.

\paragraph{The Flatness Problem} Current observations tell us that the curvature term in the Friedman equation is very close to zero. In particular, if we write the cosmic balance equation \eqref{11balan} and neglect $\Omega_\Lambda$, the cosmological parameter $\Omega(t)$ follows

\be
\Omega(t) \equiv \frac{\rho(t)}{\rho_c(t)} = 1 + \frac k{(Ha)^2} \, .
\ee

Therefore, at early times,

\be
\Omega(t_i) - 1 = (\Omega_0 - 1)\frac{(Ha)^2_0}{(Ha)^2_i} = (\Omega_0 - 1)\frac{(\dot a)^2_0}{(\dot a)^2_i} < 10^{-56} \, .
\ee

Taking into account that this difference can be interpreted as the relative difference between kinetic and potential energy in the cosmic fluid \cite{Mukh}, this means that the initial velocities must be fine-tuned to a degree of $10^{-28}$ in order to explain the current flatness of the universe! This is the flatness problem.

This fine-tuning would not be so high if the expansion rate at Planckian time would be smaller than or at least as small as the present expansion rate, which would in turn require an early accelerated expansion, as we have seen before.

\subsection{Ingredients for an alternative to $\Lambda$CDM}

The problems of $\Lambda$CDM have their origin in our attempt of extending the range of validity of our current theories of gravity and particle physics outside the region in which those theories have been experimentally tested and validated. We have no idea of what the correct theory of gravity will be at extremely high (or low) curvatures compared to the ones that have been tested; and we have arguments that point out that a quantum theory of gravity may violate some of the principles underlying GR and QFT \cite{Garay:1994en}.

At this point it becomes evident that one must find a way of extending either GR, QFT or both in order to explain the problems of cosmology at very early or recent times. This extensions can be performed either following a top-down or a bottom-up approach. In a top-down approach the fundamental principles of the theories are changed, a new theory is derived, and the phenomenological consequences are derived. Examples of top-down approaches to the problems of cosmology are ST \cite{Polchinski}(and its applied branch of String Cosmology) or LQG \cite{Rovelli:1997yv,Thiemann:2001yy} (from which Loop Quantum Cosmology is derived). The aforementioned approaches have departed from a solid theoretical and mathematical base but have grown up on the sidelines of the experimental results, and some of their implications for cosmology are in conflict with observational data. We will show a couple of examples below.

In contrast, a bottom-up approach is based on experimental and observational data, which are used to build up laws and models which can later on be incorporated to a well founded theory. In this PhD thesis we will try to show that it is possible to build a phenomenological extension to $\Lambda$CDM at the level of the cosmological evolution equations. This extension, which we have called the Asymptotic Cosmological Model (ACM), incorporates a modification of the laws of cosmic expansion such that an epoch of (asymptotically exponential) accelerated expansion arises both in the IR (corresponding to nowadays Hubble rate) and in the UV (corresponding to the very early universe). The model fits successfully the current observational data. The study of the origin of the modifications is precluded, although some possibilities ar pointed out in the following.

\paragraph{New physics in the IR}
We have mentioned that in order to solve the Cosmological Constant Problem, we need to render the UV corrections to the vacuum energy density IR. This is not possible in the frame of local EFT, due to the decoupling theorem. In fact, this is a problem not only of $\Lambda$CDM, but of GR coupled to an EFT in general. Even if there is no classical contribution to the cosmological constant/vacuum energy in the gravitational lagrangian, the renormalization of the matter lagrangian will in general induce a nonzero term. \footnote{With the remarkable exception of supersymmetric theories with unbroken supersymmetry. However, the absence of exact supersymmetry in nature discards this explanation.} The problem is that the power counting of a local EFT implies that the corrections to this term will be much greater than the observed value. Therefore a solution to the Cosmological Constant Problem will presumably require a departure from locality in QFT, which is parameterized by some new IR scale.

It has already been noticed that local EFT should fail to be a good description of nature in the presence of gravitational effects \cite{Cohen:1998zx}. In any EFT with a UV cutoff $E_{NP}$, the entropy of a system is an extensive quantity and therefore grows with its volume, $S \sim L^3  E_{NP}^3$. However blackhole physics seems to reveal that the maximum entropy contained in a certain box of size $L$ must be the Bekenstein entropy, $S < \pi E^2_p L^2$. This means that the entropy predicted by EFT can exceed the Beckenstein bound for sufficiently large sizes. Therefore it has been argued that EFT overcounts the degrees of freedom in some way \cite{'tHooft:1993gx,Susskind:1994vu}, and that the number of degrees of freedom of a given system must saturate from an extensive behavior (proper of EFT) to a proportionality with area, when the size of the region investigated is of order $L \sim E_p^2 / E_{NP}^3$. The authors of Ref.~\cite{Cohen:1998zx} propose an even more restrictive domain of validity, based on the premise that the size of the region under consideration should not exceed the Schwarzschild radius associated with their temperature $T\sim E_{NP}$ for EFT to be valid. If EFT is used to describe particles up to temperatures $T$, then the associated thermal energy of the system will be of order $U\sim L^3 E_{NP}^4$ and the associated entropy will be of order $S \sim L^3  E_{NP}^3$. Then the Schwarzschild radius will be $r_s \sim U/E_p^2$. If we want the system under description not to form a blackhole, then its size needs to be greater than this quantity, $L > r_s$, and therefore the momenta of the particles we are describing will present an IR cutoff of order $\lambda \sim 1/L$, where

\be
\lambda \sim  E_{NP}^2 / E_p \, .
\ee

This would generate, aside from corrections $\propto \frac 1{E_{NP}^n}$ to the tree level effective action, also IR corrections $\propto \lambda^2$. As a consequence, in a given experiment at energies $\sim p$ the predictions of EFT for a given quantity $Q$ will have corrections from both irrelevant terms in the action of dimension $D$ and gravitational physics, of the form

\be
\frac{\delta Q}{Q} \sim \left[ \left(\frac{p^{D-4}}{E_{NP}^{D-4}}\right) + \left( \frac{E_{NP}^4}{p^2 E_p^2}\right) \right]\, .
\ee

The maximum possible accuracy will be attained by some value $E_{NP}$ of the maximum energy scale that can be probed. As an example given in Ref~\cite{Cohen:1998zx}, if the momenta of the particles are of the order of the electroweak scale (as in the new collider experiments, like LHC), and the EFT is the standard model, new physics appears at dimension $6$ operators ($5$ if we allow for lepton number violation), the minimum uncertainty of $10^{-13}$ is achieved when the maximum scale that can be probed with EFT techniques is $10^8 GeV$. The minimum scale that could be probed would then be of the order of the $MeV$. However if new physics appears at $10 Tev$, then the SM would be valid up to that energies and down to $10^{-2} eV$, around the energy scale which is behind neutrino physics or the cosmological constant.

In the same context, if one tries to compute the quantum corrections of the matter content of the universe to the cosmological constant in the context of EFT, one should notice that for this approach to be valid at a patch of the universe of the size of the causal horizon $\sim 1/H_0$, one should introduce an IR cutoff of the order of $H_0 \sim 10^{-26} eV$ and therefore the associated UV cutoff of the EFT would be $10^{-2} eV$. As a result, quantum corrections would be of the same order of the observed value of he vacuum energy.

It has been claimed that any EFT approach to a theory including gravity will be naturally attached to a range of validity bounded by an UV and an IR scale \cite{Carmona:2000gd}. However, if the UV scale is taken to be the Planck scale, then the IR will also equal the Planck scale and the range of validity of the EFT will be constrained to a point. It may be that the effective fundamental scale of gravity in the EFT would be decreased by the existence of large extra dimensions \cite{ArkaniHamed:1998rs,Antoniadis:1998ig,Sundrum:1998ns,ArkaniHamed:1998kx}. The effective gravity scale has been proposed to lie in the order of magnitude of the few $TeV$ scale, and the IR scale (corresponding to extra dimensions of the size of a millimeter) would be of the order of the scale of the cosmological constant or neutrino physics.

Anyway, if local EFT is to be modified in the IR in order to solve the Cosmological Constant problem, it is also possible that the quantum corrections induced in the gravitational Lagrangian add not only terms with a higher number of derivatives of the metric suppressed by an UV scale, as in the standard case, but also other nonlinear or even nonlocal terms suppressed by the IR scale.

In fact, there is no reason to believe that the action of the classical theory of gravity is just given by a term proportional to the curvature scalar and a constant term, as in $\Lambda$CDM. With the premises of a metric theory with general covariance and the equivalence principle, it is possible to build an infinite plethora of theories. Einstein decided to choose the simplest one for the shake of aesthetics: he required his equations to be of second order, and this is only possible with a Lagrangian of the form $\sqrt{-g}(R-2\Lambda)$. However in modern theoretical physics we have learned that everything that is not forbidden is compulsory \cite{Weinberg:2005kr}. Additional terms in the action are not forbidden by the symmetries of the theory as long as they behave as scalars under general coordinate transformations.

Terms with a higher number of spatial derivatives can be neglected by dimensional analysis, as they must be supplied by a certain scale. If this scale is a UV scale such as the Planck scale, then the effect of these terms can be neglected at most tests, with the remarkable exception of the very early universe. Terms of these type arise naturally when one tries to renormalize a QFT in curved space \cite{BD}. In this context even if they are not originally in the classical gravitational action, terms of the following form appear in the action:

\be
S_g^{UV} = \frac{1}{16\pi G} \intxc \sqrt{-g} \left[ -2\Lambda + R + \frac 1{E_{UV}^2}\left( \alpha R^2 + \beta R_{\mu\nu}R^{\mu\nu} + \gamma R_{\mu\nu\rho\sigma}R^{\mu\nu\rho\sigma} + \delta \square R \right)\right]\, , \label{GRUV}
\ee
and there is no reason to believe that terms with even higher derivatives of the metric appear in the gravitational action, suppressed by greater powers of the inverse of the UV scale $E_{UV}$.

However, if these terms are supplied by an IR scale, like the cosmological constant, their effect should be observable at cosmological distances. Aside from a cosmological constant, terms involving an infrared scale should be at least non-polynomial or nonlocal,

\be
S_g = \frac{1}{16\pi G} \intxc \sqrt{-g} \left[ R -2\Lambda  + \alpha E_{IR}^2 \square^{-1}R + \beta\frac{E_{IR}^4}R + ... \right]\, , \label{GRIR}
\ee
where the effect of these terms can na\"{\i}vely be neglected in some local tests of gravity but heavily influence the cosmic evolution at the present time (provided that $\Lambda \sim E_{IR}^2$).

\paragraph{New physics in the UV}
We have also pointed out that, in order to solve the Horizon Problem and the Flatness Problem of the universe, some sort of new physics in the UV is required. As the universe expands it also cools down from an ``initial'' state at Planckian temperature $\sim T_p = k_B^{-1} E_p \sim 10^{32} K$, whose description requires some knowledge of the correct theory of quantum gravity. Lacking any knowledge of such a theory, the best we can do is to assume that quantum gravity provides us with some random initial conditions that can, since some initial time $t_i \gtrsim t_P$, be treated with classical gravity together with some extra corrections suppressed by an UV scale. This new physics could maybe induce an accelerated expansion just after the regime in which gravity must be described by a quantum theory.

Many models have been built in order to explain this accelerated expansion. First models resorted to the coupling of the expansion to a phase transition in the context of GUT \cite{Sato:1980yn,Guth:1980zm,Linde:1981mu,Albrecht:1982wi}. Later it was discovered that the details of the particle physics behind the accelerated expansion were not important \cite{Linde:1983gd}. The addition of terms with higher derivatives suppressed by a UV scale \cite{Starobinsky:1980te} has also been used to explain this accelerated expansion (called inflation \cite{Guth:1980zm}). As an additional consequence the quantum fluctuation of fields in that regime is enough to cause the fluctuations of the CMB spectrum, and therefore clusters, galaxies, stars, planets and life. As said, the details of the mechanism are unimportant, as several {\it a priori} very different models lead to very similar conclusions.

As an example of new physics in the UV leading to an early accelerated expansion, let us introduce the very simple model of Ref.~\cite{Starobinsky:1980te}. Assume that the Einstein-Hilbert action is just a low energy approximation to a more general theory,

\be
S = \intxc \sqrt{-g} \left[ R + \Lambda - \alpha \frac{R^2}{E_p^2} + ... \right] \, .
\ee

It has been shown \cite{BD} that such terms appear in the renormalized action of classical GR coupled to a QFT in curved spacetime. If we consider the theory modified just by the $R^2$ term, it turns out to be conformally equivalent (with a conformal factor $F = 1 + 2\alpha R/E_p^2$) to a theory of GR coupled to a scalar field $\varphi$,

\be
S = \intxc \sqrt{-g} \left[ R + \Lambda + 16 \pi G(g^{\mu\nu}\partial_\mu\varphi\partial_\nu\varphi - \frac{E_p^2}{12\alpha}\varphi^2) \right] \, ,
\ee
with $\alpha$ a few orders of magnitude above unity. The potential of the scalar field can be seen as an effective mass term with $m^2 = \frac 1{6\alpha} E_p^2$  A detailed study of the dynamical system will prove that this scenario is an example of chaotic inflation \cite{Linde:1983gd}, in which almost random initial conditions lead to an exponential expansion and to an almost flat power spectrum for the primordial fluctuations in the metric of spacetime.

Let us assume that the initial condition of the scalar field is at least locally an homogeneous condensation. The equations of motion in the homogeneous approximation (FRW metric and $\varphi \approx \varphi(t)$) can be written as:

\bea
\ddot \varphi + 3 H \dot \varphi + m^2 \varphi & = & 0 \, , \\
\frac{8\pi G}3 \left( \frac 12 \dot\varphi^2 + \frac 12 m^2 \varphi^2 \right) & = & H^2 \, ,
\eea
where we are neglecting the effect of the spatial curvature ($H^2 \gg k/a^2$) for the sake of simplicity. The effect of spatial curvature can be studied numerically. The system can be then reduced to a first order dynamical system for $\dot\varphi (\varphi)$,

\be
\frac{d \dot \varphi}{d \varphi} = -\sqrt{-12\pi G(\dot \varphi^2 + m^2\varphi^2)} - m^2\frac{\varphi}{\dot\varphi}
\ee

It should be noticed that this dynamical system possesses an attractor to which all other solutions converge with time. The attractor behaves as $\dot \varphi^2 = \frac{m^2}{12\pi G}$ and $sign(\dot\varphi)\neq sign(\varphi)$
for $\dot \varphi^2 \ll m^2 \varphi^2$ and then forms a double decaying orbit around the origin. If a solution meets the attractor at one of its flat branches, in this regime the field varies slowly with time and the universe expands almost exponentially, with

\be
H^2 \approx \frac{8\pi G}3 \left( \frac 12 m^2 \varphi^2\right) \sim const \, .
\ee
until $\varphi \sim \frac 1{\sqrt{12\pi G}}$. When the value of the field reaches the Planck scale its decrease rate is high and the field rapidly decays to $\varphi = 0$, providing the exit to inflation and transferring the energy of the scalar field to the matter content of the model (reheating).  As the original theory is related to the scalar-metric theory by a conformal factor which depends only on $R$, which is approximately constant during the inflationary regime, then in the original theory the same inflationary solution exists.

One can arbitrarily complicate this model, but the general prediction of an early accelerated expansion (in most models close to an exponential expansion) leading to a flat, thermalized universe with small perturbations obeying a scale free power spectrum is rather general. Most of the models include modifications of gravity or a set of scalar fields which mimics the previous mechanism. Noticeably, the universe is devoid of matter (or emptied by the mechanism) until the accelerated expansion stops, and then refilled by the energy released.

\paragraph{Agreement with observation}

In the past decade a vast amount of observational data has been collected, that severely constrains most of the models that have been built for cosmology, especially those derived from top-down approaches like ST or LQG. At this point, it is no longer sufficient to build models which qualitatively explain the features in the universe, but also which provide us with competitive fits of the currently gathered data and desirably with predictions to be tested by future missions.

There is a list of cosmological observations that should be accommodated in an extension of $\Lambda$CDM:

\begin{itemize}
\item
In any sensible model of inflation quantum (or thermal) fluctuations are produced, which seed the perturbations that will later produce the large scale structure of the universe, galaxies, and so on. Most inflationary models predict that the primordial scalar fluctuations will have an almost scale invariant power spectrum $P(k)$,

\be
P(k) \equiv \frac{\vert \Phi_k\vert^2 k^3}{2\pi^2} \propto k^{n_s -1} \, ,
\ee
where $\Phi_k$ is the Fourier transform of the gauge-independent scalar perturbation of the FRW metric in GR without a shear, and $n_s$ is called 'scalar spectral index' (scale invariant spectrum means $n_s \approx 1$). The Starobinsky $R^2$ model \cite{Starobinsky:1980te} and in general chaotic inflation \cite{Linde:1983gd} predict $n_s \lesssim 1$. In fact, the scalar spectral index has managed to be measured in the CMB spectrum by WMAP \cite{Dunkley:2008ie}, and it turns to be $n_s = 0.963^{+0.014}_{-0.015}$.

\item
Models for the present accelerated expansion must be confronted with a much wealthier sample of data. Given that Type I a Supernovae (SNe) can be taken as standard candles, it has been possible to determine the luminosity distance $d_L$ vs. redshift $z = a_0/a -1$ relation,

\be
d_l (z) = c(1+z) \int^z_0 \frac {dz'}{H(z')} \, ,
\ee
for a few hundred SNe \cite{Riess:2006fw,Kowalski:2008ez} ranging for $z<2$. Besides supernovae, the reduced distance to the surface of last scattering $R_{CMB}$,

\be
R_{CMB} = \Omega_m^{1/2} H_0 \int_0^{1089} dz/H(z) \, ,
\ee
can be also read from the WMAP data \cite{Komatsu:2008hk}. Both measurements involve the integral of the inverse of the Hubble parameter with respect to the redshift, but SNe observations sample the present era whereas the reduced distance $R_{CMB}$ provides an integrated information of the evolution of the universe since its youth.

A slightly different dependence is found in the distance parameter of Baryon Acoustic Oscillations (BAO) $A(z)$,
\be
A(z)= \Omega_m^{1/2} H_0 H(z)^{-1/3} z^{-2/3} \left[\int_0^z dz'/H(z')\right]^{2/3} \, ,
\ee
measured  at redshift $z=0.35$ by SDSS \cite{Eisenstein:2005su}. Although some other data can be confronted with cosmological models, these are the most widely used probes in cosmology.
\end{itemize}

In conclusion, GR can be extended to include new IR and UV phenomena, and it should be extended in order to satisfactorily explain both the initial conditions of the universe and the current accelerated expansion. It should also be extended in order to comprehend the new corrections arising from the renormalization of the matter lagrangian coupled to gravity, especially if this Lagrangian is also modified in the IR or the UV.

The effect of this extension in cosmological scales should be to provide us with an accelerated expansion in the very early universe and another accelerated expansion in the present age. These accelerated expansions are in most cases asymptotically exponential expansion. In this PhD thesis a cosmological model will be built, in which these to epochs of accelerated expansion will be phenomenologically incorporated. The resulting model will be used to fit the available observational data.

\subsection{Review on alternative models to explain the accelerated expansion}

This subsection is a brief review on the main alternatives proposed in order to explain the problems concerning the present accelerated expansion and the very early universe.

The main paradigm explaining the Horizon and Curvature problems in the very early universe is inflation; either occurring due to a modification of GR in the UV or the addition of new fields at high energies in extensions of QFT such as ST. The only real alternative to the inflationary paradigm is that of Varying Speed of Light (VSL) \cite{Albrecht:1998ir}. We will explain both alternatives further below.

With the increasing interest on the present accelerated expansion, a wide plethora of models has risen. Aside from modifications of gravity (such as $f(R)$-theories, higher dimensional models, alternatives to Dark Matter explaining the smallness of the cosmological constant), these models include new forms of matter (new fields of any type, perfect fluids with exotic equations of state,...) or rejection of spatial homogeneity as a good approximation in the present universe (effective cosmological constant arising from homogenization procedure \cite{Kolb:2005da,Wiltshire:2007zj}, local void as an explanation to the apparent accelerated expansion \cite{Celerier:1999hp,Tomita:2000jj,GarciaBellido:2008nz,Celerier:2007jc},...). Some modifications of gravity try even to unify the explanation of inflation, the present accelerated expansion and/or Dark Matter. We will briefly introduce also some models of modified gravity and of exotic sources of the gravitational field.

\paragraph{Scalar fields for inflation}
We have already mentioned that it is possible to drive the universe into an accelerated expansion era with the help of a set of scalar fields with an appropriate potential. These scalar fields might arise as fundamental fields in a QFT beyond the standard model (like in Ref.~\cite{Guth:1980zm}) or as new modes in a modification of general relativity (as in Ref.~\cite{Starobinsky:1980te}). As noticed in Ref.~\cite{Linde:1983gd}, the particular potential behind the accelerated expansion is unimportant, provided it satisfies some very general conditions, the 'slow-roll conditions'.

Since then, there has been much effort directed to find some viable inflation mechanisms in theories of quantum gravity. In particular, the string theory community has viewed the inflationary paradigm as a playground in which their ideas can be tested. Many possible realizations of inflationary scenarios have been found.

As ST lives in $10$ dimensions, it needs the compactification of $6$ extra dimensions in order to describe our world. The moduli, which are the parameters describing this compactification, are stabilized dynamically in the quantum theory, producing different vacua. There are numerous such solutions, with different values of the vacuum energy. The set of these vacua is called the moduli space of supersymmetric vacua. If we expand this set to include non-supersymmetric vacua\footnote{We are talking about $10^{500}$ possible vacua!}, we arrive to the `landscape' \cite{Susskind:2003kw}. It was first proposed as a way of solving the cosmological constant problem: if there are multiple vacua and the cosmological constant differs from one vacua to the other, and considering that human life is only possible if the value of this constant lies in a certain range, then we must be living in one of the 'anthropically permitted' vacua.

The landscape offers multiple options for inflationary model-building, such as hybrid inflation \cite{Linde:1993cn}. In this model, inflation is driven by two scalar fields, with a potential

\be
V(\phi,\chi) = \frac 12 m^2 \phi^2 +\frac 12 \chi^2 \phi^2 + \frac \lambda 4 (\chi^2 - \frac{M^2}4)^2  \, .
\ee

The main new ingredient in the hybrid inflation as compared to standard inflation is that the corresponding quantum fluctuations are non-gaussian due to the interaction. In contrast to single field inflation, hybrid inflation naturally gives $n_s \geq 1$ \cite{Bassett:2005xm}, and thus is ruled out as the mechanism realized in nature.

Other inflationary scenarios arise in the context of $Dp$-branes moving through extra dimensions. In the brane-world scenario, our spacetime corresponds to one of these $D3$-branes. The SM particles live stuck to the $D$-brane, for they are modes of open strings whose ends lie on it. Graviton and other open string modes are in contrast not confined. In 'brane-inflation' the inflaton corresponds simply to the position of this brane \cite{Dvali:1998pa}. A simple model is $D3$-brane/$\bar D3$-brane, in which brane and antibrane experiment an attractive force and finally annihilate each other, causing the end of inflation \cite{Burgess:2001fx}. Here the inflaton $\varphi$ is the relative position of brane and anti-brane and the potential $V(\varphi)$ is due to the tensions and interaction. Energy is released with the annihilation and cosmic superstrings are left as a remnant. More realistic but much more complicated is the KKLMMT scenario \cite{Kachru:2003sx}. In general, most of these models give $n_s \geq 0.98$ and predict the creation of a cosmic string network which has not been observed \cite{HenryTye:2006uv}, and are therefore ruled out.

However, despite the interest of string theorist, stringy scenarios are seen by phenomenologists and cosmologists as very degenerate and very little predictive. The idea of the landscape and the 'anthropic selection of vacua' is often received with skepticism and polemic.

\paragraph{Alternatives to inflation: VSL}
For more than a decade, inflationary cosmologies have not had any real competitor. However, in Ref.~\cite{Albrecht:1998ir} an alternative mechanism was first proposed under the name of Varying Speed of Light (VSL). The proposal is that the local speed of light, as measured by comoving observers, might have varied with time in the very early universe from a very large value to its current value. This implies obviously a departure from SR at very high energies, and therefore mediated by some UV scale.

Assume that in the preferred frame in which the universe is approximately homogeneous, the speed of light varies as $c(t)$. Therefore, $ds^2 = c(t)^2 dt^2 - a(t)^2 d\vx^2$ (although $g_{00}=1$) but the Friedman equations are still valid,

\bea
H^2 & = & \frac{8\pi G}3 \rho - \frac{k c(t)^2}{a^2} \, , \label{vslf1} \\
\frac{\ddot a}a & = & -\frac{4 \pi G}3 \left(\rho + 3\frac p{c(t)^2}\right) \, . \label{vslf2}
\eea

If the universe is radiation dominated, $p = \rho c(t)^2/3$ and thus $a \propto t^{1/2}$ as usual. As a consequence of the violation of invariance of time translations even at the local level, the energy conservation condition is modified,

\be
\dot\rho + 3H\left( \rho + c(t)^{-2}p\right) = \frac{3 k c(t) \dot c (t)}{4 \pi G a^2} \label{vslcont}
\ee

The horizon problem is straightforwardly explained within this model. The ratio of the event horizon to the causal horizon at $t = t_i$ is now given by

\be
\frac{l_i}{l_c}\sim \frac {c_0 t_0}{\int^{t_i}_0 c(t) dt }\frac{a_i}{a_0} \, ,
\ee
and thus can be made arbitrarily small with an appropriate choice of the function $c(t)$. A ``superluminal'' propagation of light at early times suffices. In the original paper \cite{Albrecht:1998ir}, a phase transition between an early value of the speed of light $c_i \sim 10^{30} c_0$ and the present one $c_0$ at $t \lesssim t_p$ is proposed.

The flatness problem can also be understood. Taking Eqs.~\eqref{vslf1},\eqref{vslf2},\eqref{vslcont} into account, we get that in a universe dominated by a perfect fluid with $p = \omega \rho c(t)^2$ the cosmological parameter $\Omega(t)$ evolves as

\be
\dot \Omega = \Omega (\Omega - 1) H (1+3\omega) +2 \frac{\dot c}c(\Omega -1) \, .
\ee

The inflationary cosmology solves the flatness problem because the inflaton field satisfies $1 + 3\omega < 0$ and thus $\Omega \approx 1$ is stable. The varying speed of light is able to solve the problem because the term proportional to $(1 + 3\omega)$ can be compensated by the term proportional to $\dot c$ as long as $\vert \frac{\dot c}c \vert > \Omega H$.

Some models providing concrete realizations of VSL can be found in Refs.~\cite{Moffat:1992ud,Albrecht:1998ir,Barrow:1999is} (see Ref.~\cite{Magueijo:2003gj} for a review). A covariant theory was presented in Ref.~\cite{Magueijo:2000zt}, as a gauged Fock-Lorentz symmetry theory in which $c$ is allowed to vary as a scalar field does in the manifold, $c(x)$. Another similar covariant realization is Rainbow Geometry \cite{Magueijo:2002xx}, in which the metric of the spacetime as seen by a test particle of a given energy depends on this energy. However, the interpretation of this scenario seems obscure. A more clear interpretation is found if the speed of light is wavelength dependent, $c(E)$, due to a modified dispersion relation. In the very early universe the wavelength of the photons becomes comparable to the Planck length or other UV scale and thus the global effect can be seen as VSL \cite{Alexander:2001ck}. This approach allows a deeper description of VSL cosmology \cite{Alexander:2001dr} and structure formation \cite{Magueijo:2002pg}. Note however that this idea is also not absent of criticism \cite{Ellis:2007ah}.

\paragraph{$f(R)$-theories}

There has been an interest in modifying GR since it was formulated \cite{Weyl:1919fi}, as a consequence of the success of GR in the description of the precession of the perihelion of Mercury. As a consequence of the difficulties in the renormalization of GR, extensions of the Einstein-Hilbert action with quadratic terms in the action (like in Eq.~\eqref{GRUV}) were studied; later it was proved that these corrections provided a mechanism of inflation \cite{Starobinsky:1980te}, as we have explained. Nonlinear corrections in the scalar curvature to the Einstein-Hilbert action have also found motivation in ST \cite{Candelas:1985en,Nojiri:2003rz}.

However, the motivation for extending the Einstein-Hilbert action can be much more mundane: there is no fundamental principle fixing the form of the gravitational action. Thus if we allow for new scales in a gravitational theory, new terms in the action will generally appear, which must respect general covariance. In this sense the gravitational action can be arbitrarily extended as in Eqs.~\eqref{GRUV} or \eqref{GRIR}. As a toy model one may consider actions which depend only on the curvature scalar,\footnote{More general models have also been considered. For instance, models which depend on the squared curvature scalars \cite{Carroll:2004de,Mena:2005ta} or which are nonlocal \cite{Wetterich:1997bz,Deser:2007jk,Capozziello:2008gu}.}

\be
 S=\int d^4 x\sqrt{-g}[\frac 1{16 \pi G}f(R) + L_m] \, .
\ee
These are called $f(R)$-theories \cite{Faraoni:2008mf}. In fact, one can distinguish between three types of $f(R)$-theories depending on how the variational principle is implemented on the action. In metric $f(R)$, the only gravitational field appearing in the action is the metric $g_{\mu\nu}$ and variation with respect to the metric provides the equations of motion. In Palatini $f(R)$ the scalar curvature is the result of the contraction of the inverse of the metric of the Ricci tensor, which in turns depends only on the connnection (treated as an independent field). The matter Lagrangian is considered to depend only on the metric, but not on the conection. In metric-affine $f(R)$, not only the Ricci tensor is made of the connection, but also the matter Lagrangian contains this field. Both of the three formalisms are equivalent if $f(R) = R + \Lambda$, but differ if $f''(R) \neq 0$. In this thesis mainly metric $f(R)$-theories will be treated.

In a metric $f(R)$ theory, the equations of motion are,

\be
 f'(R)R^\mu_\nu-\frac12
 f(R)\delta^\mu_\nu+(\delta^\mu_\nu\square-\nabla^\mu\nabla_\nu)f'(R)=8\pi G
 T^\mu_\nu \, ,
\ee
which, particularized to a FRW metric, become

\bea
8 \pi G \rho & = & -3(\dH+H^2)f'(\Ro)-\frac 12 f(\Ro) -3 H
f''(\Ro)\dRo  \, ,\\
-8 \pi G p & = & -(\dH + 3 H^2)f'(\Ro)-\frac 12 f(\Ro)
+f^{(3)}(\Ro)\dRo^2 + f''(\Ro)\ddRo  \, ,
\eea
where
\be
\Ro(t) = -6(\dH +2H^2)
\ee
is the curvature scalar of the Friedman-Robertson-Walker (FRW) metric. For a $f(R)$-type modification in the IR it is then possible to find mechanisms for the present acceleration of the universe \cite{Carroll:2003wy}; in the same way in a $f(R)$-type modification in the UV it is possible to find mechanisms for inflation \cite{Starobinsky:1980te}. It has also been tried to unify inflation and the present cosmic acceleration in the context of these theories \cite{Nojiri:2003ft,Sotiriou:2005hu}.

Metric $f(R)$ theories are formally equivalent to scalar-tensor theories (as we explained in the context of the Starobinski model), with a potential for the scalar field

\be
V(\varphi) = \frac 1{16\pi G}\frac{Rf'(R)-f(R)}{f'(R)^2}\, ,
\ee
where $R(\varphi)$ can be obtained from

\be
\varphi \equiv \sqrt{\frac 3{16\pi G}} log f'(R)\, .
\ee

The scalar presents an effective mass

\be
m^2 = \frac 12 \frac {d^2 V}{d \varphi^2} = \frac 16\left(\frac{R f'(R)-4 f(R)}{f'(R)^2} + \frac 1{f''(R)}\right)\, .
\ee

Thus, $f(R)$ can be seen as theories of a propagating massless spin-2 field (the graviton) together with a (generally massive) scalar field.\footnote{In more general modifications, more effective propagating degrees of freedom may arise.} This scalar field may act as an inflaton in the very early universe but becomes a problem when building an $f(R)$ theory consistent with solar system tests of gravity.

In order to present a viable alternative to $\Lambda$CDM, a modified gravity theory must pass a number of tests. In particular, an $f(R)$-theory must:

\begin{itemize}
\item
possess the correct cosmological dynamics. Many models present non-smooth transitions, for instance between the radiation dominated and matter dominated era, which were naively unexpected (for instance, see Ref.~\cite{Amendola:2006kh}). It has been also pointed out that some $f(R)$ theories may possess different future time singularities \cite{Abdalla:2004sw}, classified in Ref.~\cite{Nojiri:2005sx}. Future time singularities can be cured by adding $R^2$ terms to the action, as shown in \cite{Abdalla:2004sw}. Both of these problems may also be surmounted by starting from a given evolution $a(t)$ and integrating the ordinary differential equation that relates it to the family of $f(R)$ which present that solution \cite{Capozziello:2005ku}.
\item
not suffer from instabilities. The simple model in Ref.~\cite{Carroll:2003wy} was shortly found to have an instability \cite{Dolgov:2003px}. However this instability was later cured by the addition of new terms to the action relevant at high curvatures \cite{Nojiri:2003ft}, much like future time singularities. Metric $f(R)$-theories suffer from this instability if $f''(R)<0$ \cite{Faraoni:2006sy}. If this is the case the energy of the corresponding scalar field may become negative. This can be understood in terms of the effective gravitational coupling appearing in metric $f(R)$ theories $G_{eff} = G/f'(R)$. If $f''(R)<0$ then $d G_{eff}/d R >0$. Large curvature causes gravity to become stronger, so curvature further increases, causing a positive feedback mechanism. Also an unprotected curvature singularity arises in the nonlinear regime of IR modified gravity, which makes it difficult to describe the strong gravity regime of relativistic stars \cite{Frolov:2008uf}. In fact this singularities that appear in the spherically symmetric case are connected to the future singularities of the FRW spacetime mentioned above \cite{Nojiri:2009uu}, and are as well cured with the addition of new terms in the action which are relevant at high curvature \cite{Abdalla:2004sw}.
\item
be free from ghosts. Although $f(R)$ is ghost-free, more general modifications of gravity may contain ghost fields. Extensions containing also the Gauss-Bonnet scalar $\mathcal{G}= R^2 - 4 R_{\mu\nu}R^{\mu\nu} + R_{\mu\nu\rho\sigma}R^{\mu\nu\rho\sigma}$ present no ghosts \cite{Nunez:2004ji}.
\item
have the correct Newtonian and post Newtonian limit and be compatible with the data of CMB and LSS. These conditions have in common the fact that they depend on the behavior of inhomogeneous perturbations of a homogeneous background metric. In the conformally equivalent scalar-tensor theory, the problem is that the propagating scalar degree of freedom may spoil the behavior of this perturbations; it gives rise to a modification of the Newton's law and modifies cosmological perturbation dynamics \cite{Chiba:2003ir}. In most models, the mass of the scalar is naturally light ($ m < 10^{-3} \, eV$) and thus the effect on Solar System tests rules them out as viable gravity theories. However it must be taken into account that the effective mass of the scalar depends on the local curvature, and therefore differs in cosmological, Solar System and on Earth tests. This is known as the `chameleon effect' \cite{Khoury:2003aq,Khoury:2003rn}. Thus it is possible to have scalars modifying gravity at sub-millimeter scales in the Solar System (and therefore remaining undetected) while offering the long-ranged modified cosmological dynamics. Examples can be found in Refs.~\cite{Faulkner:2006ub,Hu:2007nk,Nojiri:2007as,Cognola:2007zu}. However it is not clear wether the physics of the conformally equivalent scalar-tensor theory always provides a good picture of the physics in the $f(R)$ theory \cite{Carloni:2009gp}. The behavior of scalar perturbations may change due to the redefiniton of fields done after the conformal transformation. The behavior of cosmological perturbations in metric $f(R)$-theories is still work in progress: it has been defined in the covariant approach \cite{Carloni:2007yv} and in the Newtonian gauge \cite{delaCruzDombriz:2008cp}. It is then clear that the evolution of the energy density involves a fourth order differential equation. A similar situation holds for Palatini $f(R)$-theories.
\item
have well-posed Cauchy problem. It has been shown that metric $f(R)$-theories present a well-posed Cauchy problem for several sources (perfect fluids, scalar fields) while Palatini $f(R)$-theories do not \cite{LanahanTremblay:2007sg}.
\end{itemize}

Remarkably, these problems do not only arise in modified gravity theories, but also have their equivalent in other approaches to the problem of acceleration like quintessence models or dark fluids. This can be trivially seen as one can write the effect of a mofification of gravity as a scalar-tensor theory (as shown above) or as a dark fluid model. In the same way as in singular $f(R)$ theories, the problem of singular quintessence or dark fluid models can be cured with the addition of terms with higher powers of curvature to the action, such as the ones that appear in the renormalization of field theories in curved space \cite{Nojiri:2009pf}.

To sum up, $f(R)$-theories are a simple toy model which shows the difficulties of departing from GR in a consistent way and in agreement with the experimental and observational data. This is possible under certain conditions that impose a great fine-tuning of the function $f(R)$. The origin of the constraints is the presence of new effective propagating degrees of freedom, i.e. the presence of differential equations of order greater than two.

This is a general problem of theories beyond GR. We will try to surmount this problem (at least in an effective way) in the formulation of the Asymptotic Cosmological Model (ACM) in one of the works presented in this PhD thesis, and in the treatment of cosmological perturbations in the ACM.

\paragraph{Higher Dimensional Models}
Inspired by ST, it has been proposed that our universe might be a four-dimensional hypersurface (a 3-brane) embedded in a higher dimensional bulk spacetime. The SM fields are considered to live in the brane while the gravitational field (or fields) are considered to live in the whole bulk spacetime. The question is how to recover the standard Newton's law instead of its higher dimensional version in the brane. This can be achieved through compactification of the extra dimension \cite{ArkaniHamed:1998rs,Antoniadis:1998ig,ArkaniHamed:1998kx}, whose size can be much larger than in the case of Kaluza-Klein models; thus this approach is called `Large Extra Dimensions'. This can also be achieved by embedding the brane with positive tension into an Anti-deSitter bulk \cite{Randall:1999ee,Randall:1999vf}; this approach is called `Warped Extra Dimensions'. In both cases, the extra dimension is finite and has a size of the order of the millimeter.

Dvali {\it et al.} proposed another solution: it suffices to add to the action of the brane an Einstein-Hilbert term computed with the intrinsic curvature of the brane. This term is necessary for the renormalizability of the quantum field theory in the brane interacting with the bulk gravity. Thus the standard Newton's law is recovered even for infinite size extra dimensions \cite{Dvali:2000hr}. The case of a 3-brane embedded in a 4+1 Minkowski spacetime is of special interest. We will explain this model in the lines below.

Consider the action

\be
S_{(5)} = \frac 1{2\kappa^2}\int d^5 X \sqrt{-\bar{g}}\bar{R} + \int d^4 x \sqrt{-g}(\lambda_b + l_m) +\frac 1{2\mu^2}\int d^4 x \sqrt{-g}R \, ,
\ee
where the first term corresponds to the Einstein-Hilbert action in 4+1 dimensions, with a five-dimensional bulk metric $\bar{g}_{AB}$ and curvature scalar $\bar{R}$; the third term corresponds to the Einstein-Hilbert action of the induced metric
$g_{ab}$ on the brane, with its corresponding intrinsic curvature scalar $R$; the second term corresponds to the matter action, which might include a brane tension (an effective cosmological constant) $\lambda_b$. The following quantities can be defined,

\bea
\kappa^2 & = & 8\pi G_{(5)} = M^{-3}_{(5)} \, \\
\mu^2 & = & 8\pi G_{(4)} = M^{-2}_{(4)}\, .
\eea

In Ref.~\cite{Dvali:2000hr}, it was shown that the Newtonian potential is recovered in the brane for distances smaller than $r_0 = M_{(4)}^2/2 M_{(5)}^3$, with an effective Newton's constant $G_N = \frac 43 G_{(4)}$.

The cosmology of the model was derived by Deffayet in Ref.~\cite{Deffayet:2000uy}. Consider the brane-isotropic metric

\be
ds^2 = n^2(\tau,y)d\tau^2 -a^2(\tau,y)\gamma_{ij}dx^i dx^j - b^2(\tau,y)dy^2 \, ,
\ee
where $\{\tau,\vx\}$ are coordinates on the brane ($y=0$) and $\gamma_{ij}$ is the maximally symmetric 3-metric (with spatial curvature $k$). The stress-energy tensor of the bulk is reduced to a possible bulk cosmological constant $\rho_B$, while the stress-energy tensor of the brane can be expressed as

\be
T^A_B = \frac{\delta(y)}b diag(\rho_b,-p_b,-p_b,-p_b,0) \, .
\ee

Assuming that the metric $\bar{g}_{AB}$ is symmetric under the change $y \rightarrow -y$, the corresponding Friedman equation reads

\be
\pm \sqrt{H^2 - \frac{\kappa^2}6\rho_B - \frac C{a_0^4} + \frac k{a_0^2}} = r_0 \left( H^2 + \frac k{a_0^2} \right) - \frac {\kappa^2}6 \rho_b \, ,
\ee
where $C$ is a constant of integration, the subscript $0$ means evaluation at $y = 0$, and $H = \frac{\dot a_0}{n_0 a_0}$, with dot denoting derivation with respect to the time variable $\tau$. The restriction of the fields to the brane also leads to the continuity equation for the brane, $\dot{\rho}_b + 3(\rho_b + p_b)\frac{\dot a_0}{a_0} = 0$ .

If we choose the $+$ sign and neglect $\rho_B$, $C$ and $k$ for simplicity, we arrive to the modified Friedman equation

\be
\frac{8\pi G_{(4)}}{3} \rho_b = H^2 - H/r_0 \, ,
\ee
which produces a late accelerated expansion tending asymptotically to an exponential expansion with $H = 1/r_0$, as in the case of $\Lambda$CDM. However, this modified Friedman equation is distinguishable from the one appearing in $\Lambda$CDM,

\be
\frac{8\pi G_N} {3} \rho = H^2 - \Lambda/3 \, . \label{DGPFried}
\ee

Current constraints on the expansion history from supernova luminosity distances, the CMB, and the Hubble constant exclude the simplest flat DGP model at about $3\sigma$. Even including spatial curvature, best-fit open DGP model is a marginally poorer fit to the data than flat $\Lambda$CDM. Moreover, its substantially different expansion history raises serious challenges for the model from structure formation \cite{Song:2006jk}.

Moreover, the DGP model is also plagued with anomalies \cite{Charmousis:2006pn,Gregory:2007xy}. First, it was discovered that ghosts are present in perturbation theory \cite{Nicolis:2004qq}. Second, if general bulk metrics are allowed, then the energy spectrum of the solutions is unbounded from below (negative mass blackholes are allowed) and therefore the theory is unstable \cite{Gregory:2007xy}. Furthermore there are solutions (positive mass black holes) in which `pressure singularities' are present: pressure diverges for finite energy density. Third, the DGP vacuum is unstable as it may tunnel to alternative vacua \cite{Gregory:2007xy}.

Thus, we will not consider DGP as a viable alternative to explain the current accelerated expansion. However, it is interesting how a simple modified Friedman equation such as \eqref{DGPFried} is able to produce an accelerated expansion in the absence of strange fluids.

\paragraph{Quintessence}

Aside from modifications of gravity such as the ones described above, it is also possible to explain the present acceleration of the universe as a consequence of the existence of a new form of matter, different form massless or massive particles, which presents at least at later cosmic times an effective negative pressure.

There are several ways of implementing this idea, but one of the most popular ones is to introduce phenomenologically a self-interacting scalar field. Extensions of GR including scalar fields have a long history \cite{Brans:1961sx}. This mechanism has also been used in QFT in order to explain some of the unnaturally small parameters of the SM \cite{Peccei:1977hh,Peccei:1977ur}. In analogy, it is possible to build dynamical scalar field actions which mimic the effect of a cosmological constant, and which could render its value small \cite{Wetterich:1987fk,Wetterich:1987fm}. However, this would imply a new force mediated by the scalar field, despite new interactions mediated by light scalar fields are highly constrained \cite{Turyshev:2008dr}. Thus there must be a fine-tuning in the action at least in the effective mass of the scalar or its coupling to observed matter.

Typically the action of one of such `quintessence' theories has the form

\be
S = \intxc \sqrt{-g} \left[ \frac 1{16 \pi G} R + \frac 12 g^{\mu\nu}\partial_\mu \varphi \partial_\nu \varphi - V(\varphi) +\Lag_m [g^{\mu\nu},\varphi] \right] \, ,
\ee
where $\varphi$ is the quintessence scalar field\footnote{This scalar field is given a variety of names, depending on its role in the particular model in which it is introduced. It may be the dilaton (the Goldstone mode associated to the spontaneous breaking of dilatation symmetry), the cosmon (a particular case of the dilaton), the inflaton (if we consider a model of inflation),...}, $V$ is the effective self-interaction potential, which may be induced in a more fundamental theory, and $\Lag_m$ is the matter Lagrangian, which may include couplings between ordinary matter and the quintessence.

In the homogeneous approximation, in which the metric is FRW and the quintessence and matter fields are decoupled and depend only on the time variable, the equations of motion become

\begin{eqnarray}
 \ddot{\varphi}+3 H \dot{\varphi}+V'(\varphi)&=0 \\
\frac{8\pi G_N}{3}(\rho+\frac{\dot{\varphi}^2}2 +V(\varphi))&=H^2\\
-3(1+\omega)H\rho&=\dot{\rho} \, .
\end{eqnarray}

Note that if the quintessence currently varies sufficiently slow with time ($\dot \varphi^2 \ll V(\varphi) $ ), then the quintessence is approximately frozen at its present value $\varphi_0$ and acts as an effective cosmological constant $\Lambda = 8 \pi G V(\varphi_0)$; for a sufficiently slow variation, the negative effective pressure $p_\varphi = \frac{\dot{\varphi}^2}2 - V(\varphi)$ required by the accelerated expansion is guarantied \cite{Wetterich:1987fm,Ratra:1987rm,Caldwell:1997ii}.

Some specific quintessence models deserve more attention. One of this classes is `tracker models' \cite{Zlatev:1998tr,Steinhardt:1999nw}, which try to explain the coincidence problem\footnote{Why is the value of the vacuum energy density of the same order that the matter energy density today?}. The idea underlying tracker models is the following: the energy density of quintessence component somehow tracks below the background density for most of the history of the universe, falling from a Planckian value in the very early universe with the expansion history. Only recently it grows and becomes the dominant energy density, driving the universe to its current accelerated expansion. The tracker field $\varphi$ rolls down the potential $V(\varphi)$ following an attractor-like solution. Examples of tracker potentials are exponentials $V(\varphi) = M^4 (e^{M_p/\varphi}-1)$ and inverse power laws $V(\varphi) = M^{4+n}\varphi^{-n} $.

For instance for inverse power law potentials $V(\varphi) = M^{4+n}\varphi^{-n} $ it is possible to arrive to a quintessence-dominated period when $\varphi\sim M_p$. The value of the scale $M$ must be fit by observations (typically $M \gtrsim 1 \, GeV$). For a given background component with equation of state $\omega_B \equiv \rho_B/ p_B$, the subdominant quintessence then adapts itself and adopts an equation of state

\be
\omega_\varphi \approx \frac{n \omega_B /2 -1}{n/2 +1} \, .
\ee

Thus for high $n$ if quintessence is subdominant it mimics the equation of state of radiation and its energy density decreases at the same rate. When matter takes over radiation, the pressure of quintessence becomes negative and later on begins to dominate. However for these potentials the condition $\dot \varphi^2 \ll V(\varphi) $ never holds and thus they are not able to reproduce equation of state of a cosmological constant. Thus the model is disfavored by observations. The coincidence problem is also just moved from the choice of the value of $\Lambda$ in $\Lambda$CDM to the choice of the scale $M$.

Other interesting type of quintessence models are k-essence models, in which the kinetic term of the scalar field is non-standard \cite{ArmendarizPicon:2000dh}. Such kinetic terms may arise in ST \cite{Gross:1986iv}. Typically the action of a k-essence model can be written as

\be
S = \intxc \sqrt{-g} \left[ \frac 1{16 \pi G} R + \frac 1{\varphi^2} f(X) +\Lag_m [g^{\mu\nu},\varphi] \right] \, ,
\ee
where $f(X)$ is a certain function of $X\equiv g^{\mu\nu}\partial_\mu \varphi \partial_\nu \varphi$. The energy density and pressure of the k-essence read

\bea
\rho_k & = & (2 X f'(X) - f(X) )/\varphi^2 \, ,\\
p_k & = & f(X)/\varphi^2 \, .
\eea

As shown in Ref.~\cite{ArmendarizPicon:2000dh}, k-essence models are able to reproduce the behavior of tracker models in the early universe, while showing a behavior more similar to a cosmological constant in the present universe. However, these models are quite {\it ad hoc} and do not fit the data as well as an effective cosmological constant does. Moreover they do not really solve the coincidence problem, as the transition from a matter dominated to a k-essence dominated universe occurs just as in $\Lambda$CDM.

Similar tracker behaviors arise much more naturally in vector-tensor models such as the one proposed in Ref.~\cite{Jimenez:2008au}. However, such a model presents an undesirable tension when compared to SNe, CMB and BAO observations \cite{Jimenez:2009py}.

\paragraph{Dark Energy}

When phenomenologically describing the exotic component of the universe either with a self-interacting scalar field, a k-essence or other types of fields, we are making an {\it a priori} assumption (the field content of our model) that is going to condition our following analysis. For this reason, it may be better from a phenomenological point of view to consider just the essential features of a new matter component, independently of a field theory representation. The essential ingredient of the matter content that must be characterized in GR is just the stress-energy tensor.

In the homogeneous approximation, the only two nonzero elements of the stress-energy tensor are the energy density $\rho_X$ and the pressure $p_X$. Thus, the effective description of a new fluid component amounts to the relation between this quantities, the equation of state $p_X = p_X (\rho_X)$. The most simple parametrization of this relation is a linear equation of state $p = \omega \rho$ \cite{Turner:1998ex}; \footnote{Despite it is a flagrant abuse of language, $\omega$ is often referred to as the equation of state. } for instance radiation has $\omega = 1/3$, massive matter has $\omega = 0$ (strictly speaking $\rho \gg p$) and a vacuum energy has $\omega = -1$. Naturally, this parametrization is also the most used in data analyses, such as \cite{Spergel:2003cb,Astier:2005qq,Spergel:2006hy,Dunkley:2008ie}. In contrast, an example of a nonlinear equation of state could be a gas of strings, the Chaplygin gas  $p_X=-K/\rho_X$ \cite{Kamenshchik:2001cp,Alvarenga:2001nm}, which is able to mimic the dark matter at high densities but leads to an accelerated expansion at low densities.

The relation between the pressure and the energy density does not contain all the information about the fluid if one considers the evolution of inhomogeneous perturbations \cite{Hu:1998kj}. Then, it is necessary to consider the scalar part of the non-diagonal terms of the stress-energy tensor, i.e.: the momentum density $(\rho_X + p_X)\bm u_X = (\rho_X + p_X)\bm \nabla \theta_X $ and anisotropic stress $\Pi_{Xij} = \partial_i\partial_j \Pi_X$. A complete description of the fluid should involve an expression relating the pressure and anisotropic stress as a function (or a functional) of the energy density and the momentum density of the fluid at each point.

This description is of course too general to be predictive, and one usually restricts oneself to the use of perfect fluids ($\Pi = 0$) with a certain $p(\rho)$ equation of state. However, the effect of some specific anisotropic stress has been investigated in the literature (see for instance Refs.~\cite{Hu:1998kj,Capozziello:2005pa,Ichiki:2007vn}).

In light of the vast amount of concrete models, a phenomenological description of the equation of state has been proposed as a way of parameterizing the present acceleration of the universe, such as

\be
p_X = \left[\omega_0 + \omega_a(1-a) + \mathcal{O}\left((1-a)^2\right)\right] \rho_X \, ,
\ee
where the normalization $a_0 = 1$ is taken. Although we agree with the bottom-up spirit of such parameterizations, this choice is very anthropocentric and sheds little light on the physics of the dark component or gravitational physics behind the expansion. Is it possible to make a better choice?

\subsection{The Asymptotic Cosmological Model}

The first thing one should note when thinking about an extension of the cosmological standard model is that it is difficult (impossible?) to distinguish by means of gravitational physics between modified gravitational physics and new forms of matter. This is due to the Einstein equation. Any modification of the Einstein tensor can be moved to the rhs of the equation and become a modification to the stress-energy tensor and viceversa. This term will have a null covariant divergence as a consequence of being derived from a scalar action, either from the gravitational or the matter part of the Lagrangian. Thus, we will consider the departure to be encoded in the gravitational part of the Lagrangian without loss of (phenomenological) generality.

Secondly, in order to build a concrete model we have to make a number of assumptions based on phenomenological grounds. We have mentioned that it is experimentally well established that gravity should be a metric theory \cite{Bassett:2003vu,Sahni:2006pa,DeBernardis:2006ii} and thus we will assume our model to be a metric theory. We will also consider that the cosmological principle is a good approximation at sufficiently large scales, so that the FRW metric will be a good approximation to the metric of spacetime. The FRW metric depends just on the spatial curvature of spacetime, $k$. In $\Lambda$CDM, $k$ is severely constrained by the observations to be zero, so in our model we will assume that $k = 0$.

In GR with no spatial curvature, the evolution of the universe in the homogeneous approximation is obtained by an algebraic relation between the Hubble rate and the energy density. Furthermore if we assume that $\dot H < 0$, then there is a one-to-one correspondence between $t$ and $H$, and thus we can always find an algebraic relation between the energy density (of a fluid composed of SM particles and DM) and the Hubble parameter. Therefore we are going to define our model through the relation between $H$ and $\rho$.

Third, we will consider that the standard relation between $H$ and $\rho$, the first Friedman equation \eqref{11fried}, will be valid just in a window limited by the presence of an UV and IR scale. From the point of view of the phenomenologist, every theory must be valid in the regime in which it has been tested, but might fail in the regime in which it has not. GR has been well tested in a range from the millimeter scale to the galactic scale and QFT has been tested from the $eV$ to the $TeV$ scale, and thus they might fail beyond this scales. In our model, the effect of the new UV and IR scales will be responsible of the accelerated expansion in the very early universe and the present epoch, respectively.

We will take as a basic principle that the Hubble rate is constrained to a bounded interval between an upper scale $H_+$ and a lower scale $H_-$. Therefore in the infinite past and infinite future the universe will asymptotically tend to the deSitter spacetime. We have called this model the Asymptotic Cosmological Model.

\section{Neutrino Oscillations}

The development of QFT has been the most active sector of theoretical physics in the past century. Quantum Electrodynamics (QED) has provided the most precise predictions for experiments which have later confirmed those predictions figure by figure. The Standard Model of particle physics (SM), built up through a continuous interaction with experiment, has managed to provide a framework in which to understand the electromagnetic, weak and strong interactions, and to propose a mechanism to explain the apparition of mass (yet to be confirmed).

The model consists on a $SU(3)\times SU(2) \times U(1)$ gauge theory, in which the $SU(2)\times U(1)$ symmetry is ``broken'' through a Higgs mechanism by the vacuum expectation value (vev) of a scalar field. The spinors of the theory are divided in three families (or flavors) and can be either quarks ($SU(3)$-interacting) or leptons ($SU(3)$-noninteracting), and their masses are attained by their Yukawa interactions with the scalar field.

Let us denote the fields in the standard model by

\be
\ba{ccc}
\Psi_{L \alpha}=\left(\begin{array}{c}\nu_{\alpha} \\
  l_{\alpha}\end{array}\right)_L, &  L_\alpha = (l_\alpha)_R, & \\

Q_{L \alpha}=\left(\begin{array}{c}u_{\alpha} \\
  d_{\alpha}\end{array}\right)_L, &  U_\alpha = (u_\alpha)_R, & D_\alpha = (d_\alpha)_R, \\

  \Phi=\left(\begin{array}{c}\varphi^+ \\
  \varphi^0 \end{array}\right), &
  \tilde{\Phi}=\left(\begin{array}{c}\varphi^{0*} \\
  -\varphi^- \end{array}\right), & \\

  G_\mu, & W_\mu, & B_\mu,
\ea
\ee
where in the first line we have denoted the leptons, in the second line the quarks, in the third line the Higgs boson and its conjugate, and in the last line the gauge fields carrying the $SU(3)$, $SU(2)$ and $U(1)$ interactions respectively, whose respective field strengths will be denoted $G_{\mu\nu}$, $W_{\mu\nu}$ and $B_{\mu\nu}$. The index $\alpha=1,2,3$ labels the flavor, the coupling of the gauge fields are respectively $g_3$, $g$ and $g'$. The covariant derivatives will be denoted as $\nabla_\mu$, and $A  \ddmu B \equiv A \nabla_\mu B - (\nabla_\mu A) B$. The Yukawa couplings will be denoted $Y_L$, $Y_U$, $Y_D$. Then, the Lagrangian of the SM is just the sum of the following terms:

\bea
\Lag_{lepton} & = & \frac i2 \bar{\Psi}_{L \alpha}\gamma^\mu \ddmu \Psi_{L \alpha} + \frac i2 \bar{L}_\alpha\gamma^\mu \ddmu L_\alpha\, , \\
\Lag_{quark} & = & \frac i2 \bar{Q}_\alpha\gamma^\mu \ddmu Q_\alpha + \frac i2 \bar{U}_\alpha\gamma^\mu \ddmu U_\alpha + \frac i2 \bar{D}_\alpha\gamma^\mu \ddmu D_\alpha\, , \\
\Lag_{Yukawa} & = & -[(Y_L)_{\alpha\beta}\bar{\Psi}_{L \alpha}\Phi L_\beta + (Y_U)_{\alpha\beta}\bar{Q}_\alpha\tilde{\Phi}U_\beta +(Y_D)_{\alpha\beta}\bar{Q}_\alpha\Phi D_\beta ] + h.c. \, ,\\
\Lag_{Higgs} & = & (\nabla_\mu \Phi)^\dagger \nabla^\mu \Phi + \mu^2 \Phi^\dagger \Phi - \frac \lambda{3!}(\Phi^\dagger \Phi)^2 \, , \\
\Lag_{gauge} & = & -\frac 12 Tr(G_{\mu\nu}G^{\mu\nu})  -\frac 12 Tr(W_{\mu\nu}W^{\mu\nu})  -\frac 14 B_{\mu\nu}B^{\mu\nu}\, .
\eea

In the SM, the scalar field acquires a vev through the Higgs mechanism, and thus the Yukawa terms turn into mass terms for the quarks $u$, $d$ and the electrically charged leptons $l$, but not the neutrinos $\nu$. Therefore, the neutrinos are massless electrically neutral particles which only interact weakly (exchanging $W^\pm$ or $Z$ bosons, which are combinations of the $W$ and $B$ bosons). In a given experiment involving a charged lepton $l_\alpha$, the presence of a neutrino $\nu_\alpha$ as a final state can be detected as a violation of energy-momentum conservation.

These particles are generally detected in deep underground laboratories, in which the presence of more intensely interacting particles has been damped by the Earth crust\footnote{This is due to the fact that at energies below the electroweak scale neutrinos interact very weakly and otherwise the signal to noise ratio is small. However it is possible to detect neutrinos in accelerator experiments in which the source is very luminous, of high energy and controllable. However, detection of accelerator neutrinos at installations far away the source also need to be underground in order to improve the signal to noise ratio}. Their detection usually involves the exchange of a $W^\pm$ boson and the detection of the charged lepton $l_\alpha$ associated to the incoming neutrino $\nu_\alpha$.

Therefore, if a $\nu_\alpha$ is produced somewhere in a process involving $l_\alpha$, the massless and weakly interacting neutrino can propagate huge distances until being detected in another process involving the same $l_\alpha$. This was the idea underlying
the Homestake experiment headed by Raymond Davis Jr. and John N. Bahcall \cite{Bahcall:1976zz}. The experiment designed by Davis managed to measure the flux of $\nu_e$ coming from the sun as a result of the nuclear reactions happening at its core. The aim of the experiment was to confirm the theoretical predictions, derived by Bahcall, of the Standard Solar Model. However, the experiment measured around one third of the expected neutrinos! This was named the `solar neutrino problem'. At that time it was impossible to decide whether the problem was in neutrino physics or in the current understanding of the sun.

\subsection{A conventional explanation: neutrinos do have mass!}

It was first proposed by Pontecorvo \cite{Pontecorvo:1957qd} that neutrinos may change their nature (flavor) along their propagation through long distances in the same sense the neutral kaons do. He named this process `neutrino oscillations'. This explanation fitted the results of the Homestake experiment well but has profound implications \cite{Pontecorvo:1967fh}.

This phenomenon is shocking because in the context of the SM it cannot take place. When a neutrino is produced in a certain interaction within the SM of particle physics, it is produced by definition in a state which is a flavor eigenstate of definite momentum ($\nu_e$ if the interaction involves an electron, $\nu_\mu$ if the interaction involves a muon, or $\nu_\tau$ if the interaction involves a $\tau$). In the SM, the energy of these three states $E_\alpha$ is degenerate and depends only on their momentum, $E_\alpha = p$. Thus if a neutrino is produced in some flavor eigenstate, this is also a propagation eigenstate and it will be detected in the same flavor eigenstate, i.e.: its interaction with matter will produce the same charged lepton that was involved in its creation.

For neutrino oscillations to occur, two circumstances must concur. First, it is necessary that there exist different eigenstates of the effective Hamiltonian of the neutrinos with different energies $E_i$ for the same linear momentum $\vp$. Let us call the neutrino propagation eigenstates $\{\nu_1,\nu_2,\nu_3\}$ with energies $\{ E_1(p),E_2(p),E_3(p)\}$. In a Lorentz invariant theory the only possibility of the neutrinos to have different energy eigenstates is that they have different masses $m_i$: $E_i = \sqrt{\vp^2 + m_i^2}$. Therefore it is often stated that neutrino oscillations imply that the neutrinos are massive.

Second, it is necessary that these eigenstates do not coincide with the flavor eigenstates. The reason is that the eigenstates of the effective Hamiltonian (states of definite energy for a definite momentum) coincide with the eigenstates of the propagation, but those must not coincide with the eigenstates of definite flavor for a mixing to happen. This is quite natural in the context of QFT, and we already have an example in the flavor physics of quarks, parameterized by the Cabibbo-Kobayashi-Maskawa matrix connecting mass eigenstates of negatively charged quarks with their ``flavor'' eigenstates, defined as the eigenstates resulting of their charged current weak interaction with a positively charged quark. In analogy, a unitary matrix (called the Pontecorvo-Maki-Nakagawa-Sakata or PMNS matrix \cite{Maki:1962mu}) connects the mass eigenstates of neutrinos $\nu_i$ with the flavor eigenstates $\nu_\alpha$,

\be
\nu_\alpha = U_{\alpha i} \nu_i \, ,
\ee
which are defined as the states resulting of the charged current weak interaction of a charged lepton $l_\alpha$. This state is a linear combination of states with definite and different energies, which propagate in a definite way.

However, if there exists different states with different energies, it is no longer true that the neutrino is produced in a flavor eigenstate, due to energy-momentum conservation (unless the laws of energy-momentum conservation are only an approximation for sufficiently high energy-momentum transfers). If we reject this possibility, what happens is that in the production, a $\nu_1,\nu_2$ or $\nu_3$ can appear with different probabilities of sum one, which depends of the nature of the charged lepton involved in the production. Similarly, any of this neutrinos will produce, in the interaction with a detector, one of any of the three charged leptons, with different probabilities depending on the nature of the neutrino which has been detected.

It is assumed that this mass is an infrared (IR) scale, i.e.: it is much lower than the energy of the neutrino, and therefore the difference of energy between two mass eigenstates will be much lower than the total energy of any of these states, and the oscillations are assumed to take a significant place in neutrinos propagating through very long distances. Also, the effect of the mass in the total energy of the neutrino is often neglected, and it is commonly stated that neutrinos are produced and detected in states of definite flavor.

When we define a particle in a theory with a symmetry group, it is characterized by the eigenvalue of the invariant (under the symmetry group) operators that can be defined combining elements of the symmetry group. In the Lorentz invariant case, these operators are the mass and the spin of a particle. The particles also have definite charges under the gauge group. In the case of neutrinos, these states are not the flavor eigenstates, as they do not have a definite mass, but the propagation eigenstates $\{\nu_1,\nu_2,\nu_3\}$. Therefore, although flavor eigenstates can formally be defined in the context of a theory exhibiting neutrino oscillations it is an abuse of notation to call them particles.

Therefore it has been necessary to extend the SM of particle physics to include at least mass effects. The Standard Model of Neutrino oscillations assumes that there exist a new field, the sterile neutrino $\nu^R_i$, which is a singlet of the entire gauge group and which is necessary to include perturbatively renormalizable mass terms for the neutrinos in the Lagrangian (see Ref.~\cite{GonzalezGarcia:2007ib} for a review). The new terms that are added to the SM Lagrangian are:

\be
-\Lag_{\nu_R}  =  \bar\nu^R_i \gamma^\mu \partial_\mu \nu^R_i
 + Y^D_{i\alpha} \tilde \Phi^\dagger \bar \nu^R_i \gamma^\mu \partial_\mu \Psi^L_\alpha
  + m^M_{ij} \bar \nu^R_i \nu^{R,c}_j  + h.c.\, , \label{righthand}
\ee
where $\nu^R$ is the sterile neutrino singlet, with flavor index $i$ running from $1$ to $n$; $Y^D$ is an adimensional matrix of coefficients in flavor space, which will give rise to Dirac mass terms after the Higgs acquires its expectation value; and $m^M$ is a Majorana mass matrix allowed by all the symmetries (the only of such terms in the whole lagrangian).

This terms lead to the so-called `see-saw mechanism'. Let us assume that sterile neutrinos are a signal of new physics at some UV scale $E_{NP}$ a few orders of magnitude below the Planck scale. Then, the $m^M$ matrix has components naturally of the same order of this scale, while the Dirac mass terms associated to neutrinos will be naturally of the order of the vev of the scalar field. Then, the diagonalization of the resulting total mass matrix lead to $3+n$ eigenstates with different masses. $n$ of them are mostly sterile, have a mass comparable to the UV scale, and are very weakly coupled to the charged leptons. The other three are mostly lefthanded, have a mass of order $v^2/E_{NP}$. \footnote{This idea fits well in the frame of GUT theories with symmetry group $SU(5)$, in which the GUT scale is $M_{GUT}\sim 10^{15}\, GeV$; however, the absence of proton decay \cite{Babu:1998wi} within the predicted window has ruled out the simplest of these theories.} All of these particles are Majorana fermions, i.e.: they are their own antiparticles, in contrast with Dirac fermions like the charged leptons or quarks.

In this case the SM with Majorana neutrinos is a good effective field theory (EFT) \cite{Burgess:2007pt} at energies below the almost sterile neutrino mass. In this case operators of dimension greater than $4$ cannot be omitted, but they are corrected with an appropriate negative power of the UV scale. Then the first corrections to the SM should come from terms of dimension $5$. If we compute all the $5$-order terms compatible with the Poincar\'e and gauge symmetry, the only new term which is added is

\be
-\Lag_{NP} = \frac {Z_{ij}} {E_{NP}} \bar \Psi^L_i \tilde \Phi \tilde \Phi^T \Psi^{L,c}_j +h.c. \, .
\ee

In fact, this result tells us that if we are expecting any new physics at a scale $E_{NP}$ we should expect that this new physics (not necessarily sterile neutrinos) provides the neutrinos with an effective mass term.

The proposal that massive neutrinos should suffice to explain the solar neutrino problem in terms of neutrino oscillations in the propagation of neutrinos from the Sun to the earth needed only to be contrasted with more experimental data. The so-called solar neutrino problem was further studied with similar detectors as Kamiokande, SAGE, GALLEX, Superkamiokande and the Sudbury Neutrino Observatory (SNO). SNO was the first experiment that proved that neutrino oscillations were behind the solar neutrino problem \cite{Ahmad:2002jz}, although neutrino oscillations were first detected by Superkamiokande \cite{Fukuda:1998mi}. In the latter case, the oscillation was produced in the neutrinos produced by muon decay in the atmosphere (atmospheric neutrinos). Neutrino oscillations produced in the propagation of neutrinos generated in artificial nuclear reactors \cite{Eguchi:2002dm} and in nuclear decays in the Earth's crust and mantle (geo-neutrinos) \cite{Araki:2005qa} have been also measured by KamLAND.

The aforementioned experiments have produced a vast amount of data about neutrino oscillations. Solar neutrino oscillation are affected also by the so-called MSW effect (neutrino oscillations due to its interaction with matter), so they do not provide a clean picture of vacuum neutrino oscillations. Atmospheric neutrino experiments have observed $\nu_\mu$ disappearance with $\Delta m^2 \sim 3 \times 10^{-3} eV^2$ \cite{Fukuda:1998mi}, while reactor antineutrino experiments have measured $\bar\nu_e$ disappearance at $\Delta m^2 \sim 8 \times 10^{-5} eV^2$ \cite{Eguchi:2002dm,Araki:2005qa}. The minimum joint description of solar, atmospheric and reactor neutrinos requires Dirac neutrinos with different masses

\bea
m^2_1 -m^2_2 & = & 7.9 \pm 1.1 \times 10^{-5} eV^2 \\
\vert m^2_1 -m^2_3 \vert & = & 2.6 \pm 0.6 \times 10^{-3} eV^2
\eea
and the four independent degrees of freedom in the PMNS matrix\footnote{This is true for Dirac neutrinos. In the case of Majorana neutrinos there are six independent parameters in the PMNS matrix.}, which have been constrained \cite{GonzalezGarcia:2007ib}. In fact, the phenomenology involving massive Dirac neutrinos depends only in the squared absolute value of the elements of this matrix, which we will call populations and denote by

\be
r_{i\alpha} = \vert U_{\alpha i} \vert^2 = \left(
\ba{ccc}
0.69 \pm 0.03 & 0.31\pm 0.03 & < 0.01 \\ 0.16\pm 0.07 & 0.36 \pm 0.11 & 0.46 \pm 0.11 \\ 0.14 \pm 0.08 & 0.30 \pm 0.10 & 0.52 \pm 0.12
\ea
\right) \, .
\ee

The population $r_{i\alpha}$ represent the portion of a state $\nu_\alpha$ in a state $\nu_i$. Although there exist nine populations in a three-mixing problem, the sum-one condition makes only four of them linearly independent. In the Standard Model of Neutrino Oscillations these parameters are constants, but in a more general oscillation problem they could be functions of the momentum $\vp$.

It seems that massive neutrinos fit the current experimental data well. What is then the problem of this explanation?

\subsection{Doubts arising with massive neutrinos}

\paragraph{The LSND anomaly and other suggestive experimental results}

However, not all the ongoing experiments have their results well described by this simple picture. The Liquid Scintillator Neutrino Detector (LSND) reported evidence for $\bar \nu_\mu \rightarrow \bar \nu_e$ at $\Delta m^2 \sim 1 eV^2$ \cite{Athanassopoulos:1995iw,Athanassopoulos:1996jb}. This results have later been confirmed by the same group \cite{Athanassopoulos:1997pv,Aguilar:2001ty}, which has spent a huge effort in finding sources of experimental error. The final sample tests both $\bar \nu_\mu \rightarrow \bar \nu_e$ and $\nu_\mu \rightarrow \nu_e$ oscillations in a window of $20\, MeV < p < 200 \, MeV$.

In order to confirm or deny the results of the LSND collaboration, the KARMEN and MiniBOone collaborations have been running similar experiments. The MiniBOone collaboration has reported no evidence of $\nu_\mu \rightarrow \nu_e$ oscillations for neutrino beams with $p > 475 \, MeV$ \cite{AguilarArevalo:2007it,AguilarArevalo:2008rc} but has found a tiny excess of $\nu_e$ in beams with $200 \, MeV > p > 475 \, MeV$ \cite{AguilarArevalo:2008rc} (although this excess is incompatible with the oscillation mechanism proposed by LSND). The same group has also performed a $\bar \nu_\mu \rightarrow \bar \nu_e$ search at $200 \, Mev < p < 3000 \, MeV$, but no significant signal was found \cite{AguilarArevalo:2009xn}. These results have been qualified as `inconclusive' \cite{AguilarArevalo:2009xn}.

Furthermore, if we consider that the neutrino mass arises as a consequence of EFT with new physics in the UV (maybe close to $10^{15} \, GeV$), then the corresponding dimension $5$ operator giving mass to the neutrinos is a Majorana mass term which violates lepton number by two units. This should lead to new currently unobserved phenomena related to lepton-nonconservation such as neutrinoless double-beta decay ($0\nu\beta\beta$). These experiments are also interesting because, if $0\nu\beta\beta$ is observed, it will provide a measurement of the absolute value of the neutrino mass scale (in contrast to the relative values provided by neutrino oscillations). The search for these phenomena has been performed in the last years and has become quite controversial.

A subgroup of the Heidelberg-Moscow collaboration led by H. V. Klapdor-Kleingrothaus published a series of controversial papers claiming to have observed neutrinoless double-beta decay in $^{76}Ge$ with $6\sigma$ CL \cite{KlapdorKleingrothaus:2001ke,KlapdorKleingrothaus:2004wj,KlapdorKleingrothaus:2006ff}. This ``evidence'' has been hotly argued \cite{Aalseth:2002dt}, and is not accepted by most of the scientific community. Their claim, in which the expectation value of the mass matrix in the $\nu_e$ state is found to be $m_{\beta} \sim 0.2-0.5 \, eV$ has yet to be confirmed by other experiments.
Even if the `Klapdor's evidence' turns out to be wrong, $0\nu\beta\beta$ experiments would give us a hint on the absolute value of neutrino masses in the (near?) future if neutrinos are Majorana particles: it would be just a matter of being sensitive to lower masses.

However if these experiments persist without any positive signal of $0\nu\beta\beta$, then a solution to the absence of lepton number violating phenomena is to impose lepton number conservation by hand on the model (lepton number conservation is just an accidental symmetry of the standard model, but it stops to be a symmetry if we allow for dimension 5 operators or other neutrino mass terms), so that $m^M = 0$. If $n=3$, the sterile neutrinos can be seen as the right handed components of a Dirac spinor (therefore sterile Dirac neutrinos are called `righthanded neutrinos'). Then, processes violating lepton number would be forbidden. However, the reason of the smallness of the neutrino mass would be precluded, worsening the hierarchy problem of the SM.

Apart from $\beta\beta 0\nu$, beta decay experiments in general are also able to determine the absolute scale of neutrino masses. However, the expected scale of neutrino masses are well below the sensitivities of most experiments. Recently, a handful of experiments such as Troitsk Neutrino Mass Experiment \cite{Lobashev:1999tp} and the Mainz Neutrino Mass Experiment \cite{Weinheimer:1999tn} have tried to measure the endpoint region of the tritium beta decay with enormous precision. These experiments rely on the low energy exchange in the beta decay of light nuclei ($18.6 \, keV$ in the case of tritium). Both experiments have set $m_\beta < 2.3 \, eV$, but, surprisingly, the midpoint of the confidence regions lie in the negative mass squared sector. In particular, the Troitsk Experiment gives $m_\beta = -2.3 \pm 2.5(stat.) \pm 2.0(syst.) \, eV$, while the Mainz experiment gives $m_\beta = -0.6 \pm 2.2(stat.) \pm 2.1(syst.) \, eV$. In order to work out a more precise result, the Karlsruhe Tritium Neutrino Experiment (KATRIN) \cite{Osipowicz:2001sq}, which expects to achieve a sensitivity of $m_{\beta} > 0.2 \, eV$, has already begun to gather data.

\paragraph{Are massive neutrinos a too mild assumption?}

There have been attempts of conciling the data of LSND and MiniBOone with other neutrino experiments simply by adding more sterile neutrinos to the model. At this point one might wonder, if this is how one should proceed. Taking into account that sterile neutrinos are almost undetectable particles, for which we have little evidence, is it not just a new problem of adding ``epicycles'' to the SM? Or is it just that we are driving the SM beyond its range of validity and that we should look for radically new particle physics?

In the previous section about the Cosmological Constant Problem we have argued that QFT should be modified in the IR in order to be consistent with results such as the Bekenstein entropy or the smallness of the cosmological constant. In agreement with this point of view, there is a striking coincidence that should not be overlooked. If we consider the cosmological constant as the vacuum energy of some field theory, the value of this vacuum energy is roughly $\rho_v \sim (10^{-2.5} \, eV)^4$. In comparison, the scale of neutrino oscillations (in this case, the scale driving solar and reactor neutrino oscillations in the SM with massive neutrinos) is of the order of $\Delta m_\odot^2 \sim (10^{-2.5} \, eV)^2$.

We wonder if these numbers are unrelated or if they are derived from the same extension of local relativistic QFT. It might be that the same IR physics which causes the quantum corrections to the vacuum energy to be of the order of an IR scale $\lambda \sim 10^{-2.5} eV$ also causes a modification in the neutrino propagator of the same order. If this is true, we have argued that the corresponding departure from QFT should be nonlocal, and therefore it might even be that the description of neutrino oscillations requires some kind of non-locality.

\subsection{Ingredients for an alternative to mass-driven oscillations}

Alternative mechanisms to mass-driven neutrino oscillations must share a basic characteristic: they must include some new ingredient, which must have different eigenvalues for different neutrino species, and the corresponding eigenstates must not coincide with the flavor eigenstates, i.e. the coupling of these neutrino species to the charged leptons via $W$ exchange must be non-diagonal.

We will assume that the mechanism responsible of the phenomena we have listed above is some oscillation mechanism during the propagation of neutrinos. Other possible mechanisms would be nonstandard couplings to either the gravitational potential (violation of the equivalence principle), a torsion field or even matter.

\paragraph{Lorentz Invariance Violation}
Let us think about how one could go beyond the mechanism of mass terms driving the oscillation mechanism. We have stated that an oscillation mechanism can appear only if different neutrino families have different energies $E_i(p)$ for the same momentum $p$. Lorentz invariance restricts these different energies to be a consequence of different masses. Therefore if we want to go beyond the mechanism of mass-driven oscillations, this has to be done at the cost of abandoning Lorentz invariance. Given that the CPT theorem assures that CPT is a symmetry of any local field theory with Lorentz invariance, then CPT could also be violated. Although problematic from the point of view of self-consistency, CPT violation could be very handy in order to explain the discrepancies between LSND and MiniBOone.

It has been argued that Lorentz Invariance Violation (LIV) is an expectable phenomenon in a theory of quantum gravity. It arises naturally in LQG at scales at which the discreteness of spacetime cannot be neglected; Lorentz Invariance is restored as an approximated symmetry at low energies. In this case the departure from Lorentz symmetry is characterized by the Planck scale. The dispersion relation of particles reflects the scale of the spacetime fabric, and takes the form \cite{AmelinoCamelia:2008qg}

\be
E^2 = \vp^2 + m^2 + \eta \vp^2 \left( \frac{E}{E_p} \right)^n \, , \label{dispdef}
\ee
where $m$ is the mass of the tested particle, $\eta$ is a phenomenological parameter, presumably of order $1$, and $n>0$ is assumed to be an integer. This is the kind of models that are being contrasted with experimental tests of Lorentz invariance. They are explicitly rotation invariant, as it is boost invariance what is usually assumed to be broken by quantum gravity. The reason is the presence of a new very high energy scale (typically the Planck scale $E_p$, although this might vary in particular models) which is either an invariant of the theory or a quantity in whose value every observer should agree.

Violations of Lorentz Invariance may also arise in ST as a result of Spontaneous Symmetry Breaking (SSB). In order for Lorentz SSB to occur, it would be required that some Lorentz tensor would acquire an effective negative mass term. In relativistic perturbatively renormalizable quantum field theories, it is impossible to arrive to an effective negative mass term for any vector field.

However it has been proven that this is not the case in string theory \cite{Kostelecky:1988zi}. For instance, in the open bosonic string, it is possible that a vector field $B_\mu$ acquires a nonzero vacuum expectation value due to an effective potential $V$ in the Lagrangian $\Lag_B$ of the type

\be
\Lag_B =  - \frac 14  B_{\mu\nu} B^{\mu\nu}  - V(B_\mu B^\nu \pm b^2) -  B_\mu J^\mu \, , \label{bblag1}
\ee
where $B_{\mu\nu} = \partial_\mu B_\nu -\partial_\nu B_\mu$ is the field strength, $V(x) = \frac 12 \lambda x^2 $ is the potential and $J^\mu$ is the matter current. Depending on the sign appearing in the argument of the potential, the vacuum expectation value of the Bumblebee will be timelike or spacelike (the case of a lightlike expectation value is ill-defined). It is usual to take the vacuum expectation value of the vector field, $b^\mu$, as independent of spacetime coordinates so that the broken symmetries are global Lorentz symmetries, although solitonic solutions are also plausible.

In the same sense in which Lorentz symmetry can be spontaneously broken in string theories, it is also possible to find models in which CPT symmetry is also spontaneously broken \cite{Kostelecky:1991ak,Kostelecky:1995qk}. Taking to account this spontaneous symmetry breaking, it is reasonable to consider that the low energy limit of such a quantum theory of gravity admits a description as an ordinary QFT including departures from Lorentz invariance and can be studied with the techniques of EFT.

In the Standard Model Extension (SME) \cite{Colladay:1996iz,Colladay:1998fq}, the starting point is the minimal SM of particle physics and its associated scale, the electroweak scale $m_W$. New Lorentz- and possibly CPT-violating terms\footnote{The CPT theorem ensures that a local Lorentz invariant theory preserves CPT symmetry \cite{Streater:1989vi}. Thus, a local theory violating CPT must violate Lorentz invariance. The converse is not true; Lorentz violating but CPT conserving theories exist.} arising from a spontaneous breaking of Lorentz invariance in a more fundamental theory are assumed to be suppressed at least by the ratio $\frac{m_W}{E_p}\approx 10^{-17}$. The new terms also respect the gauge invariance of the SM. A minimal standard model extension can be defined to include just power-counting renormalizable terms (as in Ref.~\cite{Colladay:1998fq}), while terms of dimensions greater than four can be included using the standard EFT technics \cite{Bluhm:2005uj}.

Both in the case of an explicit or a spontaneous Lorentz Symmetry Breaking, the symmetry group is reduced and thus a preferred frame arises. These frame is often identified with the frame in which the CMB is almost isotropic \cite{Coleman:1997xq}. However our laboratory frame does not coincide with the CMB rest frame; an obvious consequence is the need of removing the dipole in the observed CMB spectrum, as it only expresses the boost that relates both frames \cite{Jaffe:2000tx}. This boost between the preferred frame (in which the laws of physics are assumed to be isotropic) and the laboratory frame introduces a new directional effect, and therefore an anisotropy in the laws of physics in the laboratory frame. Thus one needs to consider the anisotropic couplings of the SME in order to build a realistic model.

Fortunately, the rotational and translational motions of the Earth are essentially non relativistic in the approximately inertial Sun-centered frame relevant for neutrino experiments, so in practice any direction-dependent effects are suppressed to parts in a thousand or more \cite{Bluhm:2003un}. Anisotropic effects would also be excluded if instead of a Lorentz violation a new symmetry principle (DSR) would replace Lorentz invariance \cite{AmelinoCamelia:2000mn,AmelinoCamelia:2002fw,Magueijo:2001cr,Magueijo:2002am}.

The possibility of neutrino oscillations driven or corrected by LIV has been considered in several works, which we will review below.

\paragraph{Agreement with observational and experimental data}

As it should always be in any field of science, it will be our priority to find alternatives to mass driven oscillations which are fully consistent with the experimental data. Many models of LIV driving neutrino oscillations have been proposed, although few of them have managed to explain the energy behavior that has been experimentally observed.

In order to fit the existing experimental data (aside from the controversial results of LSND and $0\nu\beta\beta$), it is sufficient for these models to meet one property: that the new effect be a correction proportional to the square of an infrared scale $\lambda \sim 1 - 10^{-3} eV$ at energies $E\gg \lambda$.\footnote{Surprisingly, the tandem model presented in Ref.~\cite{Katori:2006mz} manages to fit the experimental data without resorting to this mechanism.} This is due to the fact that all the neutrino experiments that have been carried out involve high energy neutrinos (compared to the scale $\lambda$).

It has not been easy for researchers of LIV in neutrinos to find this energy dependence in their models. In one of the works presented in this PhD thesis, we will show that it is possible to find such a behavior if we invoke nonlocal physics in the IR, as we suggested as a solution to the Cosmological Constant Problem.

Also cosmology will impose important bounds on any alternative to neutrino masses, as well as it imposes a competitive bound on the sum of the neutrino masses in the case of massive neutrinos \cite{Lesgourgues:2006nd,Seljak:2006bg}.

In order to have some information about the low energy behavior of neutrinos, it would be necessary to get information directly or indirectly about neutrinos of energies close to this scales, such as tritium decay close to the endpoint of the spectrum or detection of the Cosmic Neutrino Background (C$\nu$B), which has not yet been observed. These two experiments are completely unfeasible in the near future. However the effect of the C$\nu$B in the cosmic matter power spectrum of large scale structure might be observable.

In conclusion, although neutrino masses are a simple extension of the SM perfectly able to describe the current experimental results on neutrino oscillations (aside of the poorly understood LSND and MiniBOone), it is not the only possibility. An alternative mechanism to neutrino masses requires LIV, but it must also be consistent with the experimental and observational data. A sufficient condition is that the deformation of the dispersion relation of neutrinos involves the square of some IR energy scale.

\subsection{Review on alternatives to neutrino masses}

In this subsection I want to review some of the attempts to offer an alternative description of neutrino oscillations with massless neutrinos. As mentioned this requires to go beyond Special Relativity (SR). Neutrino oscillations have been suggested to be a good probe of physics beyond SR in Ref.~\cite{Coleman:1997xq}, maybe arising in the low energy limit of a theory of quantum gravity. A folkloric example is a quantum black hole swallowing a neutrino of a given flavor and evaporating into a neutrino of a different one \cite{Hawking:1982dj}.

In particular, it has been stated that the addition of Lorentz violating but CPT conserving terms to the SM Lagrangian would induce differences in the maximum attainable velocities of the neutrinos which may depend on their nature and helicity. It was also first stated that massless neutrinos could oscillate in vacua in the presence of LIV. More complicated models have been built in order to fit the vast amount of experimental data that has been collected in the past decade.

\paragraph{Velocity-driven oscillations}

Coleman and Glashow introduced a simple model of oscillation between electron and muon neutrino \cite{Coleman:1997xq},

\bea
\nu_\mu & = & \cos \theta \nu_1 + \sin \theta \nu_2\, , \\
\nu_e & = & \cos \theta \nu_2 - \sin \theta \nu_1\, ,
\eea
where $\nu_1$ and $\nu_2$ are propagation eigenstates with different velocities and dispersion relations

\be
E_1 = p (1 + v_1) \, , \, E_2 = p (1 + v_2) \, ,
\ee
and $\theta$ is a phase characterizing the mixing. The differences between the speed of the neutrino and the speed of light $v_i$, are assumed to be $\vert v_i \vert \ll 1$. The populations of the flavor eigenstates in the propagation eigenstates are

\be
r_{1\mu} = r_{2 e} = \cos^2\theta\, , \, r_{1 e} = r_{2 \mu} = \sin^2 \theta \, ,
\ee
and the probability of an electron neutrino of traversing a distance $R$ without changing its flavor is

\be
P_{e\rightarrow e}(R) = r_{1 e} \sin^2 \left( (v_1-v_2) p R /2 \right) \, , \label{velosc}
\ee
while the same probability, if the oscillation mechanism is driven by mass differences, is

\be
P_{e\rightarrow e}(R) = r^m_{1 e} \sin^2 \left( (m^2_1-m^2_2) R /4 p \right) \, . \label{masosc}
\ee

It is also possible that both masses and difference between velocities are behind the neutrino oscillations. In this case, the diagonalization problem would be more complicated and momentum dependent. The effective Hamiltonian would be of the type

\be
\Ham_{\alpha\beta} = p \left(\delta_{\alpha\beta} + V_{\alpha\beta} \right) + \frac 1{2p} M^2_{\alpha\beta}\,
\ee
in the base of flavor eigenstates. The diagonalization of the effective Hamiltonian should be performed for any given momentum $p$ of the neutrino, resulting in a mixing of the type

\be
P_{e\rightarrow e}(R) = r_{1 e}(p) \sin^2 \left( (E_1(p)-E_2(p)) R /2  \right) \, .
\ee

The limits of low and high momentum of this expression should coincide with the expressions for the mass driven \eqref{masosc} and velocity driven oscillations \eqref{velosc}, respectively.

The clear distinction between the behavior with momentum $p$ between \eqref{velosc} and \eqref{masosc} has enabled the experimentalists to clearly favor mass driven oscillations over velocity driven oscillations \cite{GonzalezGarcia:2007ib}, at least if the velocity of the neutrinos is independent of their frequency. In general, a deviation from the energy behavior of \eqref{masosc} would be a smoking-gun signal of LIV effects \cite{Brustein:2001ik}.

This has therefore led most of the studies of Lorentz violation in neutrino physics to consider Lorentz violation as a subleading mechanism that can be tested within the framework of mass driven oscillations. In some of these studies \cite{Coleman:1998ti,Brustein:2001ik,GonzalezGarcia:2004wg,Morgan:2008zza,Diaz:2009qk} the LIV has been studied at the level of a phenomenological deformation of the dispersion relation of the type of \eqref{dispdef}.

\paragraph{The Standard Model Extension and Neutrino Oscillations}

As we mentioned before, the Minimal SME includes all the Lorentz non-invariant terms with or without CPT violation which are perturbatively renormalizable and respect the gauge symmetries of the SM. The new terms can be divided into CPT preserving (CPT-even) or CPT violating (CPT-odd). The new terms are supposed to be suppressed with respect to the Lorentz conserving terms of the same kind by powers of the electroweak scale over the Planck scale. In this sense, some of the dimensionful terms can be interpreted as new IR scales.

As an example, we will give the explicit form of the terms in the fermion sector. In the lepton sector, the new terms are:

\bea
\Lag_{lepton}^{CPT-even} & = & \frac i2 (c_L)_{\mu\nu\alpha\beta} \bar{\Psi}_{L \alpha}\gamma^\nu \ddmu \Psi_{L \beta} + \frac i2 (c_R)_{\mu\nu\alpha\beta} \bar{L}_\alpha\gamma^\nu \ddmu L_\beta\, , \\
\Lag_{lepton}^{CPT-odd} & = & -(a_L)_{\mu\alpha\beta} \bar{\Psi}_{L \alpha}\gamma^\mu \Psi_{L \beta} -(a_R)_{\mu\alpha\beta} \bar{L}_\alpha\gamma^\mu L_\beta\, .
\eea

All the coefficients $c$ and $a$ are understood to be hermitian matrices in flavor space as well as Lorentz tensors. These new terms are of dimension $4$ and $3$, respectively. Therefore the coefficients $c_L$ and $c_R$ are dimensionless and may have symmetric and antisymmetric parts but can be assumed to be traceless. The coefficients $a_L$ and $a_R$ have dimensions of energy.   If we consider that these vectors are timelike, in the preferred frame of reference their only nonzero component will be $(a_L)_0$ and $(a_R)_0$, and will play the role of an IR scale.

In the quark sector the new terms are analogous to those appearing in the lepton sector, with coefficient matrices $c_Q$, $c_U$, $c_D$, ... Other Lorentz violating terms appear in the Yukawa, Higgs and gauge sectors of the model.

The Minimal SME has been constrained by a huge amount of experiments, among which we could stress high precision tests of QED, neutrino and neutral meson oscillations (for a living review on tests of Lorentz Invariance see Ref.~\cite{Mattingly:2005re}, for constraints on the SME parameters see Ref.~\cite{Kostelecky:2008ts}). Despite the long list of experiments, a vast sector of the Minimal SME remains yet unexplored. Meson oscillation experiments have set the CPT-violating coefficients of the quark sector  $a_\mu < 10^{-20} GeV$. The most stringent bounds come from precision tests in QED. Bounds on a spacelike vector vev $b_\mu$ are of order $10^{-25} GeV$ to $10^{-31} GeV$ depending of the experiment and the kind of particle.

The lepton sector has also been constrained with the data provided by neutrino oscillations. The possibility of LIV-induced neutrino oscillations arises in the model without sterile neutrinos, and the SME has also been extended to include the righthanded Dirac neutrinos by Kostelleck\`{y} and Mewes \cite{Kostelecky:2003cr}. The complete SME with righthanded neutrinos has many parameters and thus several simpler models have been built within this framework in order to gain predictivity. These models include neutrinos with \cite{Kostelecky:2003cr,Mewes:2004ge,Katori:2006mz,Diaz:2009qk} or without  \cite{Kostelecky:2003cr,Kostelecky:2003xn,Kostelecky:2004hg,Mewes:2004ge} mass terms, in which the Lorentz-violating coefficients of the SME are treated as subleading terms or as driving the oscillation phenomenon, respectively.

Aside from the aforementioned anomalous energy dependence of the oscillation phenomena, more new effects are expected if the SME with righthanded neutrinos is taken into account \cite{Kostelecky:2003cr}. Anisotropic terms induce periodic variations in the oscillations: those induced by the Earth's rotation and sidereal translation. These effects can be averaged away by integrating the data over time, but anisotropic effects would not be erased, as asymmetries between the east-west and north-south directions or between the ecliptic plane and the ecliptic north-south would appear. Other possible effects arising from the SME would be discrete symmetry violations, like effects violating lepton number (as neutrino-antineutrino oscillations), CPT (deviations from the relationship $P_{\nu_\alpha\rightarrow \nu_\beta} = P_{\bar\nu_\beta\rightarrow \bar\nu_\alpha}$ or both. CPT violation is extremely interesting as it could provide an explanation to the  discrepancy between the LSND \cite{Aguilar:2001ty} and MiniBoONE \cite{AguilarArevalo:2008rc} experiments.

The most general modified Dirac equation of the neutrino fields within the SME with righthanded neutrinos would be \cite{Kostelecky:2003cr}

\be
\left(i\Gamma^\nu_{\bar\alpha\bar\beta}\partial_\nu - M_{\bar\alpha\bar\beta}\right)\nu_{\bar\beta} = 0 \label{SMEDiraceq}\, ,
\ee
where $\bar\alpha$ and $\bar\beta$ run for the three possible neutrino flavors and their charge conjugates and

\bea
\Gamma^\nu_{\bar\alpha\bar\beta} & : = & \gamma^\nu \delta_{\bar\alpha\bar\beta} + c^{\mu\nu}_{\bar\alpha\bar\beta}\gamma_\mu + d^{\mu\nu}_{\bar\alpha\bar\beta} \gamma_5\gamma_\mu + e^\nu_{\bar\alpha\bar\beta} + i f^\nu_{\bar\alpha\bar\beta}\gamma_5 +\frac 1 2 g^{\lambda\mu\nu}_{\bar\alpha\bar\beta}\sigma_{\lambda\mu} \, , \\
M_{\bar\alpha\bar\beta} & := & m_{\bar\alpha\bar\beta}+ i m_{5\bar\alpha\bar\beta} \gamma_5 + a^\mu_{\bar\alpha\bar\beta}\gamma_\mu + b^\mu_{\bar\alpha\bar\beta}\gamma_5\gamma_\mu + \frac 12 H^{\mu\nu}_{\bar\alpha\bar\beta}\sigma_{\mu\nu} \, .
\eea

The coefficients $m$ and $m_5$ are a linear combination of the righthanded and lefthanded neutrino mass matrices $m_R$ and $m_L$, and thus Lorentz invariant. The coefficients $c$, $d$, $H$ are CPT conserving but Lorentz violating, while $a$, $b$, $e$, $f$, $g$ are CPT violating. Hermiticity implies that all the coefficients are Hermitian in flavor space.

The effective Hamiltonian for the neutrinos in the Minimal SME reads

\be
(\Ham_{eff})_{\bar\alpha\bar\beta} = \Ham_0 -\frac 12 \left\{\gamma^0 \delta\Gamma^0,\Ham_0\right\} -\gamma^0(i\delta\Gamma^j\partial_j-\delta M) \, ,
\ee
where $\Ham_0 = -\gamma^0(i\gamma^j\partial_j - M_0)$ is the general Lorentz conserving Hamiltonian, $M_0$ is the Lorentz conserving part of $M$ and $\delta\Gamma$, $\delta M$ are the Lorentz violating parts of $\Gamma$, $M$. The general treatment is possible but rather horrible, due to the vast amount of parameters and the fact that the coefficient matrices do not necessarily commute.

Therefore the model has to be simplified in order to gain some predictive power. It is standard to get rid off righthanded neutrinos and present a model for oscillations among active (lefthanded) neutrinos and their antiparticles. Further simplification can be achieved if oscillations between neutrinos and antineutrinos $\nu \leftrightarrow \bar \nu$ are highly suppressed. Then, the effective Hamiltonian for the lefthanded ultrarelativistic neutrinos reads

\be
\left(h_{eff} \right)_{\alpha\beta} = p \delta_{\alpha\beta} + \frac 1{2p}\left(m_L^2\right)_{\alpha\beta} + \frac 1p \left[ \left(a_L\right)^\mu p_\mu - \left(c_L\right)^{\mu\nu}p_\mu p_\nu\right]_{\alpha\beta}
\ee
and its complex conjugate for antineutrinos, with the opposite sign for the CPT-odd coefficient $a_L$. In the previous formula $m_L$ is the combination of $m$ and $m_5$ relevant for lefthanded neutrinos (see \cite{Kostelecky:2003cr} for the details), $c_L = c + d$ and $a_L = a + b$. Even simpler models can be built in order to explain the oscillations data.

`Fried-chicken' models \cite{Kostelecky:2003cr} are those in which spatial isotropy is imposed,

\be
\left(h_{eff}^{FC} \right)_{\alpha\beta} = p \delta_{\alpha\beta} + \left[ \frac 1{2p}\left(m_L^2\right) + \left(a_L\right)^0  - \frac 43 \left(c_L\right)^{00} p \right]_{\alpha\beta}
\ee
so that the study of modified dispersion relations is not complicated by direction dependent effects. Most models of Lorentz-violating neutrino oscillations fall in this class. However, isotropy in the laboratory frame is difficult to motivate in a Lorentz violating theory.

Anisotropic models, in contrast, need a well defined spacetime reference frame, which is usually taken to be the Sun-centered celestial equatorial frame of coordinates $\left\{X,Y,Z,T\right\}$, in order to report the values of its coefficients. An interesting model of this class is the so called `Bicycle Model' presented in Ref.~\cite{Kostelecky:2003xn}, in which most of the oscillations data can be explained only with two free parameters and no mass term.

It must be remembered that the $a_L$ coefficient has dimensions of mass and represents some infrared scale, presumably of order $\frac {m_W^2}{E_p}$. If the dimensionless coefficient $c_L$ is of order $\frac {m_W}{E_p}$, then for momenta well below the electroweak scale (and this happens in every single neutrino oscillations experiment), the dominant source of the oscillation in a massless model ($m_L = 0$) is the $a_L$ term. The argument of the oscillation's sine function will be $\sim \Delta a_L L$ instead of the standard $\Delta m^2 L / 2E$. This behavior is disfavored by the experimental data.

However, this na\"{\i}ve energy behavior was surmounted by the Bicycle Model. In this model, the coefficients $c_{Lee}^{TT} = \frac 32 \check{c}$ and $a_{Le\mu}^Z = a_{Le\tau}^Z = \frac{\check{a}}{\sqrt 2}$ are the only nonzero coefficients. The resulting neutrino oscillation probabilities are

\bea
P_{\nu_e \rightarrow \nu_e} & = & 1 - 4 \sin^2\theta \cos^2 \theta \sin^2\left(\Delta_{31}L/2\right) \, , \\
P_{\nu_e \rightarrow \nu_\mu} & = & P_{\nu_e \rightarrow \nu_\tau} = 2 \sin^2\theta \cos^2 \theta \sin^2\left(\Delta_{31}L/2\right) \, , \\
P_{\nu_\mu \rightarrow \nu_\mu} & = & P_{\nu_\tau \rightarrow \nu_\tau} = 1 - \sin^2\theta \sin^2\left(\Delta_{21}L/2\right) \nonumber \\
& & -\sin^2\theta \cos^2 \theta \sin^2\left(\Delta_{31}L/2\right)- \cos^2 \theta \sin^2\left(\Delta_{32}L/2\right)\, , \\
P_{\nu_\mu \rightarrow \nu_\tau} & = & \sin^2\theta \sin^2\left(\Delta_{21}L/2\right) \nonumber \\
& & -\sin^2\theta \cos^2 \theta \sin^2\left(\Delta_{31}L/2\right) + \cos^2 \theta \sin^2\left(\Delta_{32}L/2\right)\, ,
\eea
where

\bea
\Delta_{21} & = & \sqrt{(\check{c} p)^2 + (\check{a} \cos\Theta)^2} + \check{c} p \, ,\\
\Delta_{31} & = & \sqrt{(\check{c} p)^2 + (\check{a} \cos\Theta)^2} \, , \\
\Delta_{32} & = & \sqrt{(\check{c} p)^2 + (\check{a} \cos\Theta)^2} - \check{c} p \, ,\\
sin^2 \theta & = & \frac 12 \left[ 1 - \check{c}p/\sqrt{(\check{c} p)^2 + (\check{a} \cos\Theta)^2}\right] \, ,
\eea
and $\Theta$ is the angle between the celestial north pole and the direction of propagation. The same probabilities are valid for antineutrinos (which are obtained by changing $\check{a}\rightarrow - \check{a}$).

In this model, electron neutrino propagating at low energies compared with the scale $\check{a}/\check{c}$ oscillate into muon and tau neutrinos. However, at greater energies electron neutrinos do not oscillate, but there is a maximal mixing between muon and tau neutrinos, which oscillate with a wavelength $\sim \frac{\check{c}E}{\check{a}^2\cos^2 \Theta}$. Thus, high energy oscillations mimic mass driven oscillations with a direction-dependent effective mass squared difference $\Delta m^2_\Theta =  \check{a}^2 \cos^2\theta/\check{c}$. The model failed to explain the LSND anomaly, but did quite well in fitting the then-current neutrino data with $\check{c} = 10^{-19}$ and $\check{a} = 10^{-20} GeV$ . More recent tests using crossed data have ruled out the bicycle model and some simple extensions \cite{Barger:2007dc}.

The big difference between the predictions of this model and the standard mass driven oscillations mechanism is the directional dependence which induce small semiannual variations in the neutrino flux that could be observed and `compass' asymmetries. Directional variations have been tightly constrained by MINOS \cite{:2008ij}. Nevertheless, the model can be modified to have no directional dependence by changing $a_L^Z$ to $a_L^T$ with a small impact on the fits \cite{Kostelecky:2003xn}.

An extension of the Bicycle Model, the `Tandem Model' \cite{Katori:2006mz} was proposed to fit also the anomalous data of LSND. The model includes similar parameters to the ones in the Bicycle Model and a single mass term $m_{\tau\tau}$. The effective Hamiltonian can be written in the flavor basis as

\be
h^\nu_{TM} = \left(
\ba{ccc}
(1 +\check{c}) p & \check{a} & \check{a} \\ \check{a} & p & \check{a} \\ \check{a} & \check{a} & p + \check{m}^2/2p
\ea
\right) \, ,
\ee
where $\check{c} = (c_L)^{TT}_{ee}$ and $\check{a}=(a_L)^T_{e\mu} = (a_L)^T_{e\tau} = (a_L)^T_{\mu\tau}$. The corresponding effective Hamiltonian for antineutrinos is obtained by changing $\check{a} \rightarrow -\check{a}$.

The Tandem Model provides a good description of existing neutrino-oscillations data with

\bea
\frac 12 \check{m}^2 & = & 5.2 \times 10^{-3} eV^2 \, , \nonumber \\
\check{a} & = & -2.4 \times 10^{-19} GeV \, , \nonumber \\
\check{c} & = & 3.4 \times 10^{-17} \, .
\eea

Solar neutrino oscillations can be fit without the need of any MSW effect \cite{Wolfenstein:1977ue,Mikheev:1986gs}. Atmospheric neutrino oscillations can be fit as in the Bicycle Model, as it reduces to it at high energies. Long-baseline experiments like KamLAND \cite{Araki:2004mb} lie in the low energy regime in which the $\check{c}$ term can be neglected. In this regime, the CPT violation is manifest as a difference in the $\nu_e$ and $\bar\nu_e$ survival probabilities which can be tested. Short-baseline experiments like LSND \cite{Aguilar:2001ty} and MiniBOone \cite{AguilarArevalo:2008rc} can be fit due to this CPT violation. The effect lies on the edge of experimental sensitivity due to the short paths that are being tested. In fact, the tandem model predicted a low energy excess in the MiniBOone experiment prior to its discovery, although the observed excess is qualitatively greater. This discovery remarks the interest of the search for hybrid mass and LIV-driven oscillation mechanisms. Nevertheless, the observations of LSND and MiniBOone must be confirmed by other ongoing or planned short-baseline experiments.

Finally, the case of mass-driven neutrino oscillation modified by subleading LIV-effects has also been studied in the framework of the SME. The constraints of the parameters of the model have been gathered in Ref.~\cite{Diaz:2009qk}.

\paragraph{Neutrino Oscillations arising from a noncanonical quantization procedure?}

Motivated by the apparition of noncommutative spacetime \cite{connes} in ST \cite{Connes:1997cr,Seiberg:1999vs,Yoneya:2000bt}, it has been proposed that the most general extension of Noncommutative Quantum Mechanics (NCQM) \cite{Mezincescu:2000zq,Duval:2000xr,Gamboa:2000yq} to field theory is to make the fields, which are the degrees of freedom in a field theory, noncommutative. The field noncommutativity is introduced in a scalar quantum field theory at the level of the commutation relations that define the way from the classical theory to the QFT.

In this scalar quantum 'noncommutative' or noncanonical field theory (QNCFT) \cite{Carmona:2002iv,Carmona:2003kh}, the (equal-time) commutation relations read

\begin{eqnarray}
	\left[\Phi(\vx),\Phimas(\vx')\right] & = & \theta \deltat(\vx-\vx') \nonumber\, , \\
	\left[\Phi(\vx),\Pimas(\vx')\right] & = & i \deltat(\vx-\vx') \, ,\\
	\left[\Pi(\vx),\Pimas(\vx')\right] & = & B \deltat(\vx-\vx') \nonumber \, ,
\end{eqnarray}
while the free Hamiltonian remains unmodified.\footnote{This commutation relations have also been extended in Ref.~\cite{Balachandran:2007ua}, in which the Dirac deltas have been turned into Gaussian distributions. Note that this change implies that the theory becomes nonlocal.} The resulting free theory can be fully solved and proves it to be a theory of particles $a$ and antiparticles $b$ with Lorentz and CPT violating dispersion relations

\bea
\frac{\ena(\vp)}{\omp}& = & \sqrt{1+\left(\frac{B/\omp - \theta \omp}2\right)^2}+\frac{B/\omp + \theta \omp}2 \, , \\
\frac{\eb(\vp)}{\omp} & = & \sqrt{1+\left(\frac{B/\omp  - \theta \omp}2\right)^2}-\frac{B/\omp + \theta \omp}2 \, ,
\eea
where $\omp = \sqrt{\vp^2 + m^2}$. Note that the parameters introduced in the commutation relations act as an UV and an IR scale. Thus, relativistic QFT is recovered in the limit in which the energies of the particles are between but far from these values.

QNCFT was extended in Ref.~\cite{Arias:2007zz} to spinor fields. In this case, Hamiltonian of the spinor fields is the standard massless one, but the anticommutation relations are changed. In their model a massless spinor field with just two flavors is considered. The theory is quantized by imposing the (equal-time) anticommutation relations

\be
\left\{ \Psi^i_\alpha (\vx), \Psi^{j\dagger}_\beta (\vx') \right\} = \mathcal{A}_{\alpha\beta}\delta^{ij}\delta^{(3)}(\vx-\vx') \, , \label{paola}
\ee
where $i,j$ are the spinor indices and $\mathcal{A}$ is an hermitian matrix. The new parameters in $\mathcal{A}$ are in this case adimensional, and therefore the theory amounts to be a field theory description of the model introduced in Ref.~\cite{Coleman:1997xq} of velocity driven oscillations or of the SME with just a $c_L^{00}$ nonzero coefficient, and therefore is ruled out by experiment.

The authors of \cite{Arias:2007zz} try to surmount this problem by making the eigenvalues of $\mathcal{A}$ momentum-dependent, but this has not much sense as the anticommutation is just a function of the spacetime coordinates. We will try to solve this problem in one of the papers presented in this thesis.

\subsection{Neutrino Physics and new physics in the IR}

In this thesis we want to present an alternative to neutrino oscillation driven by neutrino masses which does not modify the field content of the SM, respects the gauge invariance principle and does not imply lepton number violating phenomenon. We also want this alternative to be consistent with all the present experimental and observational data and physically distinguishable from the standard mechanism with future data.

We will seek a realistic model in the framework of QNCFT, for it is a simple extension of QFT. We will try to find a modification of the dispersion relation of neutrinos which involves the square of an IR scale at momenta well above this scale.

The solution appears to invoke a nonlocal extension of the anticommutation relations \eqref{paola}, similar to the one in Ref.~\cite{Balachandran:2007ua} and to the nonlocal extension of QFT in the IR that was suggested in the previous section. We will refer to this modified neutrino physics as noncanonical neutrinos.

\subsection{Probing new neutrino physics in cosmology}

The aforementioned alternative mechanism for neutrino oscillations will be difficult to be distinguished from the mass mechanism by means of particle physics experiments. The reason is that particle physics experiments always probe neutrinos of energies well above the IR scale, and in this regime the behavior of both mechanism is the almost the same.

In order to distinguish between massive neutrinos and noncanonical neutrinos, it will be necessary to probe neutrinos of momenta comparable to or be below the IR scale, and such neutrinos can be found in the C$\nu$B. However in order to consistently explore the cosmological effect of such a Lorentz violating, nonlocal neutrinos, it will be necessary to extend the concept of QFT in curved spacetimes \cite{BD} to more general field theories.

Doing this consistently is a difficult task and is beyond the scope of this PhD thesis. However, as an first step, we will explore how the scalar QNCFT presented in Refs.~\cite{Carmona:2002iv,Carmona:2003kh} may couple to gravity. The symmetry properties of this theory have been studied in Ref.~\cite{Carmona:2009ra}. A more detailed introduction to QNCFT and its symmetry properties have been included in this work. The requirement of keeping this symmetry properties in the curved spacetime case will be essential for determining the precise form of the coupling.

\section{Objectives}

We have tried to argue that problems in cosmology might be solved by an extension of QFT in the IR and the UV, and that the effect of gravity might be behind a modification of QFT responsible of ill-understood phenomena such as neutrino oscillations. Although others have considered top-down approaches to these problems, we think that tracking the experimental data from a bottom-up perspective is the key of scientific progress. Thus we will try to explore departures from GR and QFT in the UV and the IR that might be behind the present problems in cosmology and the recently explored physics of neutrinos.

In the first of the papers presented in this thesis, we will build a cosmological model based on a phenomenological departure from the cosmological standard model both in the UV and the IR, the Asymptotic Cosmological Model. The model will be defined in the homogeneous approximation. The kinematics of the universe will be determined. The horizon problem will be automatically solved. Through a simulation using Monte Carlo Markov Chains, the model will be contrasted with the available observational data which do not depend on the behavior of inhomogeneous perturbations. Also particular realizations of the model in the homogeneous approximation will be considered among the most popular theories and models in cosmology.

We will also try to depart from the homogeneous approximation in the Asymptotic Cosmological Model. In the second paper presented in this thesis, the linearized treatment of scalar perturbations in the model will be derived with just one assumption based on phenomenological consistence. The resulting model will turn out to be distinguishable from the most popular theories and models in cosmology. The effect of the departure from the standard cosmological model in the CMB spectrum will be computed using a modification of the CAMB code \cite{CAMB}.

An alternative to neutrino masses as a mechanism for neutrino oscillations which is compatible with all the experimental data within the framework of QNCFT will be presented in the third of the papers presented in this thesis. The extension will not add new fields to the SM and respect all its gauge and discrete symmetries, but will violate Lorentz invariance and locality. This violation will be mediated by a new IR scale controlling the modification of the dispersion relation and the presence of new physics in the IR.

In order to build a theory of quantum fields in curved spacetime for particles with modified dispersion relations, it will also be important to distinguish if Lorentz symmetry is violated or deformed, as we will see in the fourth of the works presented in this PhD thesis. This distinction could be crucial in order to derive the quantum corrections to the cosmological constant, for instance. In the fouth of the papers presented in this compendium we will explain how the symmetry properties determine the formulation of an extension of QFT to curved spacetime. For the sake of simplicity we will choose the case of a QNCFT of a scalar field in the $B=0$ limit.

A comment on the implications of this mechanism of neutrino oscillations in cosmology will be included in Appendix A. Here a na\"{\i}ve calculation of the effect of a modified dispersion relation of neutrinos in the expansion of the universe is presented. 

\chapter{An extension of the cosmological standard model \\ with a bounded
  Hubble expansion rate}

%\begin{quote}
% Here there is a quote. \newline
%I want to quote the book \newline
%I read yesterday in the pool.
%\end{quote}

%\begin{abstract}
The possibility of having an extension of the cosmological standard
model with a Hubble expansion rate $H$ constrained to a finite
interval is considered. Two periods of accelerated expansion arise
naturally when the Hubble expansion rate approaches to the two
limiting values. The new description of the history of the universe is
confronted with cosmological data and with several theoretical ideas
going beyond the standard cosmological model.
%\end{abstract}

\section{Introduction}
According to General Relativity, if the universe is filled with the particles
of the Standard Model of particle physics, gravity should lead to a
deceleration of the expansion of the universe. However, in 1998 two
independent evidences of present accelerated expansion were presented
\cite{Riess:1998cb,Perlmutter:1998np} and later confirmed by different
observations \cite{Jaffe:2000tx,Spergel:2003cb,Peacock:2001gs}. On the other hand, measurements of large
scale structure \cite{Tegmark:2003ud} and CMB anisotropy \cite{Jaffe:2000tx} also
indicate that the universe evolved through a period of early accelerated
expansion (inflation).

There is no compelling explanation for any of these cosmic accelerations,
but many intriguing ideas are being explored. In the case of inflation,
the origin of the accelerated expansion can be either a modification of
gravity at small scales \cite{Starobinsky:1980te} or a coupling of the expansion of
the universe to the progress of phase transitions \cite{Kazanas:1980tx,Sato:1980yn,Guth:1980zm,Linde:1981mu,Albrecht:1982wi}.
In the case of the present accelerated expansion these ideas can be classified
into three main groups: new exotic sources of the gravitational field with
large negative pressure \cite{Turner:1998ex,Zlatev:1998tr,ArmendarizPicon:2000dh} (Dark Energy), modifications of
gravity at large scales \cite{Bekenstein:1984tv,Bekenstein:2004ne} and rejection of the spatial
homogeneity as a good approximation in the description of the present
universe \cite{Tomita:2000jj,Kolb:2005da,Wiltshire:2007zj}.

Different models (none of them compelling) of the source responsible
for each of the two periods of accelerated expansion have been
considered. Einstein equations admit a cosmological constant
$\Lambda$, which can be realized as the stress-energy tensor of empty
space. This $\Lambda$, together with Cold Dark Matter, Standard Model
particles and General Relativity, form the current cosmological model
($\Lambda$CDM). However, quantum field theory predicts a value of
$\Lambda$ which is 120 orders of magnitude higher than observed.
Supersymmetry can lower this value 60 orders of magnitude, which is
still ridiculous \cite{Weinberg:1988cp}.
In order to solve this paradox, dynamical Dark Energy models have been proposed.

This has also lead to explore the possibility that cosmic acceleration arises
from new gravitational physics. Also here several alternative
modifications of the Einstein-Hilbert action at large and small
curvatures \cite{Gasperini:1991ak,Carroll:2003wy,Sotiriou:2005hu,Hu:2007nk,Nojiri:2006ri,Nojiri:2003ft,Cognola:2007zu}, or even higher
dimensional models \cite{Deffayet:2000uy,Dvali:2000hr}, producing an accelerated
expansion have been identified.
All these analysis include an \textit{ad hoc} restriction to actions involving simple
functions of the scalar curvature and/or the Gauss-Bonnet tensor.
This discussion is sufficient to establish the point that cosmic acceleration can
be made compatible with a standard source for the gravitational field, but it is
convenient to consider a more general framework in order to make a systematic
analysis of the cosmological effects of a modification of general relativity.

In this paper we parametrize the evolution of the universe (considered isotropic and
homogeneous) with the Hubble parameter $H$. One finds that it is possible to restrict
the domain of $H$ to a bounded interval. This restriction naturally
produces an accelerated expansion when the Hubble expansion rate
approaches any of the two edges of the interval.
Therefore, we find a new way to incorporate two periods
of cosmic acceleration produced by a modification of general
relativity. But the dependence on the two limiting values
of $H$ can be chosen independently. One could even consider a Hubble
expansion rate constrained to a semi-infinite interval with a unique
period of accelerated expansion. In this sense the aim to
have a unified explanation of both periods of accelerated expansion is
only partially achieved. This simple phenomenological approach to the problem
of accelerated expansion in cosmology proves to be equivalent at
the homogeneous level to other descriptions based on modifications
of the Einstein-Hilbert action or the introduction of exotic components
in the matter Lagrangian.

In the next section we will review how a general modification of the gravitational
action leads to a generalized first Friedman equation at the homogeneous level. In
the third section we will present a specific model (ACM) based on the simplest
way to implement a bounded interval of $H$. In the fourth section we will contrast
the predictions of ACM for the present acceleration with astrophysical observations.
In the fifth section we will show that it is always possible to find modified
gravitational actions which lead to a given generalized first Friedman equation, and
present some simple examples. In the sixth section we will show that it is also always
possible to find Dark Energy models which are equivalent to a given generalized
first Friedman equation, and present some simple examples. The last section is devoted
to summary and conclusions.

\section{Action of the cosmological standard model extension}

The spatial homogeneity and isotropy allow to reduce the
gravitational system to a mechanical system with two
variables $a(t)$, $N(t)$ which parametrize the Robertson-Walker
geometry
\be
ds^2 \,=\, N^2(t) dt^2 - a^2(t) \Delta_{ij} dx^i dx^j
\ee
\be
\Delta_{ij} \,=\, \delta_{ij} + \frac{k x_i x_j}{1 - k{\bf x}^2} \, .
\ee
Invariance under parameterizations of the time variable
imply that the invariant time differential $N(t)dt$ must be used. Also
a rescaling of the spatial variables
$x^i\rightarrow\lambda x^i$ together with $a(t)\rightarrow\lambda^{-1}
a(t)$ and $k\rightarrow\lambda^{-2} k$ is a symmetry that must be kept
in the Lagrangian. The action of the reduced homogeneous gravitational
system can be then written as
\be
I_g \,=\, \int dt N \, L\left(\frac{k}{a^2}, H,
\frac{1}{N}\frac{dH}{dt},
\frac{1}{N}\frac{d}{dt}\left(\frac{1}{N}\frac{dH}{dt}\right),
...\right) \, ,
\label{Ig}
\ee
with $H \,=\, \frac{1}{aN}\frac{da}{dt}$.

If we keep the standard definition of the gravitational coupling and
the density ($\rho$) and pressure ($p$) of a cosmological homogeneous
and isotropic fluid as a source of the gravitational field, we have
the equations of the reduced system
\be
\left(\frac{8\pi G_N}{3}\right) \rho \,=\, - \frac{1}{a^3} \frac{\delta
  I_g}{\delta N(t)}
\ee
\be
- \left(8\pi G_N\right) p \,=\, - \frac{1}{N a^2} \frac{\delta
  I_g}{\delta a(t)}\, .
\ee
It is possible to choose a new time coordinate $t^{'}$ such that
\be
\frac{dt^{'}}{dt} \,=\, N(t)\, .
\ee
This is equivalent to set $N=1$ in the action (\ref{Ig}) of the gravitational
system and the evolution equations of the cosmological model reduce to a
set of equations for the scale factor $a(t)$. If we introduce the
notation
\be
H^{(i)} \,=\, \left(\frac{d}{dt}\right)^i \, H
\ee
\be
\left(\delta_i L\right)^{(j)} \,=\, \left(\frac{d}{dt}\right)^j
\left[a^3 \partial_i L\right]\, ,
\ee
with $\partial_i L$ denoting the partial derivatives of the Lagrangian
as a function of the variables $(k/a^2, H, H^{(1)}, H^{(2)}, ...)$ we
have
\be
\left(\frac{8\pi G_N}{3}\right) \rho \,=\, - L +
\sum_{i=2}^{\infty} \sum_{j=0}^{i-2} \frac{ (-1)^{i-j}}{a^3} H^{(j)} \left(\delta_i
L\right)^{(i-j-2)}
\label{rho}
\ee
\be
- \left(8\pi G_N\right) p \,=\, - 3 L + \frac{2 k}{a^2} \partial_1 L +
 \sum_{i=2}^{\infty} \frac{(-1)^i}{a^3} \left(\delta_i
L\right)^{(i-1)}   \, .
\label{p}
\ee
In the homogeneous and isotropic approximation, the vanishing of the
covariant divergence of the energy-momentum tensor leads to the
continuity equation
\be
\frac{d}{dt} \left(\rho a^3\right) \,=\, - p \frac{d}{dt} a^3
\label{cont_eq}\, .
\ee
In the radiation dominated era one has
\be
\rho \,=\ 3 p \,=\, \frac{\sigma}{a^4}
\label{rad}\, ,
\ee
where $\sigma$ is a constant parameterizing the general
solution of the continuity equation. In a period dominated by matter
one has a pressureless fluid and then
\be
p \,=\, 0 \, ,{\hskip 2cm} \rho \,=\, \frac{\eta}{a^3}\, ,
\ee
with constant $\eta$. When these
expressions for the energy density and pressure are plugged in
(\ref{rho}-\ref{p}), one ends up with two compatible differential equations for
the scale factor $a(t)$ which describe the evolution of the
universe.  From now on we will use the more common notation
$\frac{dH}{dt}\equiv\dot{H}$,...

\section{The Asymptotic Cosmological Model}
Let us assume that the Lagrangian L of the gravitational system is
such that the evolution equations (\ref{rho}-\ref{p}) admit a solution
such that $a(t)>0$ (absence of singularities), $\da >0$ (perpetual
expansion) and $\dH<0$. In that case, one has a different value of
the scale factor $a$ and the Hubble rate $H$ at each time and then one
has a one to one correspondence between the scale factor and the
Hubble rate. Since the continuity equation (together with the equation
of state) gives a relation between the scale factor and the density,
one can describe a solution of the evolution equations of the
generalized cosmological model through a relation between the energy
density and the Hubble rate, i.e. through a generalized first Friedman
equation
\be
\left(\frac{8\pi G_N}{3}\right) \rho \,=\, g(H)\, ,
\label{rho_g}
\ee
with $g$ a smooth function which parametrizes the different algebraic
relations corresponding to different solutions of different
cosmological models. Each choice for the function $g(H)$ defines a
phenomenological description of a cosmological model. Then one
can take it as a starting point trying to translate any observation
into a partial information on the function $g(H)$ which parametrizes
the cosmological model.
%%OK%%

Eq. (\ref{rho_g}) is all one needs in order to reconstruct the
evolution of the universe at the homogeneous level. The generalized
second Friedman equation is
obtained by using the continuity equation (\ref{cont_eq}) and the
expression for $\rho$ as a function of $H$. One has
\be
- \left(8\pi G_N\right) p \,=\, 3 g(H) + \frac{g^{'}(H)}{H} \dH
\label{pressure}\, .
\ee
In the matter dominated era one has
\be
\frac{\dH}{H^2} \,=\, - 3 \frac{g(H)}{H g^{'}(H)}
\label{matter}
\ee
and then the assumed properties of the solution for the evolution
equations require the consistency conditions
\be
g(H) > 0 \, ,{\hskip 1cm} g^{'}(H) >0\, .
\label{gcons}
\ee
In the period dominated by radiation one has
\be
\frac{\dH}{H^2} \,=\, - 4 \frac{g(H)}{H g^{'}(H)}
\label{radiation}
\ee
instead of (\ref{matter}) and the same consistency conditions
(\ref{gcons}) for the function $g(H)$ which defines the generalized
first Friedman equation.
%%OK%%

We introduce now a phenomenological cosmological model
defined by the condition that the Hubble rate has an upper bound $H_+$
and a lower bound $H_-$. This can be implemented through a function
$g(H)$ going to infinity when $H$ approaches $H_+$ and going to
zero when $H$ approaches $H_-$. We will also assume that there is an interval
of $H$ in which the behavior of the energy density with the Hubble parameter is,
to a good approximation, scale-free i.e. $g(H)\propto H^2$. The source
of the gravitational field will be a homogeneous and isotropic fluid
composed of relativistic and non-relativistic particles.
The Cosmological Standard Model without curvature is recovered in the
limit $H_-/H\rightarrow 0$ and $H_+/H\rightarrow\infty$ which is a
good approximation for the period of decelerated expansion.

Notice that this interpretation is independent of the underlying
theory of gravitation.
The Hubble parameter can be used to parametrize the history of
universe as long as $\dH\neq0\,\forall \,t$. The total density can be
thus expressed as a function of $H$. If $H$ is bounded, then $\rho(H)$
will have a pole at $H=H_+$ and a zero at $H=H_-$. Far from these scales,
the behavior of $\rho(H)$ can be assumed to be approximately scale-free.
%%OK%%

Under these conditions, we can parametrize the dependence of the
cosmological model on the lower bound $H_-$ by
\be
g(H) \,=\, H^2 h_{-}\left(\frac{H_-^2}{H^2}\right)
\label{g-}
\ee
and similarly for the dependence on the upper bound $H_+$
\be
g(H) \,=\, \frac{H^2}{
    h_{+}\left(\frac{H^2}{H_+^2}\right)}
\label{g+}\, ,
\ee
where the two functions $h_{\pm}$ satisfy the conditions
\be
\lim_{x\to 0} h_{\pm} (x) \,=\, \beta^{\pm 1}\, , {\hskip 1cm}
\lim_{x\to 1} h_{\pm} (x) \,=\, 0\, .
\ee
$\beta$ is a constant allowed in principle by dimensional
arguments. If $\beta \neq 1$ then it can be moved to the lhs of the
Friedman equation, turning $G_N\rightarrow\beta G_N$, and can be
interpreted as the ratio between an effective cosmological value of
the gravitational coupling and the value measured with local
tests. But $\beta\neq 1$ would be in conflict with Nucleosynthesis,
through the relic abundances of $^4 He$ and other heavy elements (for
3 neutrino species) \cite{Mukh}, so we set $\beta= 1$ . Therefore
\be
\lim_{x\to 0} h_{\pm} (x) \,=\, 1 \, ,{\hskip 1cm}
\lim_{x\to 1} h_{\pm} (x) \,=\, 0\, .
\label{hpm}
\ee

The consistency conditions (\ref{gcons}) result in
\be
h_{\pm} (x) > 0 \, ,{\hskip 1cm} h_{\pm} (x) > x h_{\pm}^{'}(x)
\label{hcons}
\ee
for the two functions $h_{\pm}$ defined in the interval $0<x<1$. Thus,
we can divide the cosmic evolution history into three periods. In the
earliest, relativistic particles dominate the energy density of the
universe and the generalized first Friedman equation shows a
dependence on the upper bound $H_+$. There is also a transition period
in which the effect of the bounds can be neglected and the rhs of the
first Friedman equation is scale-free; this period includes the
transition from a radiation dominated universe to a matter dominated
universe. In the third present period, non-relativistic particles
dominate the energy density of the universe but the dependence on the
lower bound $H_-$ must be accounted for in the generalized first
Friedman equation.

\indent From the definition of the Hubble parameter, one has
\be
\frac{\dH}{H^2} \,=\, -1 + \frac{a \dda}{\da^2}\, .
\ee
Then, in order to see if there is an accelerated or decelerated
expansion, one has to determine whether $\dH/H^2$ is greater or smaller
than $-1$.

In the period dominated by radiation one has
\be
\frac{\dH}{H^2} \,=\, - 2 \, \frac{1}{1 - \frac{x h_+^{'}(x)}{
    h_+(x)}}\, ,
\ee
with $x=H^2/H_{+}^2$, where we have used (\ref{g+}) assuming that only
the dependence on the upper bound ($H_{+}$) of the Hubble parameter is
relevant. A very simple choice for this dependence is given by
\be
g(H) \,=\, \frac{H^2}{\left(1 - \frac{H^2}{H_{+}^2}\right)^{\alpha_+}}
\label{alpha+}\, ,
\ee
with $\alpha_+$ a (positive) exponent which parametrizes the departure
from the standard cosmological model when the Hubble rate approaches
its upper bound. With this simple choice one has a transition from an
accelerated expansion for $H^2 > H_+^2/(1+\alpha_+)$ into a decelerated
expansion when $H^2 < H_+^2/(1+\alpha_+)$, which includes the domain of
validity of the standard cosmological model ($H\ll H_+$).

In the period dominated by matter (which corresponds to lower values
of the Hubble rate) we assume that only the dependence on the lower
bound ($H_-$) of the Hubble rate is relevant. Then one has
\be
\frac{\dH}{H^2} \,=\, - \frac{3}{2} \, \frac{1}{1 - \frac{x
    h_-^{'}(x)}{h_-(x)}}\, ,
\ee
with $x=H_{-}^2/H^2$. We can also consider a dependence on $H_-$
parametrized simply by an exponent $\alpha_-$
\be
g(H) \,=\, H^2 \, \left(1 - \frac{H_-^2}{H^2}\right)^{\alpha_-}
\label{alpha-}\, .
\ee
With this choice one has a transition from a decelerated expansion
for $H^2 > H_{-}^2(1+\alpha_-/2)$ leaving the domain of validity of
the standard cosmological model and entering into an accelerated
expansion when the Hubble rate approaches its lower bound for $H^2 <
H_{-}^2(1+\alpha_-/2)$.

The possibility of describing $\rho$ as a function of $H$ is independent
of the existence of spatial curvature $k$. However, in the kinematics of
observables in the expanding universe we do need to specify the value of
$k$. In the rest of the paper we will assume that the universe is flat
($k=0$), although the same analysis could be done for arbitrary $k$.

The properties of the expansion obtained in this simple example (a
period of decelerated expansion separating two periods of accelerated
expansion) are general to the class of phenomenological models with a
generalized first Friedman equation (\ref{rho_g}) with $g(H)$
satisfying the consistency conditions (\ref{gcons}) and a Hubble rate
constrained to a finite interval. The specific part of the example
defined by (\ref{alpha+},\ref{alpha-}) is the simple dependence on the
Hubble rate bounds and the values of the Hubble rate at the
transitions between the three periods of expansion. From now on we
will name this description the Asymptotic Cosmological Model (ACM).
With respect to the late accelerated expansion, ACM can be seen as a
generalization of $\Lambda$CDM, which can be recovered by setting
$\alpha_-=1$. It also includes an early period of exponential
expansion which can be seen as a phenomenological description of the
evolution of the universe at inflation in the homogeneous approximation.

\subsection{Horizon problem}
One can see that the horizon of a radiation-dominated universe can be
made arbitrarily large as a consequence of an upper bound on the
Hubble parameter and in this way one can understand the observed
isotropy of the cosmic microwave background at large angular scales.

Let us consider the effect of the modification of the cosmological
model on the calculation of the distance $d_h(t_f,t_i)$ of a source of
a light signal emitted at time $t_i$ and observed at time $t_f$
\be
d_h(t_f, t_i) \,=\, c \, a(t_f) \int_{t_i}^{t_f} \frac{dt}{a(t)}\, .
\ee
We have
\be
\frac{dt}{a} \,=\, \frac{da}{a^2 H}\,=\, \left(\frac{8 \pi G_N
  \sigma}{3}\right)^{-1/4} \frac{g^{'}(H) dH}{4H g(H)^{3/4}} \, ,
\ee
where in the first step we have used the definition of the Hubble
expansion rate $H$ and in the second step we have used the relation
between the scale factor $a$ and $H$ as given by
(\ref{rad}-\ref{rho_g}). We are considering both times $t_i$ and $t_f$
in the radiation dominated period.

The distance $d_h$ is then given by
\be
d_h(H_f, H_i) \,=\, \frac 1{4 g(H_f)^{1/4}} \int_{H_f}^{H_i} \frac{dH
  g^{'}(H)}{H g(H)^{3/4}} \label{dist}\, .
\ee
If $H_i$ is very close to $H_+$ (i.e. if we choose the time $t_i$ when
the light signal is emitted well inside the period of accelerated
expansion) then the integral is dominated by the
region around $H_i$ which is very close to $H_+$. Then one can
approximate in the integrand
\be
g(H) \approx \frac{H_+^2}{\left(1-\frac{H^2}{H_+^2}\right)^{\alpha_+}} \, .
\ee
On the other hand if the observation is made at a time $t_f$ within the
domain of validity of the cosmological standard model ($H_-^2\ll
H_f^2\ll H_+^2$) then the factor $g(H_f)^{1/2}$ in
front of the integral can be approximated by $H_f$ and then one has
\be
d_h(H_f, H_i) \approx \frac{\alpha_+}{4 \sqrt{H_f H_i}}
B_{\frac{H^2_i}{H^2_+}}(1/2,-\alpha/4) \, ,
\ee
where $B_z(m,n)$ is the incomplete Beta function, and it can be made
arbitrarily large by choosing $H_i$ sufficiently close
to $H_+$. In this way we see that a cosmological model with a finite
interval of variation for $H$ solves the horizon problem.

\section{Constraints of ACM by Observations}

In this section we will carry out a more technical analysis about how
the astrophysical observations constrain the parameter space of ACM in
the matter dominance period. This analysis is based on the use of
(assumed) standard candles, basically Type Ia Supernovae \cite{Riess:2006fw}
and CMB \cite{Spergel:2006hy}. These observations constrain the parameter
space to confidence regions in which the combination
$\Omega_m\equiv(1-H_-^2/H_0^2)^{\alpha_-}$ is constrained to be around
one quarter.
The consideration of both Type Ia SNe and CMB together favor
$\alpha_->1.5$. The results of this analysis can be seen in
figures (FIG. 1-3).
A reader not interested in technical details might well skip this section.

We center our discussion of the experimental tests of the Asymptotic
Cosmological Model
in the late accelerated expansion produced when $H$ approaches its
lower bound $H_-$. The vast amount of supernovae data collected by
\cite{Riess:2006fw} and \cite{Astier:2005qq}, the data from the SDSS  Baryon
Acoustic Oscillation \cite{Eisenstein:2005su}, the mismatch between total energy
density and total matter energy density seen at CMB anisotropies
\cite{Spergel:2006hy} and the measurements of present local mass density by
2dF and SDSS \cite{Tegmark:2003ud,Cole:2005sx} compared with the measurements of $H_0$
from the HST Cepheids \cite{Freedman:2000cf} show that the universe undergoes
a surprising accelerated expansion at the present time.

We will firstly confront the model with the Supernovae Ia data from
Riess \textit{et al.} and SNLS collaboration.
The usefulness of the Supernovae data as a test of Dark Energy models
relies on the assumption that Type Ia SNe behave as standard candles,
i.e., they have a well defined environment-independent luminosity
$\mathcal{L}$ and spectrum. Therefore, we can use measured bolometric
flux $\mathcal{F}=\frac{\mathcal{L}}{4\pi d_L^2}$ and frequency to
determine luminosity distance $d_L$  and redshift $z$.
The luminosity distance is given now by

\be
d_L(z) \, = \, c(1+z)\int_0^z \frac{dz'}{H(z')}
\, = \, \frac{c(1+z)}{3}\int^{H(z)}_{H_0}\frac{g'(H)dH}{H g^{1/3}(H_0)g^{2/3}(H)}\, .
\ee

The computed value must be compared with the one obtained
experimentally from the measured extinction-corrected
distance moduli ($\mu_0=5 log_{10}(\frac{d_L}{1 Mpc})+25$) for each
SN. The SNe data have been compiled in references \cite{Riess:2006fw}, and
we have limited the lowest redshift at $cz<7000 km/s$ in order to
avoid a possible ``Hubble Bubble'' \cite{Jha:2002af,Zehavi:1998gz}. Therefore our
sample consists on 182 SNe. We will determine the likelihood of the
parameters from a $\chi^2$ statistic,
\be
\chi^2(H_0,H_-,\alpha_-)=
\sum_i\frac{(\mu_{p,i}(z_i;H_0,H_-,\alpha_-)-\mu_{0,i})^2}{\sigma^2_{\mu_{0,i}}+\sigma^2_v}\,,
\ee
where $\sigma_v$ is the dispersion in supernova redshift due to
peculiar velocities  (we adopt $\parallel\mathbf{v_p}\parallel = 400
km/s$ in units of distance moduli), $\sigma_{\mu_{0,i}}$
is the uncertainty in the individual measured distance moduli
$\mu_{0,i}$, and $\mu_{p,i}$ is the value of $\mu_0$ at $z_i$
computed with a certain value of the set of parameters
${H_0,H_-,\alpha_-}$. This $\chi^2$ has been marginalized over
the nuisance parameter $H_0$ using the adaptive method in reference
\cite{Wang:2004xz}. The resulting likelihood
distribution function $e^{-\chi^2/2}$ has been explored using Monte
Carlo Markov Chains. We get a best fit
of ACM at $\alpha_-=0.36$ and $\frac{H_-}{H_0}=0.95$, for which
$\chi^2=157.7$. In contrast, fixing $\alpha_-=1$, we get
$\Omega_\Lambda\equiv\frac{H_-^2}{H_0^2}=0.66$ and
$\chi^2=159.1$ for the best fit $\Lambda$CDM. The confidence regions
are shown in Fig. 1 (top).

\begin{figure}
 \centerline{\includegraphics[scale=0.6,
 width=9cm]{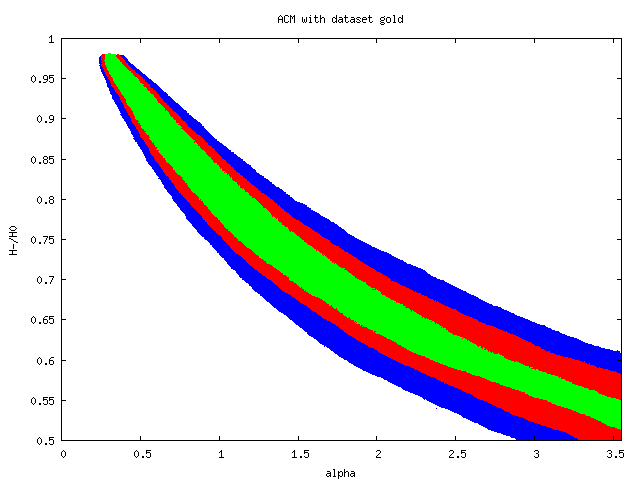}}
\end{figure}
\begin{figure}

\centerline{\includegraphics[scale=0.6,
 width=9cm]{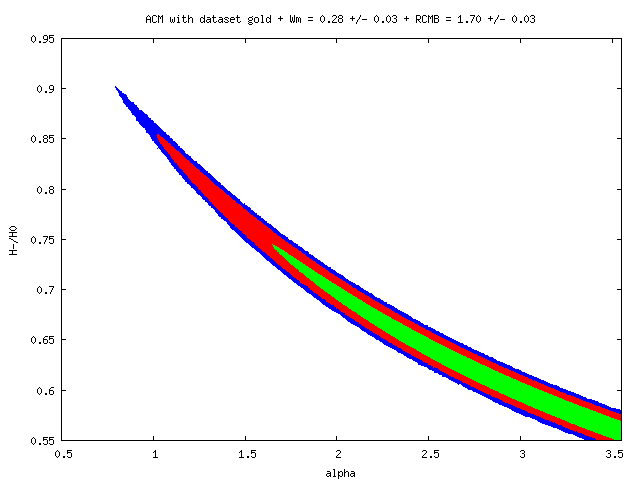}}

\label{confidence}
\caption{Confidence regions in parameter space of the Asymptotic
  Cosmological Model (ACM) with (bottom) and without (top) priors from
  CMB and estimations of present matter energy density at $1\sigma$
  (green), $2\sigma$ (red) and $3\sigma$ (blue). The $\Lambda$CDM is
  inside the 1$\sigma$ region if we consider no priors, but is moved
  outside the $2\sigma$ region when we confront the SNe data with CMB
  and estimations from present matter energy density. In this case a
  value $\alpha_-> 1.5$ is favored.}
\end{figure}

We can add new constraints for the model coming from measurements of
the present local matter energy density from the combination of 2dF
and SDSS with HST Cepheids, rendering $\Omega_m = 0.28 \pm 0.03$; and
the distance to the last scattering surface from WMAP, which leads to
$r_{CMB}\equiv\sqrt{\Omega_m}\int^{1089}_0 \frac{H_0
  dz'}{H(z')}=1.70\pm0.03$ \cite{Riess:2006fw}.
Including these priors we get a best fit of ACM at $\alpha_-=3.54$ and
$\frac{H_-}{H_0}=0.56$ (our simulation explored the region with
$\alpha_-<3.6$), for which $\chi^2=164.1$. In contrast, we get
$\Omega_\Lambda=0.73$ and $\chi^2=169.5$ for the best fit
$\Lambda$CDM, which is outside the $2\sigma$ confidence region shown
in Fig. 1 (bottom). The fits of the best fit $\Lambda$CDM and ACM
taking into account the priors to the SNe data are compared in Fig. 2.
The information which can be extracted from the data is limited. This
can be seen in Fig. 3, in which it is explicit that the data
constrain mainly the value of the present matter energy density,
$\Omega_m\equiv\frac{\rho_m}{\rho_C}=(1-\frac{H_-^2}{H_0^2})^{\alpha_-}$.

\begin{figure}
 \centerline{\includegraphics[scale=0.6, width=9cm]{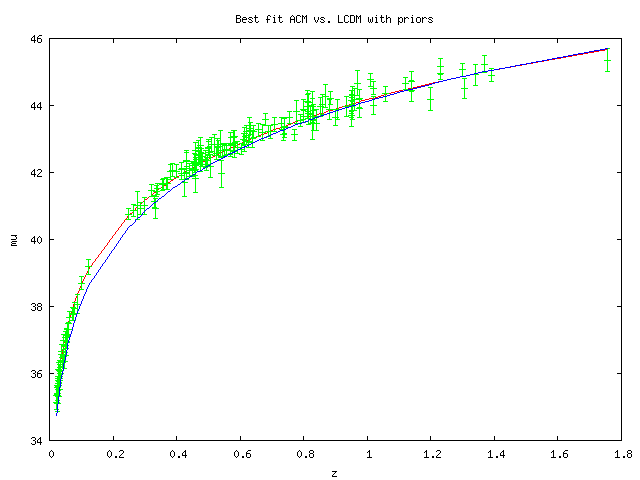}}
\label{bestfit}
\caption{Distance Moduli vs. Redshift comparison between the best
  fits of ACM ($H_0 = 66 \, Km/s MPc$, red) and $\Lambda$CDM ($H_0 =
  65 \, Km/s MPc$, blue) to the SNe data with priors. ACM fits clearly
  better the medium redshift SNe. }
\end{figure}

\begin{figure}
 \centerline{\includegraphics[scale=0.6, width=9cm]{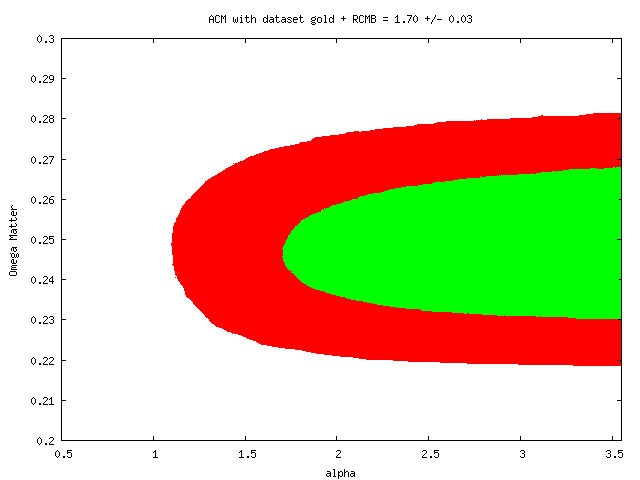}}
\label{omegabestfit}
\caption{$\Omega_m$ vs. $\alpha$ confidence regions with the CMB prior
  only. The confidence regions show $\Omega_m = 0.25 \pm 0.03$ and
  $\alpha > 1.15$ at $2\sigma$ level. }
\end{figure}

\section{Generalized First Friedman Equation and $f(R)$ gravity}

The extension of the cosmological model considered here could be
compared with recent works on a modification of gravity at large
or very short distances \cite{Gasperini:1991ak,Carroll:2003wy,Sotiriou:2005hu,Hu:2007nk,Nojiri:2006ri,Nojiri:2003ft,Cognola:2007zu}.
It has been shown that, by considering a correction to the
Einstein Hilbert action including positive and negative powers of the
scalar curvature, it is possible to reproduce an accelerated expansion
at large and small values of the curvature in the cosmological
model. Some difficulties to make these modifications of the
gravitational action compatible with the solar system tests of general
relativity have lead to consider a more general gravitational action,
including the possible scalars that one can construct with the
Riemann curvature tensor \cite{Carroll:2004de}, although the Gauss-Bonnet scalar is the only
combination which is free from ghosts and other pathologies. In fact, there
is no clear reason to restrict the extension of general relativity
in this way. Once one goes beyond the derivative expansion, one should
consider scalars that can be constructed with more than
two derivatives of the metric and then one does not have a good
justification to restrict in this way the modification of the
gravitational theory. It does not seem difficult to find an
appropriate function of the scalar curvature or the Gauss-Bonnet
scalar, which leads to a cosmology with a bounded Hubble expansion rate.

One may ask if a set of metric f(R) theories which include ACM as an
homogeneous and isotropic solution exists.
The answer is that a bi-parametric family of f(R) actions which lead to
an ACM solution exists. In the following section we will derive them
and we will discuss some examples. The derivation follows the same
steps of modified f(R)-gravity reconstruction from any FRW cosmology
\cite{Nojiri:2006gh}.

We start from the action
\be
 S=\frac 1{16 \pi G_N}\int d^4 x\sqrt{-g}[f(R) + 16\pi G L_m]\, ,
\ee
which leads to the ``generalized first Friedman equation''
\be
 -18(H \ddot H +4 H^2 \dot H)f''(R)
-3(\dot H +H^2)f'(R)
-\frac 12 f(R) = \left( 8\pi G_N \right) \rho \, ,
\ee
where
\be
R = -6(\dot{H}+2H^2)\, .
\ee
If the energy density of the universe is mainly due to matter (as in
the late accelerated expansion), we can use (\ref{matter}) and
(\ref{rho_g}) to express $\rho$, $R$, $\dH$ and $\ddH$ as functions of
$H$. Thus we get
\be
  [H^3 g g'^2-3H^2 g^2 g'+3H^3 g^2 g'']f''(R)
+\frac{[3H g g'^2-H^2 g'^3]}{18}f'(R)-\frac{g'^3}{108}f(R)=\frac{g g'^3}{18} \, , \label{difR}
\ee
where
\be
R(H) = 18 H g(H)/g'(H)-12 H^2 \label{RH}\, .
\ee
$H(R)$ can be obtained from (\ref{RH}) and set into (\ref{difR}); then
we get an inhomogeneous second order linear differential equation with
non-constant coefficients. Therefore, there will always be a
bi-parametric family of $f(R)$ actions which present ACM as their
homogeneous and isotropic solution. The difference between these
actions will appear in the behavior of perturbations, which is not
fixed by (\ref{rho_g}). Some of these actions are particularly easy to
solve. If $g(H)=H^2$ then $R=-3H^2$ and the differential equation
becomes
\begin{equation}
6 R^2 f''(R) -R f'(R)-f(R) +2 R=0\, .
\end{equation}
Its general solution is
\be
f(R)=R+c_1 R^{\frac 1{12}(7-\sqrt{73})}+c_2 R^{\frac 1{12}(7+\sqrt{73})}\, ,
\ee
which will give (\ref{rho_g}) as First Friedman equation as long as
the radiation energy density can be neglected. In general the
differential equation (\ref{difR}) will not be solvable
analytically, and only approximate solutions can be found as power
series around a certain singular point $R_0$ (a value of $R$ such that
$H(R)$ cancels out the coefficient of $f''(R)$ in (\ref{difR})). These
solutions will be of the form $f(R)=f_p(R) +c_1 f_1(R) +c_2 f_2(R)$
with
\be
\frac{f_i(R)}{R_0}=\sum_{m=0}^\infty a_m^{(i)} (\frac{R-R_0}{R_0})^{s_i+m} \, ,
\ee
where $i=p,1,2$, $a_0^{(1)}=a_0^{(2)}=1$ and the series will converge
inside a certain radius of convergence. An interesting choice of $R_0$
is $R_0 = -12 H_-^2 \equiv R_-$, which is the value of $R$ at
$H_-$. One can also find the approximate solution of the differential
equation for $\mid R\mid\gg\mid R_-\mid$, which will be of the form
\be
\frac{f_i(R)}{R}=\sum_{m=0}^\infty a_m^{(i)} (\frac{R_-}{R})^{s_i+m} \label{fR-}\, .
\ee
One can in principle assume that the action (\ref{fR-}) could be
considered as valid also in the region in which radiation begins to
dominate, but this action reproduces (\ref{rho_g}) only if matter
dominates. However, some of the new terms appearing in (\ref{fR-})
will be negligible against $R$ when radiation begins to dominate. The
others can be canceled out by setting to zero the appropriate integration
constant and therefore the Cosmological Standard Model will be
recovered as a good approximation when radiation begins to dominate.

 $f(R)$-theories are not the only modified gravity theories studied in
the literature; $f(G)$-theories \cite{Nojiri:2005jg} are also a popular field of research.
In these theories, the Einstein-Hilbert action is supplemented by a
function of the Gauss-Bonnet scalar
$G=R^2-4R_{\mu\nu}R^{\mu\nu}+R_{\mu\nu\rho\sigma}R^{\mu\nu\rho\sigma}$
(not to be confused with the gravitational coupling $G_N$).The same
approach can be used to answer what $f(G)$ actions are able to
reproduce ACM homogeneous evolution, once more in analogy with the
modified f(G)-gravity reconstruction of a FRW cosmology
\cite{Nojiri:2006je}. The form of the action will be
\be
 S=\frac 1{16 \pi G_N}\int d^4 x\sqrt{-g}[R+f(G) + 16\pi G_N L_m]\, ,
\ee
from which we derive the modified Friedman equation
\be
 \left(8\pi G_N\right) \rho = 3 H^2-\frac 12 G f'(G)+\frac 12
 f(G)+12f''(G)\dot{G}H^3 \, .
\ee
Following the same procedure as in the case of $f(R)$-theories we
arrive to the differential equation
\be
  3 g(H) = 3 H^2-\frac 12 G f'(G)+\frac 12 f(G)
+\frac{864 H^6 g}{g'^3}[9 g g'-H g'^2-3 H g g''] f''(G) \label{difG} \, ,
\ee
where
\be
G=24 H^2(H^2-3H g/g')\label{GH}\, .
\ee
For a given $g(H)$, one can use (\ref{GH}) to get $H(G)$, then set it
into (\ref{difG}) and solve the second order linear differential
equation. As in the previous case, there will be a bi-parametric family
of solutions for $f(G)$ which will have (\ref{rho_g}) as their
homogeneous isotropic solution as long as matter dominates. The
parameters will need to be fixed in order to make the contribution of
undesired terms in the Friedman equation to be negligible when
radiation dominates. Again $g(H)=H^2$ is an example which can be
solved analytically. The differential equation turns to be
\be
12 G^2 f''(G)-G f'(G)+f(G)=0\, ,
\ee
which has the solution
\be
f(G)=c_1 G + c_2 G^{1/12}\, .
\ee
In most cases this procedure will not admit an analytical solution
and an approximate solution will need to be found numerically.

A similar discussion can be made to explain the early accelerated
expansion (inflation) as a result of an $f(R)$ or $f(G)$ action,
taking into account that in this case the dominant contribution to the
energy density is radiation instead of matter. The analogue to
(\ref{fR-}) will be now a solution of the form
\be
\frac{f_i(R)}{R}=\sum_{n=0}^\infty a_n^{(i)} (\frac{R}{R_+})^{r_i+n} \label{fR+}\, ,
\ee
for $\vert R \vert \ll \vert R_+\vert$ with $R_+ = -12 H_+^2$.
Terms in the action which dominate over $R$ when $R$ becomes small
enough should be eliminated in order to recover the Standard
Cosmological model before matter starts to dominate.

Both accelerated expansions can be described together in the
homogeneous limit by an $f(R)$ action with terms
$(\frac{R_-}{R})^{s_i+m}$ with $s_i>0$ coming from (\ref{fR-}) and
terms $(\frac{R}{R_+})^{r_i+n}$ with $r_i>0$ coming from
(\ref{fR+}). The action will contain a term, the Einstein-Hilbert
action $R$, which will be dominant for $H_-\ll H\ll H_+$ including the
period when matter and radiation have comparable energy densities.

\section{Alternative Descriptions of the ACM}

In general, by virtue of the gravitational field equations, it is
always possible to convert a modification in the gravitational term of
the action to a modification in the matter content of the universe. In
particular, at the homogeneous level, it is possible to convert a
generalized first Friedman equation of the type (\ref{rho_g}) to an
equation in which, apart from the usual matter term, there is a dark
energy component with an unusual equation of state $p=p(\rho)$ and in
which General Relativity is not modified.

The trivial procedure is the following. Assume that the source of the
gravitational field in the modified gravity theory behaves as
$p_d=\omega \rho_d$ (matter or radiation). This is the case when the
modification of gravity is relevant. We can then define a dark energy
or effective gravitational energy density as

\be
 \rho_g =\frac{3}{8\pi G_N} \left( H^2-g(H)\right)  \label{dedensity}\, .
\ee
The continuity equation (\ref{cont_eq}) allows to define a pressure
for this dark energy component as $p_g=-\rho_g-\dot{\rho}_g/3 H$. The
equation of state of the dominant content of the universe leads to
\be
\dH=-3(1+\omega)\frac{H g(H)}{g'(H)} \label{dH}\, ,
\ee
which can be used to express $\dot{\rho}_g$ as a function of $H$, and
one gets the final expression for $p_g$,
\be
p_g=\frac{-3}{8\pi G_N}\left(
 H^2-2(1+\omega)H\frac{g(H)}{g'(H)}+\omega g(H)\right)
 \label{depressure}\, .
\ee
Then, for any given $g(H)$ we can use (\ref{dedensity}) to get
$H=H(\rho_g)$ and substitute in (\ref{depressure}) to get an
expression of $p_g(\rho_g;H_\pm,\omega)$ which can be interpreted as
the equation of state of a dark energy component. In the simple case
of a matter dominated universe with ACM, $\alpha_-=1$ gives obviously
a dark energy component verifying $p_g=-\rho_g$. Another simple
example is $\alpha_-=2$, for which
\be
\left(\frac{8 \pi G_N}{3 H_-^2} \right) p_g=\frac{2}{\frac{8 \pi
    G_N}{3 H_-^2}\rho_g -3}\, .
\ee

The inverse procedure is also straightforward. Suppose we have a
universe filled with a standard component $p_d=\omega\rho_d$ and a
dark energy fluid $p_g=p_g(\rho_g)$. Using the continuity equation of
both fluids one can express their energy densities as a function of
the scale factor $a$ and then use this relation to express $\rho_g$ as
a function of $\rho_d$. Then the Friedman equation reads
\be
H^2=\frac{8\pi G_N}{3}(\rho_d+\rho_g(\rho_d))\, ,
\ee
which, solving for $\rho_d$, is trivially equivalent to (\ref{rho_g}).
This argument could be applied to reformulate any cosmological model
based on a modification of the equation of state of the dark energy
component \cite{Nojiri:2005sr} as a generalized first Friedman equation
(\ref{rho_g}).

Until now we have considered descriptions in which there is an exotic
constituent of the universe besides a standard component (pressureless
matter or radiation). In these cases, pressureless matter includes both
baryons and Dark Matter. However, there are also descriptions in which
Dark Matter is unified with Dark Energy in a single constituent of the
universe. One of these examples is the Chaplygin gas $p_g=-1/\rho_g$
\cite{Kamenshchik:2001cp,Alvarenga:2001nm}. The use of the continuity equation leads to
\be
\rho_g=\sqrt{A+B a^{-6}}\, ,
\ee
where $A$ and $B$ are integration constants. The previous method can
be used to find the generalized first Friedman equation for the baryon
density in this model,
\be
\frac{8\pi
  G_N}{3}\rho_b=\frac{H^2}{k}\left(\sqrt{1+k(1-\frac{H_-^4}{H^4}})-1
\right) \label{Chap}\, ,
\ee
where $H_-=\sqrt{\frac{8\pi G}{3}}A^{1/4}$, $k=B \rho_{b0}^{-2}-1$,
and $\rho_{b0}$ is the present value of $\rho_b$. In the $H\gg H_-$
limit this model can be interpreted as a universe filled with baryons
and dark matter or as a universe filled with baryons and with a higher
effective value of $G_N$. Equation (\ref{Chap}) does not fulfill
(\ref{hpm}) because it describes the behavior of just baryon
density. In the $H\gtrsim H_-$ limit, the model can be interpreted as
a universe filled with baryons and a cosmological constant or as an
ACM model with $\alpha_-=1$ and filled only with baryons.

A similar example in the early period of accelerated expansion would
be a universe filled with a fluid which behaves as
a fluid of ultra-relativistic particles if the energy density is low
enough but whose density has an upper bound
\be
\rho_X=\frac\sigma{a^4+C}\, .
\ee
This dependence for the energy density in the scale factor follows
from the equation of state
\be
p_X=\frac 13 \rho_X-\frac {4 C}{3 \sigma}\rho_X^2 \, .
\ee
This model turns out to be equivalent to a universe filled with
radiation following eq. (\ref{alpha+}) with $\alpha_+=1$ and
$H_+^2=(\frac{8 \pi G_N}{3})\frac{\sigma}{C}$.

Another equivalent description would be to consider that the universe
is also filled with some self interacting scalar field $\varphi$ which
accounts for the discrepancy between the standard energy-momentum and
GR Einstein tensors. Given an arbitrary modified Friedman equation
(\ref{rho_g}) a potential $V(\varphi)$ can be found such that the
cosmologies described by both models are the same. The procedure is
similar to the one used to reconstruct a potential from a
given cosmology \cite{Nojiri:2005pu}. From the point of
view of the scalar field, the cosmology is defined by a set of three
coupled differential equations
\begin{eqnarray}
 \ddot{\varphi}+3 H \dot{\varphi}+V'(\varphi)&=0 \label{klein}\\
\frac{8\pi G_N}{3}(\rho_d+\frac{\dot{\varphi}^2}2 +V(\varphi))&=H^2
\label{quint}\\ -3(1+\omega)H\rho_d&=\dot{\rho_d} \label{contw}\, .
\end{eqnarray}
The solution of these equations for a certain $V(\varphi)$ will give
$H(t)$ and $\rho_d(t)$, and therefore $\rho_d(H)$ which is $g(H)$ up
to a factor $\frac{8 \pi G_N}{3}$. In this way one finds the
generalized first Friedman equation associated with the introduction
of a self interacting scalar field.
Alternatively, given a function $g(H)$ in a generalized first Friedman
equation (\ref{rho_g}) for the density $\rho_d$, one can find a scalar
field theory leading to the same cosmology in the homogeneous
limit. By considering the time derivative of (\ref{quint}) and using
(\ref{klein}) and (\ref{contw}) one gets
\be
\dot{\varphi}^2=-(1+\omega)\rho_d-\frac{\dH}{4\pi G_N} \label{dtvarphi}\, ,
\ee
where we can use (\ref{dH}) and (\ref{rho_g}) to get $\dot{\varphi}^2$
as a function of $H$. Setting this on (\ref{quint}) we get
$V(\varphi)$ as a function of $H$. On the other hand,
$\varphi'(H)=\dot{\varphi}/\dH$, so
\begin{equation}
\frac{8\pi G_N}{3}(\varphi'(H))^2=\frac{(2H
 -g'(H))g'(H)}{9(1+\omega) H^2 g(H)}
\label{phi}
\end{equation}
and
\begin{equation}
\frac{8\pi G_N}{3}V(\varphi(H))=H^2-g(H)+\frac{(1+\omega)(2H-g'(H))g(H)}{2
   g'(H)}\, .
\label{V(phi)}
\end{equation}
If $g(H)$ is such that the rhs of (\ref{phi}) is positive definite, it
can be solved and the solution $\varphi(H)$ inverted and substituted
into (\ref{V(phi)}) in order to get $V(\varphi)$. This is the case of
a function of the type (\ref{alpha-}) with $\alpha_->1$. For the
particular case of an ACM expansion with $\alpha_-=1$, the solution is
a flat potential $V=V_0$ and a constant value of $\varphi=\varphi_0$.

If $g(H)$ is such that the rhs of (\ref{phi}) is negative definite, as
it happens in (\ref{alpha-}) with $\alpha_-<1$ or in (\ref{alpha+}),
the problem can be solved by changing the sign of the kinetic term in
(\ref{quint}). The result is a phantom quintessence in the case of
(\ref{alpha-})  with $\alpha_-<1$. The case (\ref{alpha+}) is more
complicated because the energy density of the associated inflaton
turns out to be negative. Moreover, it is of the same order as the
energy density of ultra-relativistic particles during the whole period
of accelerated expansion. Therefore, it seems that this model is
inequivalent to other inflation scenarios previously studied.

In summary, there are many equivalent ways to describe the discrepancy
between observed matter content of the universe and Einstein's General
Relativity. At the homogeneous level, it is trivial to find relations
among them. The possibility to establish the equivalence of different
descriptions is not a peculiarity of the description of this discrepancy
in terms of a generalized first Friedman equation (\ref{rho_g}). The
same relations can be found among  generalized equations of state,
scalar-tensor theories and f(R) modified gravity \cite{Capozziello:2005mj}.

\section{Summary and discussion}
It may be interesting to go beyond $\Lambda$CDM in the description of
the history of the universe in order to identify the origin of the
two periods of accelerated expansion. We have proposed to use the
expression of the energy density as a function of the Hubble parameter
as the best candidate to describe the history of the universe. In this
context the late time period of accelerated expansion and the early
time inflation period can be easily parametrized.

We have considered a simple modification of the cosmological equations
characterized by the appearance of an upper and a lower bound on the
Hubble expansion rate. A better fit of the experimental data can be obtained
with this extended cosmological model as compared with the
$\Lambda$CDM fit. Once more precise data are available, it will be
possible to identify the behavior of the energy density as a function
of the Hubble parameter and then look for a theoretical derivation of
such behavior.

We plan to continue with a systematic analysis of different
alternatives incorporating the main features of the example considered
in this work. Although the details of the departures from the standard
cosmology can change, we expect a general pattern of the effects induced
by the presence of the two bounds on $H$. We also plan to go further,
considering the evolution of inhomogeneities looking for new consequences
of the bounds on $H$.

The discussion presented in this work, which is based on a new description
of the periods of accelerated expansion of the universe, can open a new way
to explore either modifications of the theory of gravity or new components
in the universe homogeneous fluid. Lacking theoretical criteria to select
among the possible ways to go beyond $\Lambda$CDM, we think that the
phenomenological approach proposed in this work is justified.

We are grateful to Paola Arias and Justo L\'opez-Sarri\'on for
discussions in the first stages of this work. We also acknowledge
discussions with Julio Fabris, Antonio Segu\'{\i}, Roberto Empar\'an,
Sergei Odintsov, Aurelio Grillo and Fernando M\'endez.
This work has been partially supported by CICYT (grant
FPA2006-02315) and DGIID-DGA (grant2008-E24/2). J.I. acknowledges a
FPU grant from MEC.

%%%%%%%%

\chapter{Linearized Treatment of Scalar perturbations in the Asymptotic Cosmological Model}

In this paper the implications of a recently proposed
phenomenological model of cosmology, the Asymptotic Cosmological
Model (ACM), on the behavior of scalar perturbations are
studied. Firstly we discuss new fits of
the ACM at the homogeneous level, including fits to the Type Ia
Supernovae UNION dataset, first CMB peak of WMAP5 and BAOs. The
linearized equations of scalar perturbations in the FRW metric are
derived. A simple model is used to compute the CMB
temperature perturbation spectrum. The results are compared with the
treatment of perturbations in other approaches to the problem of the
accelerated expansion of the universe.

\section{Introduction}
According to General Relativity (GR), if the universe is filled with the particles
of the Standard Model of particle physics, gravity should lead to a
decelerated expansion of the universe. However, in 1998 two
independent evidences of present accelerated expansion were presented
\cite{Riess:1998cb,Perlmutter:1998np} and later confirmed by different
observations \cite{Jaffe:2000tx,Spergel:2003cb,Peacock:2001gs}.

There is no compelling explanation for this cosmic acceleration,
but many intriguing ideas are being explored. These ideas can be classified
into three main groups: new exotic sources of the gravitational field with
large negative pressure \cite{Turner:1998ex,Zlatev:1998tr,ArmendarizPicon:2000dh} (Dark Energy), modifications of
gravity at large scales \cite{Bekenstein:1984tv,Bekenstein:2004ne} and rejection of the spatial
homogeneity as a good approximation in the description of the present
universe \cite{Tomita:2000jj,Kolb:2005da,Wiltshire:2007zj}.

Different models (none of them compelling) for the source responsible
of acceleration have been considered. Einstein
equations admit a cosmological constant $\Lambda$, which can be realized as
the stress-energy tensor of empty space. This $\Lambda$ together with Cold Dark
Matter, Standard Model particles and General Relativity form the current
 cosmological model, $\Lambda$CDM. However, quantum field theory predicts a
value of $\Lambda$ which is 120 orders of magnitude higher than observed.
Supersymmetry can lower this value 60 orders of magnitude, which is
still ridiculous \cite{Weinberg:1988cp}.
In order to solve this paradox, dynamical Dark Energy models have been proposed.

This has also lead to explore the possibility that cosmic acceleration arises
from new gravitational physics. Again here several alternatives for a
modification of the Einstein-Hilbert action at large and small
curvatures \cite{Faraoni:2008mf}, or even higher
dimensional models \cite{Deffayet:2000uy,Dvali:2000hr}, producing an accelerated
expansion have been identified.
All these analyses include an {\it ad hoc} restriction to actions involving simple
functions of the scalar curvature and or the Gauss-Bonnet tensor.
This discussion is sufficient to establish the point that cosmic acceleration can
be made compatible with a standard source for the gravitational field but it is
convenient to consider a more general framework in order to make a systematic
analysis of the cosmological effects of a modification of general relativity.

The Asymptotic Cosmological Model (ACM) was presented
\cite{Cortes:2008fy} as a strictly phenomenological
generalization of the Standard Cosmological Model including a
past and a future epoch of accelerated expansion. It follows from the
assumptions that GR is not a fundamental theory, but only a good
approximation when the Hubble rate $H$
  is between but far away from two
fundamental scales $H_-$ and $H_+$, which act as bounds on
$H$. A general covariant metric theory of gravity without
spatial curvature is assumed. The model is well defined in the
homogeneous approximation
and includes $\Lambda$CDM as a particular case.

In next section we review the ACM and we provide new fits
to the Type Ia Supernovae UNION dataset, first acoustic peak of CMB of
WMAP5 and BAOs. In the third section we
derive the linearized equations of the scalar perturbations of the
metric, following from
general covariance and a single new assumption on the
  perturbations. In the fourth section
we will consider how to solve the system of equations for adiabatic
perturbations in a given fluid. In the fifth section we derive the
CMB spectrum in the ACM.  In the sixth
  section we will compare the treatment of the scalar perturbations in
the ACM with other models
  which are equivalent in the homogeneous approximation. The last
section is devoted to the summary and conclusions.

\section{The Asymptotic Cosmological Model: Homogeneous Background}

The Asymptotic Cosmological Model (ACM) was introduced in
Ref. \cite{Cortes:2008fy}. In this model the universe is filled with
photons and neutrinos (massless particles), baryons (electrically
charged massive particles) and Dark Matter (electrically neutral
massive particles), but General Relativity (GR) is only a good
approximation to the gravitational interaction in a certain range of
the Hubble rate $H$, between but far from its
two bounds, $H_-$ and $H_+$. General Covariance and the absence of
spatial curvature are assumed.

The gravitational part of the action might include derivatives of the
metric of arbitrarily high order, and therefore arbitrarily high
derivatives of the scale factor should appear in the Friedman
Equations. We should start by considering a generalized first Friedman
equation
\be
8\pi G \rho(t) = 3 f(H(t),\dot{H}(t),...,H^{(n)}(t),...) \label{fh} \, ,
\ee
and the corresponding equation for the pressure will be derived using the continuity
equation
\be
\dot{\rho} = -3 H(\rho + p)\, . \label{cont}
\ee

However, as the resulting differential equations should be solved and only one of
its solutions deserves interest (the one describing the
evolution of the universe), we can use the one to one
correspondence between time $t$ and Hubble parameter $H$ (assuming $\dot{H} < 0$)
to write the modified First Friedman Equation as a bijective map linking
the total energy density $\rho$ with $H$ in our universe
\be
8\pi G \rho = 3 g(H) \label{gh} \, .
\ee
The use of the continuity equation \eqref{cont} enables us to write the modified
Friedman Equation for the pressure evaluated at the solution
corresponding to the cosmic evolution

\be
-8\pi G p = 3 g(H) + g'(H)\dot{H}/H \label{ph} \, .
\ee

The evolution of the universe will be determined by the concrete form of
the function $g(H)$, which we assume to be smooth. However, the most significant
features of the evolution at a given period can be described by some
simple approximation to $g(H)$ and the matter content. Those are a
pole at $H = H_+$ of order $\al_+$ and a zero of order $\al_-$ at $H =
H_-$. The energy density of the universe has a contribution from
both massless particles (radiation) and massive ones (matter).

For the sake of simplicity, the history of the universe can be then divided into
three periods. In the first period, $H \gg H_-$ so the effect of the lower bound
can be neglected and the universe is radiation dominated. The universe undergoes
(and exits from) an accelerated expansion which we call inflation. The simplest
parametrization is

\be
g(H) \approx H^2 \left( 1 - \frac{H^2}{H_+^2} \right)^{-\al_+} \, . \label{past}
\ee

In the second period we cannot neglect neither the effect of radiation nor the
effect of nonrelativistic matter, but $H_+ \gg H \gg H_-$. Then GR
offers a good description
of the gravitational interaction, $g(H) \approx H^2$ in this region, and the
universe performs a decelerated expansion.

In the third period, $H_+ \gg H$ so the effect of the upper bound can be
neglected, and the universe is matter dominated. This period corresponds to
the present time in which the universe also undergoes an accelerated expansion. The
simplest choice is

\be
g(H) \approx H^2 \left( 1 - \frac{H_-^2}{H^2} \right)^{\al_-} \, . \label{future}
\ee

We will assume that the transitions between these three periods are smooth
and that the details about these transitions are unimportant.

This model preserves the successes of the Cosmological Standard Model, while
giving a description of the early accelerated expansion
(inflation) and of the present one, including $\Lambda$CDM as a particular case
($\al_-= 1$).

Without any knowledge of the evolution of perturbations in this model, the
background evolution at late times can be used to fit the Type Ia Supernovae
UNION dataset \cite{Kowalski:2008ez}, the first acoustic peak in the Cosmic
Microwave (CMB) Background \cite{Dunkley:2008ie}, and the Baryon Acoustic
Oscillations (BAOs) \cite{Eisenstein:2005su}, via the parameters $H_0$
(the present Hubble parameter), $H_-$ and $\al_-$.

We use Monte Carlo Markov Chains to explore the likelihood of the fit of the
supernovae UNION dataset \cite{Kowalski:2008ez} to the ACM. The dataset provides
the luminosity distance
\be
d_l(z) = c(1+z)\int_0^z dz'/H(z')
\ee
and redshift of 307 supernovae. A $\chi^2$ analysis have been performed, where
$\chi^2$ has been marginalized over the nuisance parameter $H_0$ using the method
described in \cite{Wang:2004xz}. The resulting parameter space is spanned by the values
of $\alpha_-$ and $H_-/H_0$ (FIG. 1).

\begin{figure}
 \centerline{\includegraphics[scale=0.6, width=8cm]{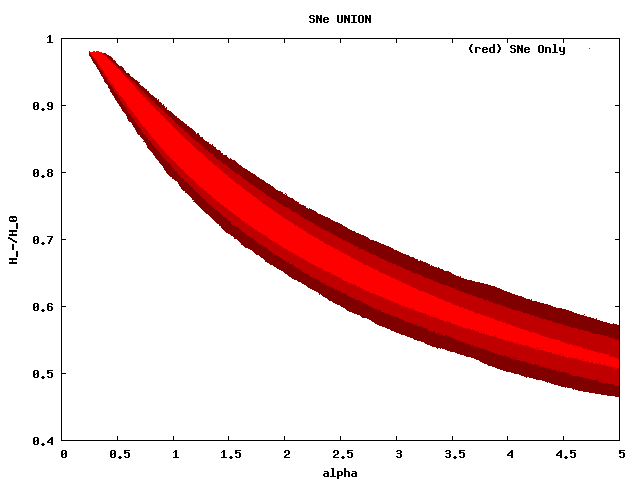}}
\label{U}
\caption{Confidence regions in parameter space of the Asymptotic
  Cosmological Model (ACM) from the fit of the supernovae UNION dataset
   without priors at $1\sigma$, $2\sigma$ and $3\sigma$. The $\Lambda$CDM is
  inside the 1$\sigma$ region. The best fit to ACM lies in
  $\alpha_- = 0.35$, $H_-/H_0 = 0.97$
($\chi^2 = 310.5$) in contrast to the best fit to $\Lambda$CDM, which lies in
$\alpha_- \equiv 1$ , $H_-/H_0 = 0.84$
($\chi^2 = 311.9$).}

\end{figure}

The constraints from the CMB data follow from the reduced distance to the surface
of last scattering at $z=1089$. The reduced distance $R$ is often written as
\be
R = \Omega_m^{1/2} H_0 \int_0^{1089} dz/H(z) \, .
\ee
The WMAP-5 year CMB data alone yield $R_0 = 1.715\pm 0.021$ for a fit assuming a
constant equation of state $\omega$ for the dark energy \cite{Komatsu:2008hk}. We
will take this value as a first approximation to the fit assuming ACM.  We can
define the corresponding $\chi^2$ as $\chi^2 =  [(R-R_0)/\sigma_{R_0}]^2$, and find
the confidence regions of the joint constraints (FIG. 2).

\begin{figure}
 \centerline{\includegraphics[scale=0.6, width=8cm]{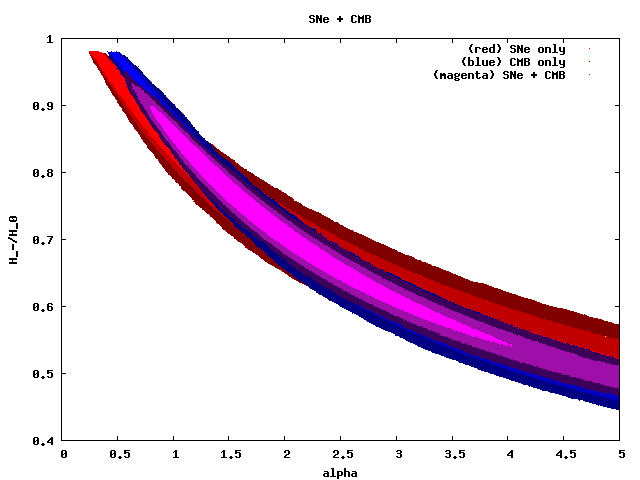}}
\label{UW}
\caption{Confidence regions in parameter space of the Asymptotic
  Cosmological Model (ACM) from the fit of the supernovae UNION
  dataset (red), of the distance to the surface of last scattering from WMAP-5
  (blue)  and from the joint fit (magenta) at $1\sigma$, $2\sigma$ and $3\sigma$.
  Significantly, the $\Lambda$CDM is still inside the 1$\sigma$ region, unlike in
  our previous study. This is due to the change in the Supernovae dataset. The
  best fit to ACM lies in $\alpha_- = 1.50$,
  $H_-/H_0 = 0.77$ ($\chi^2 = 312.5$) in contrast to the best fit to $\Lambda$CDM,
  which lies in $\alpha_- \equiv 1$, $H_-/H_0 = 0.86$
  ($\chi^2 = 313.2$).}
\end{figure}

BAO measurements from the SDSS data provide a
constraint on the distance parameter
$A(z)$ at redshift $z=0.35$,
\be
A(z)= \Omega_m^{1/2} H_0 H(z)^{-1/3} z^{-2/3} \left[\int_0^z dz'/H(z')\right]^{2/3} \, .
\ee
Ref.~\cite{Eisenstein:2005su} gives $A_0 = 0.469 \pm 0.17$. We can define the
corresponding $\chi^2$ as $\chi^2 =  [(A(z\,=\,0.35)-A_0)/\sigma_{A_0}]^2$ . The
confidence regions resulting from adding this constraint are shown in FIG. 3.

\begin{figure}
 \centerline{\includegraphics[scale=0.6, width=8cm]{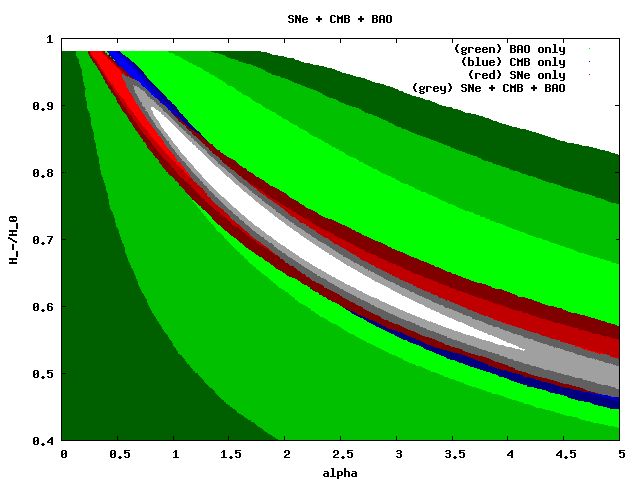}}
\label{UWB}
\caption{Confidence regions in parameter space of the Asymptotic
  Cosmological Model (ACM) from the fit of the supernovae UNION dataset (red),
  of the distance to the surface of last scattering from WMAP-5 (blue), of
  the Baryon Acoustic Oscillations peak (green) and from the joint fit
  (grey) at $1\sigma$, $2\sigma$ and $3\sigma$. Measurements of the
  BAOs peak do not add significant information with their present
  precision. The best fit to ACM lies in
  $\alpha_- = 1.50$, $H_-/H_0 = 0.77$
  ($\chi^2 = 312.7$) in contrast to the best fit to $\Lambda$CDM, which lies
  in $\alpha_- \equiv 1$, $H_-/H_0 = 0.86$
  ($\chi^2 = 313.7$).}
\end{figure}

In our previous work the use of the supernovae Gold dataset
\cite{Riess:2006fw} and of the WMAP-3 data \cite{Spergel:2006hy} led us to the
conclusion that the $\Lambda$CDM was at $3\sigma$ level in the
parameter space of the ACM. The position of the confidence regions
seems to depend very tightly on the dataset that is being
used. However, the value of the combination of parameters
\be
\Omega_m \equiv \left(1-\frac{H_-^2}{H_0^2}\right)^{\al_-} \label{omegam}
\ee
does not depend much neither on the value of $\al_-$ or the
dataset used (FIG. 4).

Moreover, we can conclude from the figures that
BAO's do not provide much information in order to constrain the
confidence regions of the ACM, unlike in other models such as
$\Lambda$CDM with nonzero spatial curvature.

\begin{figure}
 \centerline{\includegraphics[scale=0.6, width=8cm]{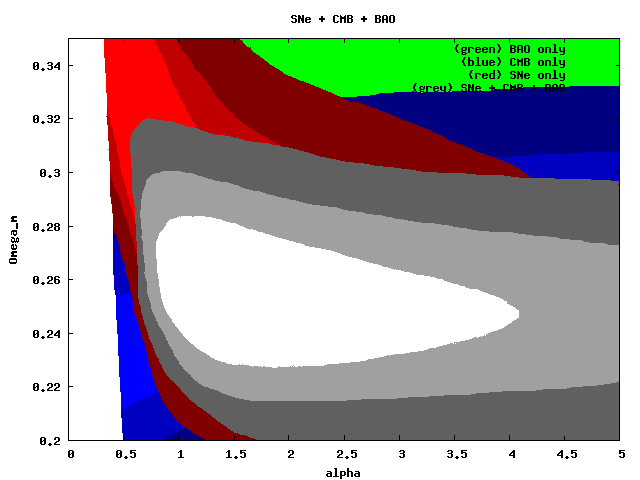}}
\label{omega}
\caption{Confidence regions in parameter space of the Asymptotic
  Cosmological Model (ACM) from the fit of the supernovae UNION dataset (red),
  of the distance to the surface of last scattering from WMAP-5
  (blue), of the Baryon Acoustic Oscillations peak (green) and from
  the joint fit (grey) at $1\sigma$, $2\sigma$ and $3\sigma$. The plot
  shows $\alpha_-$ in the horizontal axis and $\Omega_m$ in the
  vertical axis. Noticeably, $\Omega_m = 0.26 \pm 0.04$ almost
  independently of the value of $\alpha_-$.}
\end{figure}

\section{Scalar Perturbations}

The lack of an action defining the ACM is a serious obstacle in
the derivation of the equations governing the behavior of the
perturbations. Given a background behavior described by the ACM,
what can be said about the evolution of perturbations on top of this
background? We will find that general covariance together with an
additional assumption fixes completely the set of equations for the
scalar perturbations in the linearized approximation and in the region
close to $H_-$.

The key point will be the following. Knowing the exact Friedman
equations in the homogeneous approximation gives us a clue on the form
of the equations for the scalar perturbations. In particular, we can
formally describe perturbations over the FRW metric which do not depend
on the spatial coordinates, $\{\phi(\vx,t)=\phi(t),
\psi(\vx,t)=\psi(t)\}$. Thus the perturbed
metric becomes the FRW metric
written in a new coordinate frame. With a change
of coordinates one can derive, starting from the Friedman equations,
the terms containing only time derivatives of the scalar
perturbations. Next an assumption on the validity of the GR
description of scalar perturbations when $H_-\ll H \ll H_+$,
together with the relations, valid for any general covariant theory,
between terms with time derivatives and those involving spatial
derivatives, allow to derive the evolution of scalar perturbations.

After this short sketch, we will perform the derivation of the
equations for the scalar perturbations in detail. We can write the
metric of spacetime with scalar perturbations in the Newtonian gauge,
\be
ds^2 = (1+2\phi(\vx,t))dt^2 - a^2(t) (1-2\psi(\vx,t)) d\vx^2 \, ,\label{newt}
\ee
and the stress-energy tensor of the source fields will be
\be
\ba{ccl}
T^0_0 & = & \rho_{(0)}(t) + \delta \rho(\vx,t) \, , \\
T^0_i & = & (\rho_{(0)}(t) + p_{(0)}(t)) \partial_i \theta(\vx,t) \, ,\\
T^i_j & = & -(p_{(0)}(t) +\delta p(\vx,t))\delta^i_j\,+\,\partial^i\partial_j\Pi(\vx,t)\, ,
\ea \label{set}
\ee
where $\theta$ is the velocity potential of the fluid, $\Pi$ is the
anisotropic stress tensor (shear) potential, $\delta\rho$ and $\delta
p$ are small perturbations on top of the background homogeneous
density $\rho_{(0)}$ and pressure $p_{(0)}$, respectively, and
$\phi\sim\psi\ll 1$. We have used the Newtonian gauge because the
gauge invariant scalar perturbations of the metric ($\Phi$ and $\Psi$)
and gauge invariant perturbations in the stress energy tensor
coincide with the perturbations explicitly written in this gauge.

The general form of the equations for the perturbations in a metric
theory with arbitrarily high derivatives is

\begin{eqnarray}
8 \pi G \delta\rho & = & \sum_{n=0}^\infty \sum_{m=0}^\infty
\left[\frac{a_{nm}}{H^{2n+m-2}}\left(\frac\Delta{a^2}\right)^n
  \partial_t^m \phi +
  \frac{b_{nm}}{H^{2n+m-2}}\left(\frac\Delta{a^2}\right)^n
  \partial_t^m \psi \right] \label{rho1} \, ,\\
8 \pi G (\rho_{(0)} + p_{(0)}) \partial_i \theta & = &
\partial_i\sum_{n=0}^\infty \sum_{m=0}^\infty
\left[\frac{c_{nm}}{H^{2n+m-1}}\left(\frac\Delta{a^2}\right)^n
  \partial_t^m \phi +
  \frac{d_{nm}}{H^{2n+m-1}}\left(\frac\Delta{a^2}\right)^n
  \partial_t^m \psi \right] \label{vel1} \, ,\\
8 \pi G (-\delta p \,\delta^i_j + \partial_i \partial_j \Pi) & = &
\delta^i_j\sum_{n=0}^\infty \sum_{m=0}^\infty
\left[\frac{e_{nm}}{H^{2n+m-2}}\left(\frac\Delta{a^2}\right)^n
  \partial_t^m \phi +
  \frac{f_{nm}}{H^{2n+m-2}}\left(\frac\Delta{a^2}\right)^n
  \partial_t^m \psi \right] \nonumber \\
& & +\sum_{n=0}^\infty \sum_{m=0}^\infty \frac 1{a^2}(\delta^i_j
\Delta -\partial_i
\partial_j)\left[\frac{g_{nm}}{H^{2n+m}}\left(\frac\Delta{a^2}\right)^n
  \partial_t^m \phi +
  \frac{h_{nm}}{H^{2n+m}}\left(\frac\Delta{a^2}\right)^n \partial_t^m
  \psi \right] \label{p1}  \, .
\end{eqnarray}

The coefficients $a_{nm},...,h_{nm}$ are adimensional functions of the
Hubble parameter an its time derivatives, and can be turned into
functions of just the Hubble parameter using the bijection explained
in the beginning of the previous section. We have mentioned we are
going to be able to determine exactly the terms with only time derivatives, that
is, the precise form of the coefficients $a_{0m},b_{0m},e_{0m}$ and
$f_{0m}$. This shows that there is much freedom of choosing a
covariant linearized theory of cosmological perturbations, even for a
given solution of the homogeneous equations.

However, under a single assumption, it is possible to greatly reduce
this freedom. The assumption is that the standard linearized equations
of the perturbations of General Relativity are effectively recovered
in the limit $H_+ \gg H \gg H_-$ for all the Fourier modes with $H_+
\gg k > H_-$ (subhorizon modes in the present time).

Let us work out the implications of this assumption. In the equation
for the perturbation of the energy density in General relativity,
\be
8\pi G \delta \rho  =  -6 H^2 \phi -6 H \dot \psi +\frac{2}{a^2}\Delta \psi\, ,
\ee
there is only a term proportional to $\Delta \psi$. However, for
subhorizon modes ($k\gg H$) the dominant terms are those with the
highest number of spatial derivatives. If we demand
the terms with spatial derivatives of order greater than $2$ not to
spoil the behavior of these modes in the period in which GR is a good
approximation, they must be negligible at least for the observable modes.

For instance, they should
be negligible for the modes responsible of the acoustic peaks of the
CMB spectrum, and therefore for the modes which have entered the
horizon after recombination (which have even
lower $k$). It is possible that these terms
are suppressed by inverse powers of the UV scale
$H_+$, becoming irrelevant for
current tests of gravity, although they may be relevant for the
physics of quantum fluctuations in the very early universe.

Therefore, at times when $H\ll H_+$, the equations can be approximated by

\begin{eqnarray}
8 \pi G \delta\rho & = & \sum_{m=0}^\infty
\left[\frac{a_{0m}}{H^{m-2}}\partial_t^m \phi +
  \frac{b_{0m}}{H^{m-2}}\partial_t^m \psi +
  \frac{a_{1m}}{H^{m}}\frac\Delta{a^2} \partial_t^m \phi +
  \frac{b_{1m}}{H^{m}}\frac\Delta{a^2}\partial_t^m \psi \right]
\label{rho2} \, ,\\
8 \pi G (\rho_{(0)} + p_{(0)}) \partial_i \theta & = &
\partial_i\sum_{m=0}^\infty \left[\frac{c_{0m}}{H^{m-1}}\partial_t^m
  \phi + \frac{d_{0m}}{H^{m-1}}\partial_t^m \psi \right] \label{vel2}
\, ,\\
8 \pi G (-\delta p \,\delta^i_j + \partial_i \partial_j \Pi) & = &
\delta^i_j \sum_{m=0}^\infty \left[\frac{e_{0m}}{H^{m-2}}\partial_t^m
  \phi + \frac{f_{0m}}{H^{m-2}}\partial_t^m \psi \right] \nonumber \\
& & +\sum_{m=0}^\infty \frac 1{a^2}(\delta^i_j \Delta -\partial_i
\partial_j)\left[\frac{g_{0m}}{H^{m}} \partial_t^m \phi +
  \frac{h_{0m}}{H^{m}}\partial_t^m \psi \right] \label{p2} \, ,
\end{eqnarray}

where general covariance has been used to exclude the terms with
coefficients $e_{1m}$, $f_{1m}$ which are not compatible with
(\ref{rho2}), (\ref{vel2}).

We must also take into account that the modes which are responsible of
the acoustic peaks of the CMB undergo a phase in which they oscillate
as sound waves, i.e.: $\partial_t\phi~\sim~k\phi$. In order to
explain the acoustic peaks of the spectrum of the CMB it is required
that the acoustic oscillations of the modes of the gravitational
potentials, $\phi_\vk$ and $\psi_\vk$, which lead to the acoustic
peaks of the CMB spectrum, have a frequency $\sim k$. When radiation
dominates and we consider modes well inside the horizon, two of the
solutions of the system of differential equations of arbitrary order
have this property \footnotemark[1].

However, in order for nondecaying superhorizon modes to evolve
into oscillating subhorizon modes with frequency $\sim k$, a fine-tuning of
coefficients is required unless the equations are of order two in time
derivatives. Then the acoustic peaks in the CMB spectrum are
reproduced when terms with more than two derivatives in the equations
for the  perturbations can be neglected. Then

\begin{eqnarray}
8 \pi G \delta\rho & = & \sum_{m=0}^2
\left[a_{0m}H^{2-m}\partial_t^m \phi +
  b_{0m}H^{2-m}\partial_t^m \psi \right]+
a_{10}\frac\Delta{a^2}\phi + b_{10}\frac\Delta{a^2}\psi  \label{rho3}
\, ,\\
8 \pi G (\rho_{(0)} + p_{(0)}) \partial_i \theta & = &
\partial_i\sum_{m=0}^1 \left[c_{0m}H^{1-m}\partial_t^m \phi +
  d_{0m}H^{1-m}\partial_t^m \psi \right] \label{vel3} \, ,\\
8 \pi G (-\delta p \,\delta^i_j + \partial_i \partial_j \Pi) & = &
\delta^i_j \sum_{m=0}^2 \left[e_{0m}H^{2-m}\partial_t^m \phi
  + f_{0m}H^{2-m}\partial_t^m \psi \right] \nonumber \\
& & +\frac 1{a^2}(\delta^i_j \Delta -\partial_i
\partial_j)\left[g_{00}\phi + h_{00}\psi \right] \label{p3} \, ,
\end{eqnarray}

\footnotetext[1]{
As we will see
below, if the term with the highest time derivative is of order
$D$,  $a_{1,D-2} = -e_{0D}$ as a consequence of general
covariance. Well inside the horizon these are the dominant terms
appearing in the wave equation resulting when calculating the
adiabatic perturbations in the radiation domination
epoch.}

\noindent and the arbitrariness in the evolution equations for the scalar
perturbations has been reduced to twenty undetermined dimensionless
coefficients at this level.

The equations \eqref{rho3},\eqref{vel3},\eqref{p3} are in principle
valid for any perturbation mode $\{\phi_\vk,\psi_\vk\}$ as long as $k\ll H_+$.
In particular, it must be valid for the mode $\vk = \mathbf{0}$, which
corresponds formally to a perturbation with no spatial dependence. In practice,
we will be able to neglect the spatial dependence of perturbation whose spatial
dependence is sufficiently smooth, i.e.: its wavenumber $k$ is sufficiently low.
In GR it suffices for a mode to be superhorizon $k\ll H$ in order to neglect
its spatial dependence in a first approximation.

If we consider the FRW metric perturbed by one of these modes we have,
\be
ds^2 = (1+2\phi(t))dt^2 - a^2(t) (1-2\psi(t)) d\vx^2 \, . \label{superh}
\ee

By means of the invariance under time reparameterizations we can introduce
a new time variable $dt'= (1+\phi(t))dt$, a new scale factor
$a'(t')=a(t)(1-\psi(t))$ and a new energy density
$\rho'(t')=\rho_0(t)+\delta\rho(t)$
leading us back to the ACM in a homogeneous background. The Hubble
rate $H = \frac{\dot a}{a}$ and its time derivatives $H^{(n)}$ will change as
\be
\begin{array}{l}
H^{(n)'} =
\frac{d^n}{dt^{'n}}\left[\frac{1}{a'(t')}\frac{da'}{dt'}\right] =
  H^{(n)} +\delta H^{(n)}, \\
\delta H^{(n)} = -\psi^{(n+1)}
-\sum_{m=0}^n
\left(\ba{c}
n+1 \\ m+1
\ea
\right) H^{(n-m)}\phi^{(m)}.\label{Hn}
\end{array}
\ee
Noticeably, this procedure must be applied to the generalized Friedman
equation before one of its homogeneous solutions is used to write the
time variable $t$ as a function of $H$. Therefore we should start by
considering a generalized first Friedman equation in primed
coordinates
\be
8\pi G \rho'(t') = 3 f(H'(t'),\dot{H}'(t'),...,H^{(n)'}(t'),...) \label{fh'} \, ,
\ee
and the corresponding equation for the pressure will be derived using
the continuity equation. The linearized equations for the sufficiently smooth
 scalar perturbations in the Newtonian gauge~(\ref{superh}) can then be
derived with the help of~(\ref{Hn}). The result is

\begin{eqnarray}
8\pi G \delta \rho(t) & = & \sum_{n = 0}^\infty g_{\vert n}(H(t))\delta H^{(n)}(t),
\label{rho4} \\
-8\pi G \delta p & = & \frac 1H \sum_{n = 0}^\infty\sum_{m = 0}^\infty
g_{\vert nm}(H)H^{n+1}\delta H^{(m)}\nonumber\\
& &+\sum_{n = 0}^\infty g_{\vert n}(H) \left(\frac{\delta
  H^{(n+1)}}H-\frac{H^{(n+1)}}{H^2}\delta H \right)\nonumber\\
& &+ 3\sum_{n = 0}^\infty g_{\vert n} (H)\delta H^{(n)},\label{p4}
\end{eqnarray}
where
\begin{eqnarray}
 g_{\vert n} (H) & = & \left[ \frac{\partial f}{\partial
     H^{(n)}}\right]_{H^{(n)} = H^{(n)}(t(H))}\, ,\\
g_{\vert nm}(H) & = & \left[ \frac{\partial^2 f}{\partial
     H^{(n)}\partial H^{(m)}}\right]_{H^{(n)} = H^{(n)}(t(H))} \, ,
\end{eqnarray}
and the time dependence of $H$ is derived from the
homogeneous evolution \eqref{gh}. This procedure fixes exactly the
coefficients $a_{0m},b_{0m},e_{0m}$ and $f_{0m}$ in
\eqref{rho2},\eqref{vel2},\eqref{p2}, i.e. all the terms with no
spatial derivatives in the equations for the scalar perturbations, as
functions of $f(H,\dot{H},...)$ and its partial derivatives.

We can now count the number of functional degrees of freedom in the
superhorizon modes of the linearized theory. Let us assume
that there is a maximum number $D$ of time derivatives in the equations
for the perturbations~\footnotemark[2].\footnotetext[2]{We are
  forgetting here the restriction on the
number of derivatives required in order to reproduce the acoustic peaks of
CMB.} The number of derivatives, with the use of
\eqref{Hn},\eqref{rho4},\eqref{p4}, fixes $f$ to be a function of at
most $H^{(D-2)}$. When particularized to a solution of the homogeneous
Friedman equations, $f$ and its first and second partial derivatives
become functions of the Hubble rate $H$: $g(H)$, $g_{\vert n}(H)$ and
$g_{\vert nm}(H)$ respectively. That makes $1+(D-1)+D(D-1)/2=D(D+1)/2$
functional degrees of freedom. The first derivative of $g(H)$ can be
written in terms of the $g_{\vert n}(H)$, and the first derivative of the
later can be also written in terms of the $g_{\vert nm}(H)$. That makes $1 +
(D-1)$ conditions, so the result depends on $D(D-1)/2$ independent
functions of the Hubble parameter $H$ in the terms without any spatial
derivatives.

Notice however that if we restrict the number of derivatives appearing
in the equations for the perturbations to two as in
Eqs.~\eqref{rho3},\eqref{vel3},\eqref{p3}, then
Eqs.~\eqref{Hn},\eqref{rho4},\eqref{p4} tell us that we must consider
just functions $f(H,\dot H, \ddot H,...)\, =\, f(H)\, = \, g(H)$, at
least as an approximation at times $H\ll H_+$ and modes $k \ll
H_+$. Therefore the number of functional degrees of freedom in the
terms with no spatial derivatives is just one: the homogeneous
evolution $f(H) = g(H)$.

Our assumption might be relaxed and we could impose that general
relativity should be valid for all the modes $H_-<k<k_{obs}$ that have
been observed in the spectrum of CMB and matter perturbations. This
could lead to new terms in the equations for the perturbations
(suppressed not necessarily by the UV scale $H_+$) which have been
negligible for the observed modes but that could lead to ultraviolet
deviations from the spectrum derived in the general relativistic
cosmology which have not yet been observed. However, terms with more
than two spatial derivatives will be very tightly constrained by solar
system experiments.

In this article we will restrict ourselves to the system of second order
 differential equations (\ref{rho3}), (\ref{vel3}),
(\ref{p3}).
Let us now derive the linearized equations for
the rest of the modes under this assumption. In the $(t',\vx)$
coordinate system, the metric is Friedman Robertson Walker, and we
know that for this metric, we can use equation (\ref{gh}). Thus,
\be
8 \pi G \rho' = 3 g(H') \, ,
\ee

with
\be
H' = \frac 1 {a'} \frac{d a'}{d t'} = H(1-\phi)-\dot{\psi}
\ee
at linear order in perturbations. Therefore, we can deduce from
(\ref{gh}) that in the $(t,\vx)$ coordinate system

\begin{eqnarray}
8 \pi G \delta\rho & = & -3 g'(H)\left(H \phi\right) +\dot{\psi})+
a_{10}\frac\Delta{a^2}\phi + b_{10}\frac\Delta{a^2}\psi  \label{rho5}
\, ,\\
8 \pi G (\rho_{(0)} + p_{(0)}) \partial_i \theta & = &
\partial_i\sum_{m=0}^1 \left[\frac{c_{0m}}{H^{m-1}}\partial_t^m \phi +
  \frac{d_{0m}}{H^{m-1}}\partial_t^m \psi \right] \label{vel5} \, ,\\
8 \pi G (-\delta p \,\delta^i_j + \partial_i \partial_j \Pi) & = &
\delta^i_j  \left[g'(H)\left(3 H \phi +3
  \dot{\psi}-\frac{\dot{H}}{H^2}\dot{\psi}+\frac{\dot{H}}H \phi
  +\dot{\phi} + \frac 1 H \ddot{\psi}\right) \right. \nonumber\\
& & \left. +\frac{\dot{H}}H g''(H)(H\phi+\dot{\psi}) \right] \nonumber \\
& & +\frac 1{a^2}(\delta^i_j \Delta -\partial_i
\partial_j)\left[g_{00}\phi + h_{00}\psi \right] \label{p5} \, ,
\end{eqnarray}

and the twelve coefficients of terms with no spatial derivatives
are fixed by the function $g(H)$ which defines the homogeneous cosmological
model.

The last requirement comes again from general covariance. If the
tensor $T_{\mu\nu}$ comes from the variation of a certain matter
action $S_m$ with respect to the inverse of the metric $g^{\mu\nu}$,
and $S_m$ is a scalar under general coordinate transformations, then
$T^\mu_\nu$ must be a 1-covariant, 1-contravariant divergenceless
tensor, i.e.  $\nabla_\mu T^\mu_\nu = 0$. As the stress-energy tensor
is proportional to $G^\mu_\nu$, the later must also be
divergenceless. In the linearized approximation $G^\mu_\nu =
G^\mu_{(0)\nu}+\delta G^\mu_\nu$ and the Christoffel symbols
$\Gamma^\lambda_{\mu\nu} = \Gamma^\lambda_{(0)\mu\nu}+\delta
\Gamma^\lambda_{\mu\nu}$  , and we can write the linearized version of
this requirement as

\be
\ba{c}
\delta G^0_{\mu\vert 0} +\delta G^i_{\mu\vert i} + \delta
G^\lambda_\mu \Gamma^\nu_{(0)\lambda\nu} - \delta G^\lambda_\nu
\Gamma^\nu_{(0)\lambda\mu} \\
+ G^\lambda_{(0)\mu} \delta\Gamma^\nu_{\lambda\nu}
- G^\lambda_{(0)\nu} \delta\Gamma^\nu_{\lambda\mu} = 0 \, .
\ea
\ee

Let us study this condition in order to see if we can further limit
the number of independent coefficients of
the equations for scalar perturbations. For $\mu = i$ the term
$\delta G^0_{i\vert 0}$ will give at most a term proportional to
$d_{01}\ddot{\psi}_{\vert i})$, which will not be present in other
terms except $\delta G^j_{i \vert j}$. This term will give at most a
term $g'\ddot{\psi}_{\vert i}/H$ which fixes the
exact value of $d_{01}$ for all $H \ll H_+$ . For the same reason,
the term $c_{01}\ddot{\phi}_{\vert i}$ in $\delta G^0_{i\vert 0}$
can not be canceled and the term $c_{00}H\dot{\phi}_{\vert i}$ can
only be canceled by the term $g'\dot{\phi}_{\vert i}$ in
$\delta G^j_{i \vert j}$, which means that $c_{01}=0$ and fixes
 the exact value of $c_{00}$ for all $H\ll H_+$. The value
of $d_{00}$ can be fixed in
terms of $d_{01}$ and the terms proportional to $\dot{\psi}$ in
(\ref{p5}). This fixes (\ref{vel5}) completely in terms of the
function describing the homogeneous evolution, $g(H)$:

\be
8 \pi G (\rho_0+p_0)\partial_i\bar{\theta}  =
\frac{g'(H)}{H}\partial_i \left (H \Phi + \dot{\Psi} \right )
\label{vdef} \, .
\ee

In the previous equation and from now on we will refer directly to the
gauge invariant counterparts of the variables in the Newtonian gauge,
which we will denote by $\Phi,\Psi,\bar{\theta},\bar{\delta\rho}$ and
$\bar{\delta p}$ ($\Pi$ is already gauge invariant). Now we can use
the divergenceless condition for $\mu=0$. The term $\delta G^0_{0\vert
  0}$ will give at most a term proportional to $a_{10}\Delta
\dot{\phi}$ and a term proportional to $b_{10}\Delta \dot{\psi}$,
which will not be present in other terms except $\delta G^j_{0 \vert
  j}$. This term will give at most a term proportional to $\Delta
\phi$ and a term proportional to $\Delta \dot{\psi}$. Therefore
$a_{10}=0$  and the value of $b_{10}$ is set completely in terms of
$g(H)$:

\be
8 \pi G \bar{\delta \rho} = -3 g'(H)(H \Phi +\dot{\Psi})+ \frac
{g'(H)}{a^2 H}\Delta \Psi
\label{rhodef} \, .
\ee

The complete knowledge of (\ref{rhodef}) and (\ref{vdef}) when $H \ll
H_+$ gives a complete knowledge of (\ref{p5}) in terms of $g(H)$ when
$H \ll H_+$:

\be
\ba{ccl}
8 \pi G (\bar{\delta p} \, \delta_{ij} + \partial_i \partial_j \bar{\Pi})
& = &
\delta_{ij}\left\{
g'(H)\left[3 H \Phi +3
  \dot{\Psi}-\frac{\dot{H}}{H^2}\dot{\Psi}+\frac{\dot{H}}H \Phi
  +\dot{\Phi} + \frac 1 H \ddot \Psi \right]
+\frac{\dot H}H g''(H)\left[H \Phi + \dot \Psi \right]
\right\}\\
& & +\frac 1{a^2}\left(\delta_{ij}\Delta -\partial_i\partial_j\right)
\left( \frac{g'(H)}{2 H}(\Phi-\Psi) + \frac{\dot H}{2 H^3}(g'(H) - H
g''(H))\Psi \right)
\ea\label{pdef} \, ,
\ee

and we have finally a set of equations for the scalar perturbations
(\ref{rhodef}), (\ref{vdef}), (\ref{pdef}) which are determined by the
function $g(H)$ which defined the ACM in the homogeneous approximation.

Had we relaxed our assumption, we could work with the system of
equations~\eqref{rho1},\eqref{vel1},\eqref{p1} restricted to a maximum
number of derivatives $D$ as an approximation. Together with the
condition $G^\mu_{\nu;\mu} =0$, the system of equations for the
perturbations define a set of coupled differential equations for the
coefficients of the terms with at least one spatial derivative. The
set of equations coming from $G^\mu_{i;\mu}=0$ define the terms
proportional to $e_{mn}$ and $f_{mn}$ in \eqref{p1} as a function of the
terms in \eqref{vel1}. The set of equations coming from
$G^\mu_{0;\mu}$ define the rest of the terms in \eqref{p1} as a
function of the terms in \eqref{vel1} and \eqref{rho1}. Therefore, the
only freedom, for what concerns linearized perturbations, is that of
choosing the set of functions $\{a_{nm}, b_{nm}, c_{nm}, d_{nm}\}$,
with some of them fixed by the homogeneous dynamics \eqref{fh}.  An
interesting case is found if $D=4$ is imposed. This includes a
description of $f(R)$-theories \cite{Faraoni:2008mf}, bigravity
theories \cite{Damour:2002wu}, and other of the most studied modified
gravity theories. We will study further this case
in the subsection devoted to the comparison of the model with $f(R)$
theories.

The rhs of the equations derived are the equivalent to the components
of the linearized Einstein tensor derived in a general covariant
theory whose Friedman equation is exactly (\ref{gh}). Aside from
General Relativity with or without a cosmological constant, i.e. for
$\alpha_-\neq 1, 0$, it will be necessary to build an action depending
on arbitrarily high derivatives of the metric in order to derive a
theory such that both the equations in the homogeneous approximation
and the linearized equations for the scalar perturbations are of
finite order.

Noticeably, the nonlinear equations for the perturbations will include
derivatives of arbitrary order of the metric perturbations, with the
exceptional case of General Relativity with a cosmological
constant. This makes problematic the consistency of the ACM beyond the
linear approximation. This issue will be further studied in a future
work.

To summarize, what we have shown in this section is
that the behavior of scalar perturbations in a general covariant
theory is intimately connected to the background evolution.
Except in the very early universe, the
linearized equations for the scalar perturbations are determined by
the equations in the homogeneous approximation if one assumes that
there are no terms with more than two spatial derivatives. This
assumption could be relaxed in order to include more general theories.

There will be a subset of general covariant theories with a background
evolution given by (\ref{gh}) that verify these
equations for the perturbations. We may be able to distinguish among
these at the linearized level by means of the vector and tensor
perturbations.

\section{Hydrodynamical Perturbations}

Vector perturbations represent rotational flows which decay very
quickly in the General Relativistic theory. As we expect a small
modification of the behavior of perturbations just in the vicinity of
the lower bound $H_-$, we will assume that the vector perturbations
have decayed to negligible values when the scale $H_-$ begins to play
a role and therefore they will be ignored.

Tensor perturbations correspond to gravitational waves, which at the
present have been not observed, their effect being far beyond the
resolution of current observations.

Present measurements restrict their attention to perturbations in the
photon sector (CMB) and the matter sector (matter power spectrum, both
dark and baryonic). These observations can be computed taking into
account only the effect of scalar perturbations. Therefore, we can use
the result of the previous section to study the deviations from
$\Lambda$CDM in the spectrum of CMB. A comprehensive study of the
matter power spectrum predicted by the ACM would require a detailed
knowledge of the nonlinear regime, which we lack at present.

Let us start by deriving the behavior of scalar perturbations in a
universe whose gravitation is described by the ACM and filled with a
perfect fluid ($\Pi = 0$) in (\ref{set}). Therefore, (\ref{pdef}) with
$i\neq j$ fixes $\Phi$ as a function of $\Psi$,
\be
\Phi = \left\{1\,+\,\frac{\dot{H}}{2 H^2}\left(1 - \frac{H
  g''(H)}{g'(H)}\right)\right\}\Psi \label{Phi}\, .
\ee
Then, this expression can be used to substitute $\Phi$ in the
remaining three differential equations for $\bar{\delta\rho}$,
$\bar{\delta p}$ and $\bar{\theta}$. Given the equation of state of
the fluid $p = p(\rho, S)$, with $S$ the entropy density, the
perturbation of the pressure can be written as
\be
\bar{\delta p} = c_s^2 \bar{\delta \rho} + \tau \delta S \, \label{ppert},
\ee
where $c_s^2 \equiv (\partial p/\partial \rho)_S$ is the square of the
speed of sound in the fluid and $\tau \equiv (\partial p/\partial
S)_\rho$. Substituting the pressure and the energy density
perturbations for the corresponding functions of $\Psi$ and its
derivatives into (\ref{ppert}), we arrive at the equation for the
entropy density perturbation $\delta S$. The perturbations we are
interested in are adiabatic, i.e.: $\delta S = 0$ and therefore, the
equation for the entropy perturbations turns into an equation of
motion of the adiabatic perturbations of the metric. If we want to
consider the effect of the lower bound of the Hubble rate, $H_-$, on
the evolution of perturbations, we must consider a matter dominated
universe ($c_s^2 = 0$),
\be
\ba{c}
g'(H)\left[3 H \Phi +3
   \dot{\Psi}-\frac{\dot{H}}{H^2}\dot{\Psi}+\frac{\dot{H}}H \Phi
   +\dot{\Phi} + \frac 1 H \ddot{\Psi}\right] \\
+\frac{\dot{H}}H g''(H)\left[H\Phi+\dot{\Psi}\right] \; = \; 0 \, .
\ea \label{Psi}
\ee
This equation is a second order linear differential equation with
non-constant coefficients. Thus it is useful to work with the Fourier
components of the metric perturbation
$\Psi_k$. The equation of motion for the Fourier components is just a
second order linear ODE with non-constant coefficients, which will be
analytically solvable just for some simple choices of $g(H)$.
However, in general, it will be mandatory to perform a numerical
analysis.

\section{The CMB Spectrum in ACM versus $\Lambda$CDM}

The main observation that can be confronted with the predictions of
the theory of cosmological perturbations at the linearized level is
the spectrum of temperature fluctuations of the Cosmic Microwave
Background (CMB), from which WMAP has recorded accurate measurements
for five years \cite{Komatsu:2008hk}.

The temperature fluctuations $\frac{\delta T}T$ are connected to the
metric perturbations via the Sachs-Wolfe effect \cite{Sachs:1967er},
which states that, along the geodesic of a light ray
$\frac{dx^i}{dt}=l^i(1+\Phi+\Psi)$, characterized by the unit
three-vector $l^i$, the temperature fluctuations evolve according to

\be
(\frac \partial{\partial t} + \frac {l^i} a \frac \partial{\partial
  x^i})(\frac{\delta T}T+\Phi) = \frac \partial {\partial  t}(\Psi +
\Phi) \, .
\ee

Neglecting a local monopole and dipole contribution, taking
recombination to be instantaneous at a certain time and assuming
that the reionization optical depth is negligible, the present
temperature fluctuation of a distribution of photons coming from a
given direction of the sky $l^i$ can be related to the temperature
perturbation and metric perturbation $\Psi$ in the last scattering
surface plus a line integral along the geodesic of
the photons of the derivatives of $\Phi$ and $\Psi$ (Integrated Sachs
Wolfe effect, ISW).

These relations are purely kinematical and remain unchanged in the
ACM. In order to take into account all the effects involved in the
calculation of the CMB spectrum the code of CAMB \cite{CAMB} has been
adapted to the ACM. (see the appendix for details).

We have compared the late time evolution of the perturbations
described by the ACM \eqref{future} for several values of $\al_-$ (taking into
account that $\al_- = 1$ corresponds to $\Lambda$CDM) for constant
$H_0$ and $\Omega_m$ defined in Eq.~\eqref{omegam}. We have
assumed a Harrison-Zel'dovich scale invariant spectrum of scalar
perturbations as initial condition ($n_s = 0$) in the region in
which the universe is radiation dominated but $H\ll H_+$ . Thus the study of the
effect of the scale $H_+$ is postponed. We have taken $H_0 = 72 \pm 8\,
km/sMpc$ from the results of the Hubble Key Project
\cite{Freedman:2000cf}\footnotemark[3] and $\Omega_m = 0.26 \pm 4$ from our previous
analysis of the evolution at the homogeneous level. The effect of ACM
as compared to $\Lambda$CDM is twofold (FIG. 5).

\begin{figure}
 \centerline{\includegraphics[scale=0.6, width=9cm]{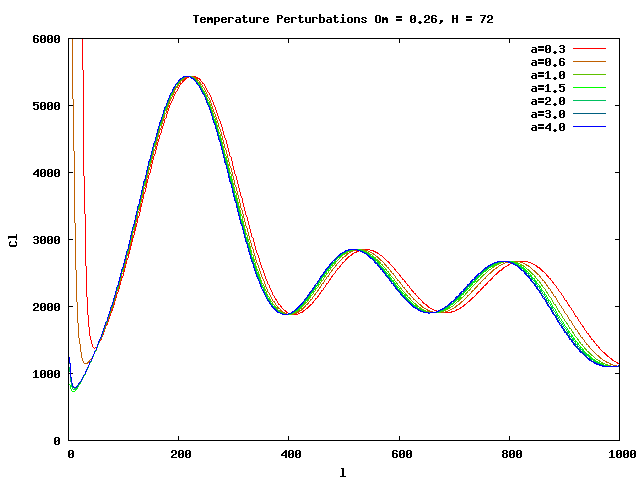}}
\label{spectrum}
\caption{Spectrum of temperature fluctuations of the Cosmic Microwave
  Background for $\Omega_m = 0.26$, $H_0 = 72 km/sMpc$ and varying
  $\alpha_- = 0.3$ (red), $0.6$, $1.0$, $1.5$, $2.0$, $3.0$, $4.0$
  (blue) . The lowest values of $\alpha_-$ show extreme ISW effects
  and therefore will be ruled out also by the CMB
  spectrum. Unfortunately, the cosmic variance masks the effect of
  $1.0 < \alpha_- < 4.0$ at large angular scales and therefore we can
  obtain little information about the ACM from the late ISW effect.}
\end{figure}

There is a shift in the peak positions due to the difference in the
distance to the last scattering surface for different values of
$\alpha_-$. This shift increases with the peak number. In particular,
more precise measurements of the position of the third peak could be
used to estimate the value of $\alpha_-$.

There is also an increase of the lower multipoles due to the late
Integrated Sachs-Wolfe effect, which is particularly extreme for
$\alpha_- \lesssim 1.0$ (which corresponds to $H_-\simeq H_0$ for
constant $\Omega_m$). This large deviation is clearly not present in
the experimental data. The much smaller deviation for $\alpha_-
\gtrsim 1.0$ is within the error bands due to the cosmic variance, and
therefore the value of the lower multipoles cannot be used to exclude
values of $\alpha_-$ greater than but of the order of one.

\section{Comparison with the treatment of perturbations in other models}

\subsection{Perturbations in f(R) theories}
In our previous work \cite{Cortes:2008fy} we found that given an
expansion history parameterized by a modified Friedman equation in a universe
filled by a given component, it was always possible to find a biparametric
family of $f(R)$ theories which had the same homogeneous evolution as
a solution of their equations of motion. We
wonder now if these theories have also the same linearized equations
for the scalar perturbations of the metric.

An $f(R)$ theory is defined by its action and therefore the
equations for the perturbations are uniquely determined. The action is
given by
\be
 S=\frac 1{16 \pi G}\int d^4 x\sqrt{-g}[f(R) + 16\pi G L_m] \, ,
\ee
where $R$ is the curvature scalar and $f(R)$ is an arbitrary function
of this scalar. The equations governing the evolution of the
perturbations will be derived from the Einstein equations of the
$f(R)$ theory,
\be
 f'(R)R^\mu_\nu-\frac12
 f(R)\delta^\mu_\nu+(\delta^\mu_\nu\square-\nabla^\mu\nabla_\nu)f'(R)=8\pi G
 T^\mu_\nu \, .\label{EinfR}
\ee

At the homogeneous level, it is possible to write the Friedman
equations of the $f(R)$ theory,

\begin{eqnarray}
8 \pi G \rho_{(0)} & = & -3(\dH+H^2)f'(\Ro)-\frac 12 f(\Ro) -3 H
f''(\Ro)\dRo \label{fRrho} \, ,\\
-8 \pi G p_{(0)} & = & -(\dH + 3 H^2)f'(\Ro)-\frac 12 f(\Ro)
+f^{(3)}(\Ro)\dRo^2 + f''(\Ro)\ddRo \label{fRp} \, ,
\end{eqnarray}

\footnotetext[3]{ Recently the SHOES team has provided a
more precise measurement, $H_0 = 74.2 \pm 3.6 \, km/sMpc$,\cite{Riess:2009pu}}

\noindent where
\be
\Ro(t) = -6(\dH +2H^2)
\ee
is the curvature scalar of the Friedman-Robertson-Walker (FRW) metric
(zeroth order in perturbations). The solution of the system of
equations \eqref{fRrho},\eqref{fRp} together with the equation of
state of the dominant component of the stress-energy tensor, defines
$\Ro$ and its derivatives as functions of the Hubble parameter $H$
\cite{Cortes:2008fy}.

If we expand \eqref{EinfR} in powers of the
scalar perturbations of the FRW metric \eqref{newt}, the first order
term gives the linearized equations for the
perturbations. This will be a system of differential equations of
fourth order, but it will be possible to turn it into a system of
differential equations of second order, as we will see below.

If the stress-energy tensor has no shear, the Einstein equation for
$\mu = i\neq \nu = j$ reduces to

\be
f''(\Ro) \delta R +f'(R_0) (\Phi-\Psi) = 0 \label{fRGij} \, ,
\ee
where $\delta R$ is the perturbation of the curvature scalar

\be
\ba{c}
\delta R \,=\, 12(\dH + 2 H^2) \Phi + 6 H \dot{\Phi} + 24 H \dot{\Psi}
\\ + 6 \ddot{\Psi} + \frac 2{a^2}\Delta(\Phi-2\Psi))\, .
\ea \label{delR}
\ee

Eq.~\eqref{fRGij} is a second order differential equation for the scalar
perturbations except in the case  of General Relativity ($f''=0$)
where it becomes an algebraic equation ($\Phi=\Psi$).
Eq.~\eqref{fRGij} can be used to turn the remaining components
of the equations of $f(R)$ theories \eqref{EinfR} into second order
equations.

In the case of adiabatic perturbations in a matter dominated epoch one has

\be
\ba{c}
-8 \pi G \bar{\delta p} =  f'(\Ro)(a^{-2}\Delta(\Phi-2\Psi)+\ddot
\Psi +6 H \dot\Psi + H\dot\Phi +2(\dH + 3 H^2)\Phi -\frac 12 \delta R)
\\ +  f''(\Ro)(-(\dH+3H^2)\delta R -2\dRo \dot \Psi -8 H \dRo
\Psi -4 H \dRo \Phi -2 \ddRo \Phi) - 2
f^{(3)}(\Ro)\dRo^2\Phi  \\
 -\partial^2_t\left[ f'(\Ro)(\Phi-\Psi) \right]-2 H
\partial_t\left[ f'(\Ro)(\Phi-\Psi) \right] = 0 \, .
\ea \label{fRadiab}
\ee

If we compare the equations governing the behavior of adiabatic
perturbations in a $f(R)$ theory \eqref{fRGij}, \eqref{fRadiab} with
those coming from the ACM, \eqref{Phi},\eqref{Psi}, we find that they
describe very different
behaviors. In one case we have a system
of two coupled second order differential equations for $\Phi$ and
$\Psi$. In the ACM case we got a single second order differential
equation and an algebraic equation between the two gravitational
potentials. For a given $f(R)$ theory with an associated $g(H)$
homogeneous behavior, we find that the behavior of linearized
perturbations differs from the one defined by the linearized
perturbations in the ACM.  This means that $f(R)$ theories break in
general our assumption in the third section (there are terms with more than
two derivatives in the equations for the perturbations, and therefore
the behavior of perturbations in General Relativity
is not recovered when $H_+ \gg H \gg H_-$ for all modes).

Let us see this in a simple example. We will choose the easiest
biparametric family of $f(R)$ theories: the one that, under matter
domination, gives the same background evolution as General Relativity
without a cosmological constant \cite{Cortes:2008fy},

\be
f(R)=R+c_1 \vert R \vert^{\frac 1{12}(7-\sqrt{73})}+c_2 \vert R
\vert^{\frac 1{12}(7+\sqrt{73})} \, \label{fRGR}.
\ee
These kind of models were introduced in Ref.~\cite{Nojiri:2003ft}.

Let us first check if the solutions of the equations for the
perturbations in General Relativity in the presence of pressureless
matter,

\be
\begin{array}{ccl}
\Psi = \Phi & = & const \, ,\\
\Psi = \Phi & \propto & t^{-5/3} \, , \label{solsGR}
\end{array}
\ee
are solutions of the equations of motion in the case of the
biparametric family of $f(R)$ theories (\ref{fRGR}).
The easiest is to verify if Eq.~\eqref{EinfR} with $\mu = i \neq \nu = j$,
\be
\begin{array}{c}
2 f''(\Ro) (3 H^2 \Phi + 3 H \dot{\Phi} + 12 H \dot{\Psi} + 3
 \ddot{\Psi} + \frac 1{a^2}\Delta(\Phi-2\Psi))\\
 +f'(\Ro) (\Phi-\Psi) = 0 \label{fRGijGR} \, ,
\end{array}
\ee
where $\Ro (H) = -3 H^2$ is the value of the homogeneous curvature
scalar as a function of $H$ in this family of theories, is also
fulfilled by the solutions (\ref{solsGR}). The result is obviously
not.

The second order differential equations for the perturbations in the
$f(R)$ theories (\ref{fRGR}) in the presence of matter can be found by
substituting (\ref{fRGijGR}) and its derivatives into the other
equations of the system \eqref{EinfR}. The equation for the pressure
perturbations then gives
\be
\begin{array}{c}
-f'(\Ro)[\ddot{\Psi}+\ddot{\Phi}+ 4 H
  \dot{\Psi}+4H\dot{\Phi}+3H^2\Phi] \\
-27H^3f''(\Ro)\dot{\Phi}+\frac{f(\Ro)}2(\Psi-3\Phi)\,=\,8\pi G
  \bar{\delta p} \, = \, 0 \, \label{fRGiiGR} \, .
\end{array}
\ee
It is obvious that the $i\neq j$ equation of the cosmologic
perturbations in General Relativity,
\be
\Phi - \Psi = 0
\ee
is not recovered from (\ref{fRGij}) in the $H \gg H_-$ limit for all
$k\ll H_+$. In fact the deviation would be significant for modes with
$k \gtrsim H a \left(\frac{H}{H_-}\right)^{\frac 1{12}(7-\sqrt{73})}
\sim H a$. Therefore, the assumption we have made in order to derive
the equations for the perturbations in the ACM is broken.

It is a known problem that $f(R)$ theories are unable to pass
cosmological and astrophysical tests involving perturbations \cite{Chiba:2003ir},
unless the function involved is properly fine-tuned. In particular, if the
conformal equivalence between $f(R)$ theories and scalar-tensor theories is
used, it is necessary that the effective mass acquired by the new scalar
degree of freedom is unnaturally large \cite{Faulkner:2006ub}.
Some $f(R)$ theories that pass cosmological and solar system tests have been
proposed \cite{Hu:2007nk,Nojiri:2007as,Cognola:2007zu}.
However, it is also subject of debate if the conformal equivalence can
be used in order to extract physical predictions from this models, especially
predictions which involve perturbations \cite{Carloni:2009gp}. A mathematically
rigorous treatment of perturbations in $f(R)$-gravity can be found in
Ref.~\cite{Carloni:2007yv}.

\subsection{Quintessence Models}

Another class of models used to describe the acceleration of the
universe are those in which a so called quintessence field, typically
of scalar type $\varphi$, is added as a component of the universe
\cite{Wetterich:1987fm}. The only functional degree of freedom in most
models is just the scalar potential $V(\varphi)$, which can be tuned
to fit the homogeneous expansion of the universe. In our previous work
\cite{Cortes:2008fy} we found the correspondence between a given
homogeneous evolution parameterized by $g(H)$ and the potential for
the quintessence which drives under General Relativity this
evolution. We now wonder if the behavior of perturbations in
quintessence models also resembles the behavior in ACM.

The field action is given by

\be
S_\varphi = \int d^4 x \sqrt{-g} \left[ g^{\mu\nu}\partial_\mu \varphi
  \partial_\nu \varphi -V(\varphi) \right] \, . \label{lagquint}
\ee

Let us assume that the field can be split in two components: one which
is only time dependent and which drives the homogeneous evolution of
the universe, $\varphi_{(0)} (t)$, and a small perturbation which is
inhomogeneous, $\delta\varphi({\vx},t)$. We will consider linearized
perturbations of the metric, the field, and the other components of
the universe. The resulting linearized stress-energy tensor of the
field is

\begin{eqnarray}
\delta T^\varphi_{00} & = & \dot{\varphi}_{(0)}
\dot{\delta\varphi}-\Phi V(\varphi_{(0)}) + \frac 12
V'(\varphi_{(0)})\delta\varphi \, , \\
\delta T^\varphi_{0i} & = & \dot{\varphi}_{(0)} \partial_i \delta\varphi \, ,\\
\delta T^\varphi_{ij} & = & a^{-2}
\delta_{ij}\left[\dot{\varphi}_{(0)}^2 (\Phi-\Psi)
  +\dot{\varphi}_{(0)} \dot{\delta\varphi} \right. \nonumber \\
& & \left. +\Psi V(\varphi_{(0)}) - \frac 12 V'(\varphi_{(0)})\delta\varphi
\right] \, .
\end{eqnarray}

By virtue of the Einstein's equations, it is always possible to turn a
modification in the Einstein tensor into a new component of the
stress-energy momentum tensor of the sources of the gravitational
field. We wonder if it is possible to account for the modification of
the effective Einstein tensor described in the third
section with a new component described by this quintessence
field. However, it is not hard to see that the effective Einstein
tensor that we are proposing has a modified $i\neq j$ component, while
the $i\neq j$ component of the stress-energy tensor of the quintessence field
$\varphi$ is zero. Therefore, it is not possible to describe the
evolution of perturbations in the ACM as driven by an effective scalar
field component~(\ref{lagquint}).

\subsection{Dark Fluid Models}

A deformation of the gravitational physics can be also made equivalent
to the addition of a non-standard fluid component to the cosmic pie at
the homogeneous level. The fluid component used to explain the present
accelerated expansion of the universe is typically taken to be a
perfect fluid with large negative pressure.

One of the most popular parameterizations of this fluid is the so
called equation of state $\omega = p / \rho$ \cite{Turner:1998ex}. The
simplest models are $\omega = -1$, which is equivalent to a
cosmological constant, or a constant $\omega$, but recently a possible
time dependence of $\omega$ has been considered
\cite{Bassett:2004wz}. Needless to say, these {\it ad hoc}
parameterizations are subsets of the more general equation of state of
a perfect fluid, $p = p(\rho,s)$, where $s$ is the entropy density. It
is also questionable why must we restrict ourselves to perfect fluids
and not include a possible anisotropic stress tensor, possibly
depending on the energy and entropy densities
\cite{Capozziello:2005pa}.

As in the previous subsection, it is always possible to use the
Einstein's equations to turn a modification in the gravitational
physics into a new fluid component of the universe. In the case of the
equations derived in the third section, the
equivalent Dark Energy Fluid would have the following properties:

{\small
\begin{eqnarray}
 8\pi G \bar{\delta \rho}_X & = & (2H-g')\left[\frac 1{a^2 H}\Delta
 \Psi -3(H\Phi + \dot{\Psi})\right], \\
 8\pi G \bar{\theta}_X & = & \frac{(2H-g')}H(H\Phi + \dot{\Psi}), \\
 8\pi G \Pi_X & = & \frac
 1{a^2}\left\{\frac{2H-g'}{2H}(\Phi-\Psi)-\frac{\dot{H}}{2H^3}(g'-Hg'')\Psi\right\},
 \\
8\pi G \bar{\delta p}_{X} & = & (2H-g')(3H\Phi +3\dot\Psi-\frac{\dot
 H}{H^2}\dot \Psi + \frac{\dot H}{H}\Phi + \dot \Phi + \frac 1H
 \ddot\Psi)\nonumber \\
& & +\frac{\dot H}{H}(2-g'')(H\Phi+\dot\Psi)-8\pi G \Delta\Pi_X.
\end{eqnarray}
}

The resulting fluid is not an ideal fluid ($\Pi_X = 0$) or even a
Newtonian fluid ($\Pi_X \propto \bar\theta_X$). With the use of
\eqref{gh}, \eqref{Phi} and \eqref{Psi}, it will be possible to write
$\bar{\delta\rho}_X$ and $\bar{\delta p}_X$ as a combination of $\bar
\theta_X$, $\Pi_X$ and their derivatives,

\begin{eqnarray}
 \bar{\delta\rho}_X & = & f_1(\rho_{(0)X}) \Delta \Pi_X +
 f_2(\rho_{(0)X}) \bar\theta_X \, ,\\
\bar{\delta p}_X & = & f_3(\rho_{(0)X})\Pi_X -\Delta\Pi_X +
 f_4(\rho_{(0)X})\bar\theta_X \, .
\end{eqnarray}

These very non-standard properties show that, although it is formally
possible to find a fluid whose consequences mimic the ones of such a
modification of the gravitational physics, this fluid is very exotic.

\section{Summary and Conclusions}

In view of recent data, an updated comparison of cosmological
observations with a phenomenological model proposed in a recent work
has been presented.

An extension of this phenomenological model (ACM)
beyond the homogeneous approximation has been introduced allowing us to
describe the evolution of scalar perturbations at the linear level.

A comparison with the
spectrum of thermal fluctuations in CMB has been used to explore the
possibility to determine the parameters of the ACM through its role in
the evolution of scalar perturbations. The results of this comparison
does not further restrict the parameters of the model, due to the masking
of the associated late ISW effect by the cosmic variance. However, better
measurements of the position of the third acoustic peak should improve
the constraints significatively.

It has been shown that the equivalence of different formulations of
the accelerated expansion of the universe in the homogeneous
approximation is lost when one considers inhomogeneities. In
particular we have shown that the general structure of the evolution
equations for scalar perturbations in the ACM differs from the
structure of the equations corresponding
to modified $f(R)$ theories of gravity, to quintessence models or to a
dark fluid with standard properties.

The possibility of going beyond
the linearized approximation for the scalar perturbations and to
consider vector and tensor perturbations will be the subject of a
future work.

We would like to thank Marco Bruni for inspiration. We would also like
to thank Valerio Faraoni, Daniel G. Figueroa, Troels Hagub\o{}lle,
Sergei Odintsov and
Javier Rubio for fruitful discussions. J.I. also thanks the
hospitality of the Benasque Center for Science and the Scuola
Internazionale Superiore di Studi Avanzati (SISSA) during the
development of this work.

This work has been partially supported by CICYT (grant
FPA2006-02315) and DGIID-DGA (grant2008-E24/2). J.I. acknowledges a FPU
grant from MEC.

\section*{Appendix: Covariant Perturbation Equations for the ACM}

The code of CAMB \cite{CAMB} makes use of the equations for the
perturbations of the metric in the covariant approach. The quantities
can be computed in a given ``frame'', labeled by a 4-velocity
$u^\mu$. In particular, CAMB uses the dark matter frame, in which the
velocity of the dark matter component is zero (the dark matter
frame).
Furthermore, it parameterizes the time evolution with the conformal
time $a(\tau)d\tau = dt$.

In order to apply CAMB to the ACM model it is necessary to
identify frame invariant quantities, and then relate them to
gauge invariant quantities \cite{Gordon:2002gv}. The following
comoving frame quantities are used in the CAMB code: $\eta$ (the
curvature perturbation), $\sigma$ (the shear scalar), $z$ (the
expansion rate perturbation), $A$ (the acceleration, $A=0$ in the dark
matter frame), $\phi$ (the Weyl tensor perturbation), $\chi^{(i)}$
(the energy density perturbation of the species i), $q^{(i)}$ (the
heat flux of the species i), and $\Pi^{(i)}$ (the anisotropic
stress of the species i). All of these quantities are defined in
\cite{Challinor:1998xk}.  These variables are related to the gauge
invariant variables via the following dictionary,

\be
\begin{array} {rcl}
-\frac \eta 2 -\frac{\mH\sigma}{k} & \equiv & \Psi \, ,\\
- A + \frac{\sigma'+\mH\sigma}{k} & \equiv & \Phi \, ,\\
\Ji + \frac{\rho'\sigma}{k} & \equiv & \bar{\delta\rho} \, ,\\
\rho q + (\rho + p)\sigma & \equiv & \frac k a (\rho + p) \bar{\theta}\, ,
\end{array}
\ee
where prime denotes derivatives with respect to conformal time,
except when acting on $g(H)$ where it denotes a derivative with
respect to the Hubble rate $H$, and $\mH = a'/a = aH$. On the other hand
$z = \sigma + \frac{3}{2k}(\eta'+2\mH A)$ and $\phi = (\Phi +
\Psi)/2$. The anisotropic stress $\Pi$ is already frame invariant.

Written in the dark matter frame, the equations for the scalar perturbations read

\be
\begin{array}{rcl}
g'\left(\frac{k^2 \eta}{2 \mH} + k z\right) & = & 8\pi G a \Sigma_i \Ji^{(i)} \, ,\\
\frac{g' k^2}{3\mH}(\sigma - z) = -\frac{g'k}{2\mH} \eta' & = & 8\pi G
a \Sigma_i \rho_i q^{(i)}\, ,\\
-\mH k g'(z' + \mH z) & = & 8 \pi G a \Sigma_i\left[
  \mH^2 (1+3c_s^{(i)2}) - (\mH'-\mH)^2(1-\frac{\mH g''}{a g'}) \right]\Ji^{(i)}\, ,
\\
\frac {g'}\mH \left(\frac{\sigma'+\mH \sigma}{k}-\phi
\right)-\frac{\mH'-\mH^2}{2 \mH^3}(g'-\mH g''/a) \left(\frac\eta
2+\frac{\mH\sigma}k \right)& = & - 8\pi G a \Sigma_i\Pi^{(i)}/k^2 \, .
\end{array}
\ee

The following combination of the constraint equations is also useful:

\be
k^2\phi = -\frac{8\pi G a \mH}{g'}\Sigma_i\left[ \Pi^{(i)}
  +(1-\frac{\mH'-\mH^2}{2 \mH^2}(1-\frac{\mH g''}{a g'}))(\Ji^{(i)}
  +3\mH \rho_i q^{(i)}/k)\right]\, .
\ee

These equations are plugged into the Maple files provided with the
CAMB code and run to get the ISW effect \cite{CAMB}.

%%%%%%%%%%%%%%%%%%%%%%%%%%%%%%%%%%%%%
\chapter{Interpretation of neutrino oscillations based on new physics in the infrared}
%%%%%%%%%%%%%%%%%%%%%%%%%%%%%%%%%%%%%

An interpretation of neutrino oscillations based on a modification of
relativistic quantum field theory at low energies, without the need to
introduce a neutrino mass, is seen to be compatible with all observations.

\section{Introduction}

The now well-established observation of deficit of solar neutrinos,
atmospheric neutrinos, and neutrinos from reactors and accelerators finds
a coherent interpretation in terms of neutrino
oscillations between three neutrino flavors of different masses~\cite{GonzalezGarcia:2007ib}.
In the minimal standard model (SM), and in contrast with the rest
of the matter particles, the neutrino is assumed to be a zero mass,
left-handed fermion. Therefore neutrino oscillations is our first
glimpse of physics beyond the SM.

Massive neutrinos are introduced in extensions of the SM which
normally invoke new physics at high energies. In particular, one can
consider a Majorana mass term for the neutrino, generated by a
five-dimensional operator in the SM Lagrangian which would be
suppressed by the inverse of a certain high-energy scale.
Another possibility is to enlarge the field content of the SM with
a right-handed neutrino, which allows mass to be generated by
the usual Higgs mechanism. One has to account however for the smallness
of the neutrino mass, which is achieved by the see-saw mechanism~\cite{Minkowski:1977sc,Ramond:1979py,Gell-Mann,Yanagida:1979as,Mohapatra:1979ia},
again invoking a grand-unification scale.

The presence of new physics at high energies has been explored in
several attempts to find alternatives to the standard neutrino
oscillation mechanism.
This new physics might include Lorentz and/or CPT
violations. These two low-energy symmetries are being questioned at
very high energies in the framework of quantum gravity and string theory
developments~\cite{AmelinoCamelia:1997gz,Gambini:1998it,Alfaro:1999wd,Alfaro:2001rb,Alfaro:2001gk,Jacobson:2001tu,Kostelecky:2000mm}, and in fact simple models with Lorentz and/or
CPT violations are able to generate neutrino oscillations,
even for massless neutrinos~\cite{Coleman:1997xq,Kostelecky:2003cr,Kostelecky:2003xn}. Some of them are considered in
the context of the Standard Model Extension (SME)~\cite{Colladay:1996iz,Colladay:1998fq}, which
is the most general framework for studying Lorentz and CPT violations
in effective field theories.

However, all these alternative mechanisms involve new energy dependencies
of the oscillations which are in general disfavored over the standard
oscillation mechanism by experimental data,
which also put strong bounds on the contribution of new physics to this
phenomenon~\cite{Fogli:2003th,GonzalezGarcia:2004wg}. This seems to indicate that our understanding
of neutrino oscillations as driven by mass differences between neutrino flavors
is indeed correct.

In this letter we want to argue that this might not be the case. We will
present an example of new physics to the SM able to generate neutrino
oscillations and which is essentially different from previously considered
models in one or several of the following aspects: it does not necessarily
add new fields to those present in the SM, it may be completely indistinguishable
from the standard oscillation mechanism in the energy ranges where the
phenomenon has been studied (for neutrinos of medium and high energies), so
that automatically satisfies all constraints which are already fulfilled by
the standard mechanism, and finally, it predicts new physics in the infrared,
so that future neutrino low energy experiments could
distinguish this mechanism from the standard one.

Our example will be based on the so-called theory of noncommutative
quantum fields, which has recently been proposed
as an specific scheme going beyond quantum field
theory~\cite{Carmona:2002iv,Carmona:2003kh,Mandanici:2004ht,Balachandran:2007ua}.
The consequences for neutrino oscillations of a simple model with
modified anticommutators for the neutrino fields, which can be
identified as an example of a SME in the neutrino sector, has very
recently been explored in Ref.~\cite{Arias:2007zz}.
We will see however that it is possible to introduce a generalization
of the anticommutation relations of fields in a more general way than
that studied in Ref.~\cite{Arias:2007zz}, going beyond the effective field
theory framework of the SME,
which is the key to reproduce the oscillation results without the
need to introduce a neutrino mass, and with new consequences at
infrared energies.

\section{Noncanonical fields and neutrino oscillations}

The theory of noncommutative fields was first considered in
Refs.~\cite{Carmona:2002iv,Carmona:2003kh}. It is an extension of the usual canonical
quantum field theory in which the procedure of quantification of a
classical field theory is changed in the following way:
the quantum Hamiltonian remains the same
as the classical Hamiltonian, but the canonical commutation
relations between fields are modified. In the case of the scalar
complex field this modification leads to the introduction of two
new energy scales (one infrared or low-energy scale, and another one
ultraviolet or high-energy scale), together with new observable
effects resulting from the modification of the dispersion relation
of the elementary excitations of the fields~\cite{Carmona:2003kh}. If one is
far away from any of these two scales the theory approaches the
canonical relativistic quantum field theory with corrections involving
Lorentz invariance violations which can be expanded in powers of the
ratios of the infrared scale over the energy and the energy over the
ultraviolet scale. By an appropriate choice of the two new energy
scales one can make the departures from the relativistic theory
arbitrarily small in a certain energy domain.

We will now explore the relevance of the extension of relativistic
quantum field theory based on noncanonical fields in neutrino
oscillations. In particular, we will show that it is possible to
obtain oscillations with the observed experimental properties just
by considering a modification of the anticommutators of the fields
appearing in the SM, without the need to introduce a right-handed
neutrino or a mass for this particle.

With the left-handed lepton fields of the SM
\begin{equation}
\Psi_{L \alpha}=\left(\begin{array}{c}\nu_{\alpha} \\
  l_{\alpha}\end{array}\right)_L ,
\end{equation}
where $\alpha$ runs the flavor indices, $(\alpha=e,\mu,\tau)$, the
simplest way to consider an analog of the extension of the canonical
quantum field theory for a complex scalar field proposed in
Refs.~\cite{Carmona:2002iv,Carmona:2003kh} is to introduce the modified anticommutation
relations
\be
\{\nu_{L\alpha}(\xbf),\nu^{\dagger}_{L\beta}(\ybf)\} =
\{l_{L\alpha}(\xbf),l^{\dagger}_{L\beta}(\ybf)\}
= \left[\delta_{\alpha\beta} + A_{\alpha\beta}\right]
\,\delta^3(\xbf-\ybf).
\label{leptoncomm}
\ee
A particular choice for the matrix $A_{\alpha\beta}$ in flavor space which
parametrizes the departure from the canonical anticommutators
corresponds to the new mechanism for neutrino oscillations proposed
in Ref.~\cite{Arias:2007zz} which, however, is not compatible with the
energy dependence of the experimental data.

In order to reproduce the observed properties of neutrino
oscillations~\cite{GonzalezGarcia:2007ib} one has to go beyond this extension
and consider an anticommutator between fields at different
points. This can be made compatible with rotational and translational
invariance and with $SU(2)\times U(1)_Y$ gauge symmetry by making use
of the Higgs field
\begin{equation}
  \Phi=\left(\begin{array}{c}\varphi^+ \\
  \varphi^0 \end{array}\right),
  \tilde{\Phi}=\left(\begin{array}{c}\varphi^{0*} \\
  -\varphi^- \end{array}\right).
\end{equation}
The modified anticommutators of the left-handed lepton fields that we
consider in this work are
\be
\{\Psi_{L\alpha}(\xbf), (\Psi_{L\beta})^\dagger(\ybf)\}  =
\delta_{\alpha\beta}\,\delta^3(\xbf-\ybf)
 + \tilde{\Phi}(\xbf)
\tilde{\Phi}^\dagger(\ybf) B_{\alpha\beta}(|\xbf-\ybf|),
\label{phicomm}
\ee
where $B_{\alpha\beta}$ are now functions of $|\xbf-\ybf|$ instead
of constants. Note that Eq.~(\ref{phicomm}) is compatible with gauge
invariance since $\tilde{\Phi}(\xbf)$ has the same $SU(2)\times U(1)_Y$
quantum numbers as $\Psi_{L\alpha}(\xbf)$.

After introduction of spontaneous symmetry breaking ($\langle
\varphi^0 \rangle = v/\sqrt{2}$),
and neglecting effects coming from the fluctuation of the scalar field
which surely is a good approximation for neutrino oscillations, the
only anticommutators that are changed are those of the neutrino fields
\begin{equation}
\{\nu_{L\alpha}(\xbf),\nu^{\dagger}_{L\beta}(\ybf)\}=
\delta^3(\xbf-\ybf)\,\delta_{\alpha\beta}+C_{\alpha\beta}(|\xbf-\ybf|),
\label{neutrinocomm}
\end{equation}
where
\begin{equation}
C_{\alpha\beta}(|\xbf-\ybf|) = \frac{v^2}{2}
B_{\alpha\beta}(|\xbf-\ybf|).
\end{equation}

One can suspect that the new anticommutation relations Eq.~(\ref{neutrinocomm})
introduce a source of mixing between flavors that will affect neutrino oscillations.
We will see in the next Section that this is indeed the case.

\section{Solution of the free theory}

In order to study the neutrino oscillations induced by the modified
anticommutators of fields in the neutrino sector,
one needs to solve the free theory given by the Hamiltonian
\begin{equation}
H=\sum_{\alpha}\int d^3\xbf \left[i\,\nu_{L\alpha}^\dagger
\left(\bm{\sigma}\cdot\bm{\nabla}\right) \nu_{L\alpha}\right]
\label{ham}
\end{equation}
(where $\bm{\sigma}$ are the $2\times 2$ Pauli matrices),
and the anticommutation relations
showed in Eq.~(\ref{neutrinocomm}).

Let us introduce a plane wave expansion for the neutrino field
\be
\nu_{L\alpha} (\xbf)  =  \int \frac{d^3\pbf}{(2\pi)^3} \, \frac{1}{\sqrt{2p}}\,
\sum_{i} \left[\, b_i(\pbf)\, u_{L\alpha}^i (\pbf) \, e^{i
    \pbf\cdot\xbf}\right.  +
\,\left. d_i^\dagger(\pbf)\, v_{L\alpha}^i (\pbf) \, e^{-i \pbf\cdot\xbf} \right],
\label{neutrino}
\ee
where $p=|\pbf|$, and $(b_i(\pbf),d_i^\dagger(\pbf))$ are the annihilation
and creation operators  of three types of particles and antiparticles
(expressed by subindex $i$) with momentum $\pbf$. We now use the
following \emph{ansatz} for the expression of the
Hamiltonian~(\ref{ham}) as a function of the creation-annihilation operators:
\bea
H  =  \int \frac{d^3\pbf}{(2\pi)^3} \,\sum_{i}
\left[\ E_i(p)\, b_i^\dagger(\pbf)\,b_i(\pbf)\right.  +
\left. \bar{E}_i(p)\, d_i^\dagger(\pbf)\,d_i(\pbf)\right].
\label{ham2}
\eea
This corresponds to the assumption that the free theory describes
a system of three types of free particles and antiparticles for each
value of the momentum, with energies $E_i(p), \bar{E}_i(p)$, respectively.

Now, computing $[H,\nu_{L\alpha}]$ by two procedures: firstly by using Eq.~(\ref{ham})
for the Hamiltonian and the anticommutators~(\ref{neutrinocomm}), and secondly, by using
the expressions~(\ref{neutrino}) and~(\ref{ham2}), and equalling both results, one
obtains the following simple result for the energies and the coefficients in the
plane wave expansion of the field:
\bea
E_i(p) & =  & \bar{E}_i(p) = p\,[1+\tilde{c}_i(p)], \label{E} \\
u_{L\alpha}^i(\pbf) & = & v_{L\alpha}^i(\pbf)=e_\alpha^i(p)\,\chi^i(\pbf), \label{uv}
\eea
where $\chi^i(\pbf)$ is the two component spinor solution
of the equation
\begin{equation}
(\bm{\sigma}\cdot \pbf)\, \chi^i(\pbf)=-p\, \chi^i(\pbf)
\end{equation}
with the normalization condition
\begin{equation}
\chi^{i\dagger}(\pbf)\, \chi^i(\pbf)=2 E_i(p),
\end{equation}
$\tilde{c}_i(p)$ are the three eigenvalues of
$\tilde{C}_{\alpha\beta}(p)$,
the Fourier transform of $C_{\alpha\beta}(|\xbf-\mathbf{y}|)$ in Eq.~(\ref{neutrinocomm}),
and $e_{\alpha}^i(p)$ are the components of the normalized eigenvectors
of $\tilde{C}_{\alpha\beta}(p)$.

From Eq.~(\ref{E}), we see that the model presented here contains violation
of Lorentz invariance, but preserves $CPT$ symmetry.

\section{New IR physics and neutrino oscillations}

Since in the free theory solution
there are three types of particles and antiparticles with different energies,
and a mixing of creation and annihilation operators of different kinds of
particle-antiparticle in the expression of each field, it is clear that the
nonvanishing anticommutators of different fields will produce neutrino
oscillations, even for massless neutrinos. This observation was
already present in Ref.~\cite{Arias:2007zz}.
The probability of conversion of a neutrino of flavor $\alpha$
produced at $t=0$ to a neutrino of flavor $\beta$, detected at time
$t$, can be directly read from the propagator of the neutrino
field (\ref{neutrino}). This probability can be written as

\begin{equation}
{\cal P}\left(\nu_{\alpha}(0)\to \nu_{\beta}(t)\right) \,=\, \left|\sum_i
e^i_{\alpha}(p)^* \, e^i_{\beta}(p) \, e^{-i\,E_i(p)t}\right|^2 .
\end{equation}

This is the standard result for the oscillation between three states
with the unitary mixing matrix elements $U^i_{\alpha}$ replaced by the
coefficients $e^i_{\alpha}(p)$ of the plane wave expansion of the
noncanonical neutrino fields and the energy of a relativistic
particle $\sqrt{p^2 + m_i^2}$ replaced by the energy $E_i(p)$ of the
particle created by the noncanonical neutrino fields.

Let us now make the assumption that the modification of the anticommutators
is a footprint of new physics at low energies, parametrized
by an infrared scale $\lambda$. Although the introduction of corrections
to a quantum field theory parametrized by a low-energy scale has not been
so well explored in the literature as the corrections produced by
ultraviolet cutoffs, there are several phenomenological and
theoretical reasons that have recently lead to think on the necessity to
incorporate a new IR scale to our
theories~\cite{ArkaniHamed:1998rs,Antoniadis:1998ig,Sundrum:1998ns,ArkaniHamed:1998kx,Libanov:2005vu,Cohen:1998zx,Carmona:2000gd,Carmona:2005we}.

If the modifications of the anticommutation relations
are parametrized by an infrared scale $\lambda$ then it is reasonable
to assume an expansion in powers of $\lambda^2/p^2$ so that
\begin{equation}
\tilde C_{\alpha\beta}(p)\approx \tilde C_{\alpha\beta}^{(1)}
\,\frac{\lambda^2}{p^2} \text{ for } p^2\gg\lambda^2,
\label{approximation}
\end{equation}
and then
\begin{equation}
\tilde{c}_i(p)\approx \tilde{c}_i^{(1)}\,\frac{\lambda^2}{p^2},
\quad e_\alpha^i(p)\approx e_\alpha^{i(1)},
\label{eigenvalues}
\end{equation}
where $e_{\alpha}^{i(1)}$ ($\tilde{c}_i^{(1)}$) are eigenvectors
(eigenvalues) of $\tilde C_{\alpha\beta}^{(1)}$,
independent of $p$.

But in this approximation, the description of neutrino oscillations produced
by the new physics is completely undistinguishable from the conventional
description based on mass differences ($\Delta m_{ij}^2$) with a
mixing matrix ($U^i_{\alpha}$) between flavor and mass
eigenstates, just by making the correspondence
\begin{equation}
\Delta m_{ij}^2= 2 \,(\tilde c_i^{(1)}-\tilde c_j^{(1)})\,\lambda^2,
\quad U^i_{\alpha}=e_{\alpha}^{i(1)}.
\end{equation}

One should note that when $\lambda \neq 0$ the different energies for
different states select a basis in the Fock space and the mixing of
Fock space operators in the fermionic fields is unavoidable. It is
only when one considers the energy splitting of the
different particles that one has a physical consequence of the mixing
of different creation-annihilation Fock space operators in each
fermionic field. On the other hand, in the case $\lambda=0$
(corresponding to unmodified anticommutation relations) there is
an arbitrariness in the construction of the Fock space. One could make
use of this arbitrariness to choose a basis such that each fermionic
field contains only one annihilation and one creation operator (which
is equivalent to saying that the $e_{\alpha}^i(p)$ are indeed the
$\delta_{\alpha i}$) and therefore no oscillation phenomenon is
produced.

\section{Conclusions}

We have seen in the previous Section that the observations of neutrino
oscillations are compatible with their interpretation as a footprint of
new physics in the infrared. As far as we are aware, this is the first time
that an interpretation of neutrino oscillations coming from new physics,
without the need to introduce neutrino masses, and compatible with
all experimental results, is presented.

This is achieved because of the indistinguishability of the new mechanism
from the conventional one in the range of momenta $p^2 \gg \lambda^2$.
In order to reveal the origin of the oscillations it is necessary to go
beyond the approximation Eq.~(\ref{approximation}), which requires the
exploration of the region of small momenta ($p^2 \approx
\lambda^2$). To get this result it has been crucial to introduce a new
infrared scale through a nonlocal modification of the anticommutation
relations of the neutrino field. Gauge invariance forbids a similar nonlocal
modification for the remaining fields due to the choice of quantum
numbers for the fermion fields in the standard model. In fact the
possibility to have the modified anticommutators (\ref{neutrinocomm}) for the
neutrino fields is related to the absence of the right-handed neutrino
field.

The model of noncanonical fields presented in this work has to be considered
only as an example of the general idea that new infrared physics may be
present in, or be (partially) responsible of, neutrino oscillations, and that
the conventional interpretation may be incomplete. In fact
an extension of relativistic quantum field theory based on the
modification of canonical anticommutation relations
of fields might not be consistent. We have not examined the associated
problems of unitarity or causality beyond
the free theory. But it seems plausible
that the consequences that we have obtained in the neutrino sector will be
valid beyond this specific framework.

In conclusion, in this work we have shown that the experimentally observed
properties of neutrino oscillations do not necessarily imply the
existence of neutrino masses. In fact, future experiments
attempting to determine the neutrino mass, such as KATRIN~\cite{Osipowicz:2001sq},
may offer a window to the
identification of new physics beyond relativistic quantum field theory
in the IR. At this level it is difficult to predict specific
observational effects due to the lack of criteria to select a choice
for $\tilde C_{\alpha\beta}(p)$ in this specific model. A simple
example, however, would be the presence of negative eigenvalues
of this matrix, which could be reflected in an apparent negative
mass squared for the neutrino (see Eq.~(\ref{eigenvalues}))
in the fits from the tail of the tritium spectrum.
Effects on cosmology could also be possible, again depending
on the exact modification of the neutrino dispersion
relation in the infrared. All we can say is that if
the origin of neutrino oscillations is due to new physics in the
infrared then experiments trying to determine the absolute values of
neutrino masses and/or cosmological observations might have a
reflection of the generalized
energy-momentum relation Eq.~(\ref{E}) for neutrinos.

This work has been partially supported by CICYT (grant
FPA2006-02315) and DGIID-DGA (grant2006-E24/2). J.I. acknowledges a FPU
grant from MEC. 
%*************************************************************************
\chapter{The Theory of a Quantum Noncanonical Field in Curved Spacetimes}
%*************************************************************************

Much attention has been recently devoted to the possibility that quantum gravity effects could lead to departures from Special Relativity in the form of a deformed Poincar\`e algebra. These proposals go generically under the name of Doubly or Deformed Special Relativity (DSR). In this article we further explore a recently proposed class of quantum field theories, involving noncanonically commuting complex scalar fields, which have been shown to entail a DSR-like symmetry.
An open issue for such theories is whether the DSR-like symmetry has to be taken as a physically relevant symmetry, or if in fact the ``true'' symmetries of the theory are just rotations and translations while boost invariance has to be considered broken.
We analyze here this issue by extending the known results to curved spacetime under both of the previous assumptions. We show that if the symmetry of the free theory is taken to be a DSR-like realization of the Poincar\'e symmetry, then it is not possible to render such a symmetry a gauge symmetry of the curved physical spacetime. However, it is possible to introduce an auxiliary spacetime which allows to describe the theory as a standard quantum field theory in curved spacetime. Alternatively, taking the point of view that the noncanonical commutation of the fields actually implies a breakdown of boost invariance,  the physical spacetime manifold has to be foliated in surfaces of simultaneity and the field theory can be coupled to gravity by making use of the Arnowitt-Deser-Misner prescription.

\section{Introduction}

Quantum gravity has historically suffered from  lack of observational support and problems with conceptual issues  common to all of the most studied proposals \cite{GreSchWit,Polchinski2,Rovelli:1997yv,Thiemann:2001yy,Ashtekar:2001ir,Smolin:2003rk} (e.g.\/  the ``problem of time" and the ``background-independence problem" \cite{stachashtbook}). However, over the past few years there has been a growing interest in possible low energy, observable, effects of quantum gravity scenarios, ranging from TeV-scale quantum gravity to high energy departures from Local Lorentz invariance of spacetime (see e.g.\/ the discussion in Ref.~\cite{Mattingly:2005re,AmelinoCamelia:2008qg}). In particular, it was recently suggested that the Planck length $L_p$ ($L_p \sim 10^{-33} \text{cm} $) should be taken into account in describing the rotation/boost transformations between inertial observers.

In alternative to the standard approach of considering Planck suppressed Lorentz (and possibly CPT) invariance violations in effective field theory (EFT) (see e.g. Ref.~\cite{Mattingly:2005re,Mattingly:2007zz}), it was conjectured that the relativity principle could be preserved via a deformation of special relativity, in the form of the so-called ``Doubly or Deformed special relativity" (DSR) \cite{AmelinoCamelia:2000mn,AmelinoCamelia:2002wr,Judes:2002bw}. While this proposal is formulated in momentum space, a coordinate space definitive formulation is still lacking (and several open issues are still unresolved, see e.g.\/ Ref.~\cite{Schutzhold:2003yp,Liberati:2004ju}).

Although some interpretation of DSR in commutative spacetime have been proposed \cite{Liberati:2004ju,Aloisio:2006nd,Aloisio:2005qt},  there has been a growing activity \cite{KowalskiGlikman:2002we} toward connecting it  with noncommutative spacetime \cite{Majid:1994cy, Lukierski:1993wx,KowalskiGlikman:2004qa}, which also arise in 3d-Quantum Gravity \cite{Freidel:2005me} and String Theories \cite{connes}.
However, the connection between quantum gravity and noncommutative spacetime is in principle problematic: in fact the latter seem to entail a ``background-independence problem" given that the commutation relations among spacetime coordinates promote them to the role of {operators}, despite they should be treated as mere labels in a background independent theory (see e.g.\/ Ref.~\cite{Westman:2007yx}).

An alternative to modify the commutation relations between coordinates, while preserving background independence, is to modify the commutation relations of field operators on a spacetime manifold. If the manifold is flat spacetime, the Quantum Theory of Noncanonical Fields (QNCFT) has been studied \cite{Carmona:2002iv,Carmona:2003kh}\footnote{{These references named the theory Quantum Theory of Noncommutative Fields, but Quantum Theory of Noncanonical Fields provides indeed a more precise description.}} and it can be seen as a natural extension to field theory of Non Commutative Quantum Mechanics \cite{Mezincescu:2000zq, Duval:2000xr, Gamboa:2000yq}.  Noticeably, it has been shown that there is a link between the one particle sector of QNCFT in flat spacetime and DSR~\cite{Carmona:2009ra} .

Here we are going to use the results of Ref.~\cite{Carmona:2009ra} to study the case of the QNCFT in Curved Spacetimes. In the Section II we will review the main results of Ref.~\cite{Carmona:2003kh}. In the Section III we will discuss the connection between QNCFT and DSR as derived in Ref.~\cite{Carmona:2009ra}. In the Section IV we will derive the Lagrangian, the equation of motion, and the internal product of solutions of the QNCFT in Curved Spacetimes in different cases depending on {which are the} symmetries of the free theory in flat spacetime that we want to promote to gauge symmetries of the curved spacetime. The Section V is devoted to some closing remarks.

\section{QNCFT in Flat Spacetime}

Let us consider the theory of a complex scalar field, given by the Hamiltonian density:
\be
\Ham = \Pimas(\vx)\Pi(\vx)+\mathbf{\nabla}\Phimas(\vx)\cdot\mathbf{\nabla}\Phi(\vx)+m^2\Phimas(\vx)\Phi(\vx) \label{ham1} \, ,
\ee
the Hamiltonian being the integral to spacelike coordinates of $\Ham$.
The field is quantized by imposing the noncanonical commutation relations:
\begin{eqnarray}
	\left[\Phi(\vx),\Phimas(\vx')\right] & = & \theta \deltat(\vx-\vx') \nonumber\, , \\
	\left[\Phi(\vx),\Pimas(\vx')\right] & = & i \deltat(\vx-\vx') \label{com1}\, ,\\
	\left[\Pi(\vx),\Pimas(\vx')\right] & = & B \deltat(\vx-\vx') \nonumber \, ,
\end{eqnarray}
where $\theta$ and $B$ are considered to be ``sufficiently small", so that the canonical quantum field theory should be seen as a good approximation to this theory for a certain range of energies or momenta.  It should be noticed that $1/\theta$ plays the role of an UV scale and $B$ plays the role of an IR scale \cite{Carmona:2003kh}.
We will forget for the moment about the IR scale and set $B=0$ in our calculations (the effect of the scale $B$ turns out to be, when $\theta = 0$, a redefinition of the mass $m$ and a constant energy gap between n-particle and (n+1)-particle states)~\cite{Carmona:2003kh}.
We shall see now that the just given set of definitions completely specifies the theory.

Let us start by noticing that the momentum operator
\be
\mathbf{P}=\intxt \left[\Pimas(\vx)\mathbf{\nabla}\left(\Phi(\vx)-\frac{i\theta}2\Pi(\vx)\right)+ \,h.c.\,\right]\, ,
\ee
commutes with the Hamiltonian, and hence the field can be expanded in terms of creation-annihilation operators of particles and antiparticles with a given momentum.
\begin{eqnarray}
\Phi(x)& =&  \intpt \left(\udp(\vx,t) \, a_\vp + \vdp^*(\vx,t) \, b^\dagger_\vp\right)\label{field}\, ,\\
\udp(\vx,t) & = & \sqrt\frac{\ena}{\eb\left(\ena+\eb\right)} e^{-i(\ena t- \vp \cdot \vx)}\label{possol}\, ,\\
\vdp(\vx,t) & = & \sqrt\frac{\eb}{\ena\left(\ena+\eb\right)} e^{-i(\eb t- \vp \cdot \vx )}\label{negsol} \, .
\end{eqnarray}

Equations (\ref{ham1}) and (\ref{com1}) describe a theory of particles with energy-momentum relation
\be
\frac{\ena(\vp)}{\omp}=\sqrt{1+\left(\frac{\theta \omp}2\right)^2}+\frac{\theta \omp}2 \label{disp_a}\, ,
\ee
and their antiparticles, with energy-momentum relation
\be
\frac{\eb(\vp)}{\omp}=\sqrt{1+\left(\frac{\theta \omp}2\right)^2}-\frac{\theta \omp}2 \label{disp_b}\, ,
\ee
where $\omp = \sqrt{\vp^2 + m^2}$. The difference of energies between particles and antiparticles with the same momentum is a consequence of the CPT violation of the theory.
Notice also that  $\ena(-\theta) =\eb (\theta)$, as it should be, due to the symmetry $a\leftrightarrow b$, $\Phi\leftrightarrow \Phimas$, $\theta \rightarrow -\theta$. Therefore $\vdp(\vx,t;\theta)=\udp(\vx,t;-\theta)$.

The equation of motion of the field can be obtained by using the Heisenberg equation for the evolution of operators in the Heisenberg picture. The result is
\be
L(x,\partial) \Phi(x)=\left[\partial_0^2+(m^2-\Delta)(1+ i \theta \partial_0)\right] \Phi(x) = 0 \label{motion1}\, .
\ee
If we define the following internal product in the space of solutions of (\ref{motion1}),
\begin{eqnarray}
\left(\varphi_1(\theta),\varphi_2(\theta)\right) & = &  -i\intxt
 \left(\varphi_1(\theta)\frac {\textstyle 1}{\textstyle 1-i\theta\partial_0}\partial_0 \varphi_2^*(\theta) \right.\nonumber \\
& & \left. -  \varphi_2^*(-\theta)\frac {\textstyle 1}{\textstyle 1-i\theta\partial_0}\partial_0 \varphi_1(-\theta)
 \right)\, ,
\label{prod}
\end{eqnarray}
then $\udp(\vx,t)$ and $\vdp^*(\vx,t)$ form an orthonormal (in the sense of Ref.~\cite{BD}) basis of solutions of the equation of motion.

As a side remark let us notice that equations (\ref{disp_a}), (\ref{disp_b}), (\ref{motion1}), show for $m=0$ a striking similarity with the dispersion relation
\be
\omega = \pm \sqrt{c^2 k^2 -\left(\frac{2 \nu k^2}{3} \right)^2} -i \frac{2\nu k^2}3\, ,
\ee
 and equation of motion
\be
\partial_t^2 \psi_1 \, = \, c^2\mathbf{\nabla}^2 \psi_1 + \frac 43 \nu \partial_t \mathbf{\nabla}^2 \psi_1
\ee
of sound waves propagating through a viscous fluid of viscosity $\nu$ at rest ($\vec{v}_0 = 0$) \cite{Visser:1997ux}. They turn out to be the same, taking into account that the frequency associated to a b-mode of energy $\eb$ is $-\eb$, and redefining $i\theta = \frac {4}{3} \nu$.
The meaning of a purely imaginary viscosity is far from clear, although the main consequence is obviously that the dissipative effects associated to viscosity are turned into dispersive ones, the dispersion relation becoming real.

\section{Symmetries of QNCFT in Flat Spacetime}

A symmetry of a theory is defined as a transformation which leaves the action of the theory invariant. A symmetry also has the property of transforming solutions of the equation of motion into solutions of the equations of motion. Thus, if we want to search for the symmetries of the theory, a good starting point would be to find the operators $\mathcal{O}(x,\partial,...)$ such that if $\Phi(x)$ is a solution of \eqref{motion1}, then also $\mathcal{O}(x,\partial,...)\Phi(x)$ is. {This implies that the action of the commutator of $L$ and $\mathcal{O}$ on solutions of the field equations has to vanish}
\be
\left[L(x,\partial,...),\mathcal{O}(x,\partial,...)\right]\Phi(x) = 0 \label{prop}\, .
\ee
The above program is carried out more easily if one Fourier transforms the field $\Phi(x)=\intpc e^{-i p\cdot x}\Phi(p)$.
Then, the equation of motion turns out to be \cite{MasterDiego},
\be
L \Phi(p) = \left\{ -p_0^2+(m^2+\vp^2)(1+\theta p_0) \right\}\Phi(p) = 0\, ,
\label{eq:disp}
\ee
and the operators  verifying the property (\ref{prop}) are only
\begin{eqnarray}
\mathcal{P}_i & = & p_i \, , \label{eq:sy1}\\
\mathcal{P}_0 & = & p_0\, , \label{eq:sy2}\\
\mathcal{M}_{ij} & = & p_i \frac \partial {\partial p^j}-p_j \frac \partial {\partial p^i}\, , \label{eq:sy3}\\
\mathcal{M}_{0i} & = & \frac {p_0}{\sqrt{1+\theta p_0}} \frac \partial {\partial p^i}-p_i \frac{(1+\theta p_0)^{3/2}}{1+\theta p_0/2} \frac \partial {\partial p^0}\, , \label{eq:sy4}
\end{eqnarray}
or arbitrary functions of them. Therefore the previous operators can be considered the generators of the symmetry group.

It is now evident that while the generators of translations $\mathcal{P}_i ,\mathcal{P}_0$ and rotations $\mathcal{M}_{ij}$ are left unchanged, a new set of generators for the boosts $\mathcal{M}_{0i}$ is induced by the noncanonical commutation of the field.
Notice however, that if one  restricts $p_0$ so that $p_0>-1/\theta$, then under the change of variables
\be
\tilde{p}_0 = \frac {p_0}{\sqrt{1+\theta p_0}}\, , \qquad \tilde{p}_i = p_i \, ,\label {aux}
\ee
the generators of the symmetry group can be rewritten as:
\begin{eqnarray}
\mathcal{P}_i & = & \tilde{p}_i \, , \\
\mathcal{P}_0 & = & \tilde{p}_0\left(\sqrt{1+\theta^2 \tilde{p}_0^2/4}+ \theta \tilde{p}_0/2 \right)\, ,\\
\mathcal{M}_{ij} & = & \tilde{p}_i \frac \partial {\partial \tilde{p}^j}-\tilde{p}_j \frac \partial {\partial \tilde{p}^i}\, ,\\
\mathcal{M}_{0i} & = & \tilde {p}_0 \frac \partial {\partial \tilde{p}^i}-\tilde{p}_i \frac \partial {\partial \tilde{p}^0}\, \label{gens}.
\end{eqnarray}

This shows that the generators of rotations and deformed boosts are isomorphic to the generators of the Lorentz group and it is only the action of the boosts on the four-momentum that it is deformed due to the presence of $\theta$ (reducing to its standard form in the limit of $\theta\to 0$). This is in striking similarity with what is normally conjectured in DSR scenarios \cite{AmelinoCamelia:2000mn, AmelinoCamelia:2002wr, Judes:2002bw} \footnote{Note that in the DSR literature as well as in Ref.~\cite{Carmona:2009ra} Greek letters $(\pi_0,\mathbf{\pi})$ are used to refer to the auxiliary variables $(\tilde{p}_0,\tilde{p}_i)$. We chose here a different notation in order to render the expressions in the rest of the paper more understandable. } and hence it is no surprise that also in this case the whole group of symmetries can be made isomorphic to the Poincar\'e group when $\mathcal{P}_0$ is substituted by $\tilde{\mathcal{P}}_0 = \tilde{p}_0$, which is also the generator of a symmetry. This clearly hints at the possibility to recast the theory in an auxiliary spacetime associated with momenta $(\tilde{p}_i ,\tilde{p}_0)$ in which the action of the Poincar\'e group will be standard. We shall explore this possibility in what follows but before a few remarks about the nature of the just found symmetry are in order.

 Symmetries in field theories can be of two types. Internal symmetries are transformations in the field space which do not mix fields at different spacetime points. These transformations will have the form $\Phi(x)\rightarrow\Phi'(x)$ with $\Phi'(x)=\mathcal{O}(x)\Phi(x)$. We will call instead external symmetries those symmetries which do mix fields at different spacetime points. For example this can happen if the transformation affects also spacetime itself. Indeed, spacetime symmetries are transformations in the spacetime background which affect the field. These transformations will have the form $x\rightarrow x'(x)$ and $\Phi(x)\rightarrow \Phi'(x')$ where, in the case of the scalar field, $\Phi'(x')=\Phi(x)$. Therefore spacetime symmetries are external symmetries. However, the opposite is not always true. In particular the symmetry associated to (\ref{eq:sy1}), (\ref{eq:sy2}), (\ref{eq:sy3}), (\ref{eq:sy4}),  is not a proper spacetime symmetry because of its non-local nature.

 Coming back to our study of the symmetries of the theory in flat spacetime we now want to explore the possibility to recast the whole field theory in a auxiliary spacetime in which the Poincar\'e group acts in the usual way. We shall do so by looking for transformations which leave the action invariant. While a rigorous treatment can be found in Ref.~\cite{Carmona:2009ra} we shall provide here a more concise derivation for the sake of simplicity.

The first step would be to find the action $S$, which will be the spacetime integral of some Lagrangian density $\Lag$. In order to achieve this, we will write (\ref{ham1}) in terms of field variables whose commutation relations are canonical,
\be
\Phi_c = \Phi-\frac{i\theta}2 \Pi \label{phic}\, .
\ee
Then, writing the Hamiltonian in these variables and performing the canonical transformation
\begin{eqnarray}
\bar{\Phi}  & = & \left(1-\frac{i\theta}2 \sqrt{m^2-\Delta}\right)^{-1}\Phi_c \, ,\nonumber \\
\bar{\Pi} & = & \left(1-\frac{i\theta}2 \sqrt{m^2-\Delta}\right) \Pi \nonumber \, ,
\end{eqnarray}
the action can be written, up to boundary terms, as
\begin{eqnarray}
S & = & \intxc \Lag (x) \nonumber \\
& = & \intxc \left(\Pimas \dot{\Phi}_c + \dot{\Phimas}_c \Pi - \Ham (\Phi(\Phi_c),\Pi)\right) \nonumber \\
& = & \intxc \left(\bar{\Pi}^\dagger \dot{\bar{\Phi}} + \dot{\bar{\Phi}}^\dagger \bar{\Pi} - \Ham (\Phi(\bar{\Phi}),\Pi(\bar{\Pi}))\right) \nonumber \\
& = & \intxc \, \bar\Phi^\dagger \left( -\partial_0^2-(1+i\theta\partial_0) (m^2-\Delta)\right)\bar{\Phi} \label{lag1} \, ,
\end{eqnarray}
where in the last line we have substituted the momentum as a function of the field and its derivatives via the Hamilton-Jacobi equations.

 We want now to show that the action \eqref{lag1} can be expressed as a standard scalar field action in flat spacetime. We can start by introducing the field redefinition
\be
\phi(x) = \sqrt{1+i \theta \partial_0}\bar{\Phi}(x)\, ,
\ee
 so that the action takes the simpler form,
\be
S = \intxc \, \phi^\dagger(x) \left( \frac{-\partial_0^2}{1+i\theta\partial_0}-(m^2-\Delta)\right)\phi (x)\, .
\ee
Let us look at this action in momentum space by using the Fourier transform of $\phi(x)$, $\phi(p)$
\be
S = \intpc \phi^\dagger(p) \left( \frac{p_0^2}{1+\theta p_0}-(m^2+\vp^2)\right)\phi (p)\, \label{lagfrac},
\ee
where the integral over $p_0$ goes from $-\infty$ to $+\infty$. Now we can split the action into two terms $S^\theta$ and $\bar{S}^\theta$, with the integral over $p_0$ ranging respectively from $-1/\theta$ to $+\infty$ and from $-\infty$ to $-1/\theta$. All the solutions of the classical equations of motion lay in $S^\theta$, but the $\bar{S}^\theta$ term can affect the quantum behavior of the theory~\cite{Carmona:2009ra}. Changing in $S^\theta$ the integration variable $p_0$ for $\tilde{p}_0$, defined in (\ref{aux}),  this can be rewritten as
\be
S^\theta = \intpt \frac{d\tilde{p}_0}{2 \pi}  \frac{\partial p_0}{\partial \tilde{p}_0}\phi^\dagger(p(\tilde{p})) \left( \tilde{p}_0^2-(m^2+\vp^2)\right)\phi (p(\tilde{p}))\, .
\ee
 Finally we can define the auxiliary field
 \be
 \tilde{\phi}(\tilde{p})=\sqrt{\frac{\partial p_0}{\partial \tilde{p}_0}}\,\phi (p(\tilde{p})) \, ,
\label{auxfield}
 \ee
 and inverse Fourier transform, so to arrive to the standard Klein-Gordon-action in some ``auxiliary spacetime".
\be
S^\theta =\intxt \, d\tilde{x}_0\, \tilde{\phi}^\dagger (\tilde{x})\left(-\eta^{\mu\nu}\partial_\mu\partial_\nu - m^2 \right) \tilde\phi(\tilde{x})\, . \label{actaux}
\ee

Now the properties of the symmetry group of the theory can be understood. When written in the auxiliary variables $(\tilde{x}_0,\vx)$ (auxiliary frame), the symmetry group is the Poincar\'e one associated to the flat auxiliary spacetime symmetries of $S^\theta$.
Reversely, if we assume that the Poincar\'e group acts trivially on $\bar{S}^\theta$, its action on the real spacetime and the original field variables (original frame) can be derived by pulling back the operations that we have applied above. In this case the action of the group is found to be realized in some non-local way. Indeed, this setting is similar to the one proposed in the DSR framework~\cite{AmelinoCamelia:2000mn}.\footnote{Note that the above treatment does not depend on the detailed form of the field dispersion relation nor on its origin. In principle, the discussion may be applied to more general dispersion relations.}

However, it has to be noticed that the necessary splitting of the action into two terms implies that our initial theory is not exactly equivalent to a relativistic field theory in flat space, at the quantum level \cite{Carmona:2009ra}. As a matter of fact, one might argue that the symmetry we have found is a fake one, because it invokes non-locality (see e.g.\/ the discussion about symmetries of Maxwell equations in \cite{Giulini:2006uy}).

Indeed spacetime symmetries are local because the transformed field in a spacetime point is a function of the value of the field in just one spacetime point. In the case we are considering, the $\mathcal{M}_{0i}$ are generating a family of external transformations which cannot be interpreted as spacetime symmetries, at least in the usual way, due to their intrinsic non-locality. Indeed such a symmetry transforms a whole field configuration, which is a solution of the equation of motion, into a different field configuration which is also a solution.

Finally, one must also take into account that the Hamiltonian of the theory, the space integral of Eq.~(\ref{ham1}),  does not belong to the set of generators of this Poincar\'e-like symmetry group, although time translations are a symmetry of the free theory. The problem is that the commutator of the Hamiltonian with the generator of boosts would end up defining a new generator and hence the algebra would not be not closed.

 An infinite set of generators of the symmetry group of the free theory in flat spacetime
may be obtained by iteration of this procedure. However, the whole symmetry group forbids
the addition of new terms in the action or the appearance of a spacetime dependence of
the coefficients of the existing terms. Therefore the symmetry group must be reduced
(broken into subgroups) in order to extend the theory.

These considerations force us to take a decision. We either try to generalize the theory to curved spacetime by promoting the non-local implementation of the Poincar\'e group to a gauge symmetry, and therefore time translations are taken as an accidental symmetry of the theory in flat spacetime  (or at low energies where $\theta$ corrections are supposed to be negligible), or we can take an alternative point of view and consider as the physical symmetries only those realized in a local way while boost invariance is broken.

The first approach has been shown to suffice in the construction of a renormalizable interacting theory in flat spacetime~\cite{Carmona:2009ra}, but, as previously discussed, some of the intuitive notions of what a symmetry is can be spoiled by the non-locality of the implementation. While more intuitive, the second approach will imply a preferred time direction and preferred slices of simultaneity in spacetime.

\section{QNCFT in Curved Spacetimes}

The aforementioned alternative points of view regarding the actual symmetries  imply that  two different approaches in constructing a QNCFT in Curved spacetimes can be taken depending on which is the symmetry of the free field theory in flat spacetime we would like to promote to a gauge symmetry of the spacetime. Hence, we will perform them separately in the following subsections.

\subsection{A spacetime with a gauge DSR-like Invariance?}

One of the ways of formulating General Relativity (GR) consists in turning the global Poincar\'e symmetry of Minkowski spacetime into a gauge symmetry of any curved geometry.\footnote{ Indeed the Strong Equivalence Principle (which selects GR among relativistic theories of gravitation with spin-2 gravitons) entails Local Lorentz invariance as a fundamental postulate (together with universality of free fall and local position invariance of experiments, including gravitational ones)\cite{Will:2001mx}.} This procedure provides also a link between Quantum Field Theory (QFT) in flat spacetimes and QFT in curved backgrounds. We will try to follow the same steps by gauging the symmetry we have found for the noncanonical field.

A global symmetry in the noncanonical field has been found, whose generators are:
\begin{eqnarray}
\mathcal{P}_i & = & -i\partial_i \, ,\nonumber\\
\mathcal{\tilde{P}}_0 & = & i\frac{\partial_0}{\sqrt{1+i\theta\partial_0}} \, ,\nonumber\\
\mathcal{M}_{ij} & = & x_i\partial_j - x_j\partial_i \, ,\nonumber\\
\mathcal{M}_{0i} & = & x_i\frac{\partial_0}{\sqrt{1+i\theta\partial_0}}-x_0\partial_i \frac{(1+i\theta\partial_0)^{3/2}}{1+i\theta\partial_0/2} \label{gen}\, .
\end{eqnarray}
These operators satisfy the Poincar\'e Algebra. We want to generalize our considerations about the symmetry to curved spacetime without loosing this connection. We can do this by introducing the tetrad field formalism  i.e.\/ by constructing a set of normal coordinates $y^a_X$ at each point $X$ of the spacetime manifold (for example for time-like observers one can pick up as a preferred axis the one associated to the tangent vector to the observer worldline and the other three axis as an orthogonal basis in the spacelike hypersurface orthogonal to such vector).

The relationship between these coordinates and the general coordinates in the manifold defines the tetrad field $e^a_\mu (x) = \frac{\partial y^a_X}{\partial x^\mu}\vert_{x=X}$.  In terms of the $y^a_X$ the metric at $X$ is simply $\eta_{ab}$ (which is used to lower latin indices), and can be transformed to the metric expressed in terms of the general coordinates $g_{\mu\nu}(x)$ (which is used to lower greek indices) with the use of the tetrad.

In the standard case, we must introduce the covariant derivative, an operator which turns global Poincar\'e and Lorentz invariant tensors into local Poincar\'e and general coordinate transformations  invariant tensor.
\be
D_\mu = \partial_\mu - e^a_\mu (x) \mathcal{P}_a -\frac 12 \omega^{ab}_\mu (x) \mathcal{M}_{ab} \label{covar}\, ,
\ee
where $e^a_\mu (x)$ and $\omega^{ab}_\mu (x)$ (the spin connection) play the role of gauge fields.  This leads from QFT in Flat Spacetime to QFT in Curved Spacetimes when partial derivatives $\partial_a$ are replaced by covariant derivatives $D_a = e_a^\mu D_\mu$ and Lorentz tensors are replaced by generalized tensors built up with the tetrad.

However, if this procedure is applied to the case of the noncanonical field, changing the partial derivatives by covariant derivatives in the action (\ref{lag1}) and inserting the tetrad field does not make the gauge transformation (\ref{gen}) a gauge symmetry. The connection between the QNCFT and the relativistic QFT (\ref{actaux}) is broken.

 In order to see this, let us try to follow the same procedure as in the standard case. The action will have the form
\begin{eqnarray}
%S & = & \intxc \Lag (x) \nonumber\\
S & = & \intxc \sqrt{-g(x)} \left[\Pimas e_0^\mu(x) (D_\mu \Phi) + e_0^\mu(x) (D_\mu \Phimas) \Pi \right.\nonumber\\
&&- \Pimas\Pi - \delta^{ij}e_i^\mu(x) (D_\mu \Phimas) e_j^\nu(x) (D_\nu \Phi) - m^2\Phimas\Phi\nonumber\\
&&\left.  -\frac{i\theta}{2} \left(\Pimas e_0^\mu(x) (D_\mu \Pi) - e_0^\mu(x) (D_\mu \Pimas) \Pi \right) \right].\label{eq:break}
\end{eqnarray}
This action is explicitly general covariant and in the flat spacetime limit does reproduce the second line of Eq.~(\ref{lag1}). However, it is easy to see that there is no transformation enabling one to write \eqref{eq:break} as the curved spacetime generalization of the last line of Eq.~(\ref{lag1}) due to the explicit dependence of the tetrad field on the spacetime coordinates. Indeed, the derivation of Eq.~(\ref{lag1}), required infinite integrations by parts. If these were done in the curved spacetime case, an infinite series of new terms would appear in the action, involving covariant derivatives of the tetrad fields. Furthermore, it is also possible to check that the infinitesimal gauge transformations generated either by $\mathcal{\tilde{P}}_0$ or by $\mathcal{M}_{0i}$ do not leave the Lagrangian associated with \eqref{eq:break} invariant even at linear order in $\theta$.

These evident problems lead us to the conclusion that there seems to be a fundamental obstruction in gauging external symmetries generated by non-local operators and in particular in making the symmetry generated by (\ref{gen}) a gauge symmetry of the field action.

However, in the previous section we have shown that the non-locally implemented symmetry can be seen as a proper (i.e.~local) spacetime symmetry of a new auxiliary $\tilde{x}$-spacetime. Though non-locality prevents the gauging of symmetries in $x$-spacetime, the procedure of gauging spacetime symmetries and its results are well known. Therefore we can argue that the proper way to proceed is to gauge global DSR-like symmetries in $\tilde{x}$-spacetime. Of course, the procedure of $\tilde{x}$-gauging the global Poincar\'e symmetry leads to the standard QFT in curved spacetime, but now the curved spacetime is the auxiliary one.

In those regions in which the auxiliary spacetime is asymptotically flat, the mapping between auxiliary and noncanonical fields (\ref{auxfield}) can be used and the propagator in the  noncanonical field, in physical spacetime, can be found. However, a connection between the auxiliary field and the noncanonical field in a point of auxiliary spacetime in which the metric is not asymptotically flat seems precluded, at least in the present treatment.

Let us then reformulate all the theory in the auxiliary frame, in which everything is standard. The {part of the action which contains the solutions of the equations of motion} in this DSR-like approach would be
\be
S^\theta=\intauxx \gaux \tilde{\phi}(\tilde{x})\left(-\tilde{g}^{\mu\nu}\tilde{\nabla}_\mu\tilde{\nabla}_\nu - m^2 +\xi \tilde{R}(\tilde{x})  \right) \tilde\phi(\tilde{x})\, ,
\ee
where $\tilde{\nabla}$ means covariant derivative, $\tilde{R}(\tilde{x})$ is the Ricci scalar, and the line element in the $\tilde{x}$-spacetime is given by
\be
d\tilde{s}^2 = \tilde{g}_{\mu\nu} d\tilde{x}_\mu d\tilde{x}_\nu \nonumber\, .
\ee

The theory in the auxiliary variables is ordinary QFT in Curved Spacetime \cite{BD}. The equation of motion of the field in auxiliary spacetime is,
\be
\left(\tilde{g}^{\mu\nu}\tilde{\nabla}_\mu\tilde{\nabla}_\nu + m^2 +\xi \tilde{R}(\tilde{x}) \right) \tilde\phi(\tilde{x})=0 \label{mov2}\, .
\ee
Its solutions are spanned by a basis $\left\{\tilde{u}_\vp(\tilde{x}),\tilde{u}^*_\vp(\tilde{x})\right\}$ which is orthonormal under the internal product
\be
\left(\varphi_1,\varphi_2\right)=  -i\int_{\tilde{\Sigma}} d\tilde{\Sigma}\sqrt{-\tilde{g}_{\tilde{\Sigma}}(\tilde{x})}\tilde{n}_\mu
\left(\varphi_1 \tilde{\partial}_\mu \varphi_2^*-\varphi_2^*\tilde{\partial}_\mu \varphi_1 \right)\, ,
\ee
where $\tilde{\Sigma}$ is some spacelike hypersurface, $\tilde{n}$ the future-oriented timelike vector orthonormal to it. The value of the internal product is independent of the choice of $\tilde{\Sigma}$.

From the above discussion the standard approach to QFT in curved spacetime straightforwardly follows. In particular standard problems like particle creation from the vacuum can be approached with the usual Bogoliubov techniques. Of course, one may wonder how far one could trust such calculations given that the correspondence between the original, physical, spacetime and the auxiliary one (in which one ends up working) is only via the $S^\theta$ part of the real action. In this sense it is important to notice that the virtual modes belonging to the $\bar{S}^\theta$ term of the action do not affect the Bogoliuvov transformation between modes in the $S^\theta$ term, as these virtual modes ``do not see" the auxiliary spacetime.

The conclusion is that working in this framework the resulting theory will give then the same results as the standard QFT in curved spacetime, at least insofar one is working in the auxiliary spacetime. It still remains open the issue of  mapping back the results to the original spacetime which, as we said, seems doable only in asymptotically flat regions and hence not generically.

\subsection{Foliating spacetime with noncanonical commutation relations}
Let us adopt the alternative point of view that meaningful symmetries are local implementations of a certain symmetry group, in this case rotations in space and translations in space and time, {and that the DSR-like Poincar\'e group is just an accidental symmetry of the free action in flat spacetime}. Therefore, we would like to curve physical spacetime in a way such that just spatial rotations and spacetime translations are preserved. It should be clear that in this case, a preferred arrow of time emerges in which the fields appearing in the commutation relations are simultaneous. Each set of simultaneous events defines a spacelike hypersurface in the manifold of spacetime, foliating it, while the orthogonal, timelike direction defines the evolution in time. In those hypersurfaces no spacelike coordinates are preferred, and thus the commutation relations of the field in flat spacetime (\ref{com1}) have to be  suitably generalized in order to make them covariant under general coordinate transformations on the spacelike hypersurfaces,
\begin{eqnarray}
\left[\Phi(\vx,t),\Phimas(\vx',t)\right]& =&\frac \theta{\sqrt{h(t)}} \deltat(\vx-\vx') \nonumber\\
\left[\Phi(\vx,t),\Pimas(\vx',t)\right]&=&\frac {i\hbar}{\sqrt{h(t)}} \deltat(\vx-\vx') \nonumber\\
\left[\Pi(\vx,t),\Pimas(\vx',t)\right]&=&0\,
\end{eqnarray}
where $h(t)$ is the determinant of the three-metric $h_{ij}(t)$ associated to each spacelike hypersurface of simultaneity. This foliation of spacetime into slices of simultaneity naturally leads to the Arnowitt Deser Misner description of GR (ADM) \cite{Arnowitt:1960es,Arnowitt:1962hi}.
 We briefly review it here for completeness and for fixing the notation.

In the ADM formalism one takes a foliation of spacetime in spacelike hypersurfaces, which implies a split of the whole spacetime metric, the four-metric, into its spacelike-spacelike components $g_{ij}(\vx,t)$ --- which coincide in this frame of reference with the ones of the induced metric on the hypersurfaces (the three-metric $h_{ij}$) --- a three-vector shift function $N^i(\vx,t)$ and a three-scalar lapse function $N(\vx,t)$. The components of the four-metric in this frame are then
\be
\begin{array}{c}
g_{\mu\nu}
\end{array}
=
\left(
\begin{array}{cc}
g_{00} & g_{0j}\\
g_{i0} & g_{ij}
\end{array}
\right)
=
\left(
\begin{array}{cc}
N^2-N^i N^k h_{ik} & -h_{kj} N^k \\
-h_{ik} N^k & -h_{ij}
\end{array}
\right)\, .\label{ADM}
\ee
It has to be noticed that the formalism is still covariant under general spacelike coordinate transformations. Spacelike indices $i$,$j$,$k$,... are lowered with the three-metric $h_{ij}$ and raised with the inverse of the three-metric, $h^{ij}$. The inverse of the four-metric turns out to be
\be
\begin{array}{c}
g^{\mu\nu}
\end{array}
=
\left(
\begin{array}{cc}
1/N^2 & -N^j/N^2 \\
-N^i/N^2 & -h^{ij}+N^i N^j/N^2
\end{array}
\right)\, ,
\ee
and the determinant of the four-metric is $g = -N^2 h$. The {timelike} vector which is orthonormal to the spacelike hypersurfaces is
\be
n^\mu = (1/N, -N^i/N) \, . \label{time}
\ee

The covariant four derivative is constructed in the usual way with the affine connections which are furthermore assumed to be the Christoffel symbols $\Gamma^{(4)\lambda}_{\mu\nu}$ associated with the four-metric $g_{\mu\nu}$.
The extrinsic curvature $K_{ij}$ is defined with the covariant four-derivative of the normal to the hypersurfaces ($K_{ij}= N \Gamma^{(4)0}_{ij})$. The covariant three-derivative is defined as the projection of the covariant four-derivative on the spacelike hypersurface, $\nabla^{(3)}_i V^j \equiv V^j_{\vert i} = \partial_i V^j + \Gamma^{(3)j}_{ik}V^k$, where $\Gamma^{(3)j}_{ik} = \Gamma^{(4)j}_{ik} + K_{ik}N^j/N$ coincides with the Christoffel symbols built up with the three-metric. (From now on the superscript $(3)$ can be dropped in order to simplify the notation.)  The intrinsic curvature $R^i_{jkl}$ is built in the usual way with the three-metric $h_{ij}$ and its derivatives, and can be related to the four-curvature and the extrinsic curvature.

Finally, this notation has also a link to the tetrad notation introduced in the previous subsection. The tetrad $e_a^\mu$ associated with the metric (\ref{ADM}) is
\be
\begin{array}{ccc}
e_0^0 \, = \, 1/N & & e_0^i \, = \,N^i/N \\
e_i^0 \, = \, 0 & & e_j^i \, = \,e^{(3)j}_i \\
\end{array}\, ,\label{tetradADM}
\ee
where $e^{(3)i}_a$ is the dreibein associated to the $h_{ij}$ three-metric.

We have now to consider how can we couple the noncanonical field to the physical spacetime metric (\ref{ADM}). The result was already given in the previous subsection in Eq.~(\ref{eq:break}). Written in the notation of the ADM prescription with the use of (\ref{tetradADM}), the expression of the action, Lagrangian density and Hamiltonian density of the field are
\begin{eqnarray}
S_\Phi & = & {\textstyle \intxc }\sqrt{h} \Lag_\Phi \, ,\label{actionADM}\\
\Lag_\Phi & = & \Pimas\dot{\Phi}_c + \dot{\Phi}^\dagger_c \Pi-N \Ham_\Phi \, , \label{lagADM} \\
\Ham_\Phi & = & \Pimas \Pi+h^{ij}\partial_i\Phimas\partial_j\Phi+m^2\Phimas\Phi \nonumber\\
&& +\left(\frac{N^i}N \Pimas \partial_i\Phi_c \,+\, h.c.\,\right)\, . \label{hamADM}
\end{eqnarray}
where $\Phi_c$ is given by (\ref{phic}). The stress energy tensor is defined as
\be
T_{\mu\nu}  =  \frac 2{N\sqrt{h}}\frac{\delta S_\Phi}{\delta g^{\mu\nu}}\, ,
\ee
and turns out to be
\begin{eqnarray}
T_{00} & = & N\Pimas\dot{\Phi}_c +N N^i\Pimas \partial_i\Phi_c+N^i N^j \partial_i \Phimas \partial_j \Phi\, +\,h.c.\nonumber\\
& & -(N^2-N_i N^i)\Lag_\Phi \nonumber\\
T_{0i} & = & N^j\partial_i\Phimas \partial_j\Phi+N\Pimas\partial_i\Phi_c \,+\,h.c.\,+N_i \Lag_\Phi \nonumber\\
T_{ij} & = & \partial_i\Phimas \partial_j \Phi \, +\, h.c.\, +h_{ij} \Lag_\phi\, .
\end{eqnarray}

Now we have all the information required about the coupling of the noncanonical field to the gravitational potential in curved physical spacetime. In particular, we can derive from the action (\ref{actionADM}) the equation of motion of the field. These equations can be rewritten in a more compact way if we write them in terms of the Lie derivative of the fields along the timelike direction (\ref{time}),
\be
 \Lie\Phi = \frac 1 N \dot{\Phi} - \frac{N^i} N \partial_i\Phi\, .
\ee
Varying the action with respect to $\Pimas$ we get
\be
N\Lie\Phi-i\theta N\Lie\Pi-N\Pi-\frac{i\theta}4 \frac{\dot h}h\Pi + \frac{i\theta}2 N^i_{\vert i}\Pi\,=\, 0 \label{PiADM}
\ee
and varying the action with respect to $\Phimas$,
\be
-N\Lie\Pi-\frac{\dot{h}}{2 h}\Pi-N m^2\Phi+(N \partial_i \Phi)^{\vert i}+N^i_{\vert i}\Pi\,=\, 0\, .\label{PhiADM}
\ee
These equations define a system of coupled differential equations which should be solved provided the metric of spacetime is known and we neglect the back-reaction of the field on the metric. We should remark that the general covariance of the whole spacetime is broken by the preferred choice of a time variable, but the general covariance on the spacelike hypersurfaces is guaranteed by construction. This setting reminds the context of analogue gravity models \cite{Barcelo:2005fc}.

{The formulation of the quantum theory of a field in curved spacetime requires the definition of a scalar product in the space of solutions of the field equations. We are now able to define an inner product in the space of solutions. The inner product essentially carries the information of the commutation relations. Given two solutions, $\Phi = \varphi_{1,\theta}$ and  $\Phi = \varphi_{2,\theta}$, of the  equations of the field (\ref{PiADM}), (\ref{PhiADM}) we} can define the internal product $\left(\, , \,\right)$ as
\be
\left(\varphi_1, \varphi_2\right)  =  -i\intxt \sqrt{h} \left[\varphi_{1,\theta}\varphi^{\Pi\,*}_{2,\theta}-\varphi^*_{2,-\theta}\varphi^\Pi_{1,-\theta}\right]\, ,
\label{intprod}
\ee
where the integral is evaluated at any of the spacelike hypersurfaces in which spacetime becomes foliated and
\be
\varphi_{r,\theta}^\Pi  =  \left[ 1+\frac{i\theta}N\left(N\Lie+\frac{\dot{h}}{4h}+\frac{N^i_{\vert i}}{2}\right) \right]^{-1} \Lie\varphi_r (\theta)\, ,
\ee
is the conjugate momentum solution $\Pi = \varphi_{r,\theta}^\Pi$ associated to the field solution $\Phi = \varphi_{r,\theta}$ with $r=1,2$. This internal product preserves the symplectic structure of the theory and is independent of the choice of the hypersurface by construction \cite{Wald}.

Let $\{\Phi \, = \, \udp(\vx,t;\theta)\}$ be an orthonormal basis of solutions of the system of differential  equations (\ref{PiADM}), (\ref{PhiADM}) with positive frequency with respect to the preferred time direction (\ref{time}), labeled by the index $\vp$. Let $\{\Phi \, = \, \vdp^*\}$ be an orthonormal basis of solutions of negative frequency of the same system of differential equations. Then the field can be expanded in terms of annihilation operators of the $\udp(\vx,t;\theta)$ and creation operators of the $\vdp^*(\vx,t;\theta)$
\be
\Phi(\vx,t)=\intpt\left[\udp(\vx,t;\theta) \,a_\vp\,+\,\vdp(\vx,t;\theta)^* \,b^\dagger_\vp\right]\, .
\ee
 If $\vdp^*$ is a solution of negative frequency of the system of field equations (\ref{PiADM}), (\ref{PhiADM}), then $\vdp$ is a solution of positive frequency of the complex conjugate of the system, which coincides with the result of changing $\theta\,\rightarrow\, -\theta$ in the system. Therefore $\vdp(\vx,t;\theta) \propto \udp (\vx,t;-\theta)$. A more careful analysis of the commutation relations shows that $\vdp(\vx,t;\theta) = \udp (\vx,t;-\theta)$. This is  due to the symmetry $\theta\,\rightarrow \,-\theta, \, a\,\leftrightarrow \, b, \,\Phi \,\leftrightarrow \, \Phimas, \, \Pi\,\leftrightarrow \,\Pimas $.

As usual, there is non-uniqueness in the choice of an orthonormal basis $\{\udp,\vdp^*\}$ of solutions of the system of field equations if the metric induces a loss of the time translation symmetry. We must then resort to Bogoliubov techniques in order to relate the associated non equivalent vacua. Let us expand the field in terms of two different basis of solutions of the equation of motion
\begin{eqnarray}
\Phi(x) & = & \intpt \left[ \udp(x)\,  a_\vp +\vdp^*(x) \, b^\dagger_\vp \right]\nonumber\\
& = & \int \frac{d^3 \vp'}{(2\pi)^3} \left[ \bar{U}_{\vp'}(x) \bar{a}_{\vp'} +\bar{V}_{\vp'}^*(x) \bar{b}_{\vp'}^\dagger \right]
\label{expansion}\, .
\end{eqnarray}
As both sets of solutions of the equation of motion are basis, we can define the Bogoliubov transformation as a change of basis in the space of solutions of the equation of motion,
\be
\udp =  \int \frac{d^3 \vp'}{(2\pi)^3} \left[ \bar{U}_{\vp'}(x) \alpha_{\vp' \vp} +\bar{V}_{\vp'}^*(x) \beta_{\vp' \vp} \right] \, ,\label{Bogo}
\ee
where
\begin{eqnarray}
\alpha_{\vp' \vp} =  \left( \udp , \bar{U}_{\vp'} \right)\, , \qquad \beta_{\vp' \vp} & = & -\left( \udp , \bar{V}^*_{\vp'} \right)
\, .
\end{eqnarray}
We define a state $\vert 0 \rangle$ as the vacuum in the $\{\udp,\vdp^*\}$ basis. We want to compute the expectation value in this state of the number operators $\bar{N}_{a,\vp'}$, $\bar{N}_{b,\vp'}$ defined in the $\{\bar{U}_{\vp'},\bar{V}_{\vp'}^*\}$ basis. Inserting (\ref{Bogo}) in (\ref{expansion}) in order to get the Bogoliubov transform of the creation-annihilation operators and plugging them in the expression of the number operators, the result will be
\begin{eqnarray}
\langle 0 \vert \bar{N}_{a,\vp'} \vert 0 \rangle & = &  \int d^3 \vp \vert \beta_{\vp' \vp}(-\theta) \vert ^2\, ,\\
\langle 0 \vert \bar{N}_{b,\vp'} \vert 0 \rangle & = &  \int d^3 \vp \vert \beta_{\vp' \vp}(\theta) \vert ^2\, .
\end{eqnarray}
Noticeably, the internal product (\ref{intprod}) has two terms which differ not only in sign but also in magnitude, unlike in the standard QFT in curved spacetime or the QNCFT in Curved auxiliary Spacetime. This two terms will in general be complex and oscillating with different phases.  Hence, when computing the number of particles and antiparticles observed by a detector in the vacuum of some other detector (like for example in the case of Rindler observers in flat spacetime or static observers at infinity in a black hole spacetime), one then finds that the corresponding spectrum follows the pattern of interference of the terms in the Bogoliubov coefficient $\beta$. This result is essential for example in the studying black hole evaporation by Hawking radiation, something which we plan to address elsewhere.

\section{conclusions}

Recent works \cite{Carmona:2009ra} have offered a mechanism to identify new symmetries in the theory of QNCFT \cite{Carmona:2003kh}. This mechanism goes beyond this particular example and can be applied in principle to more general theories with modified dispersion relations. The new symmetry transformations will be non-local, but the action can be rewritten in terms of an auxiliary spacetime in which the symmetry transformations are local and the action is Poincar\'e invariant. There is a link between the conjugate four-momenta in physical and auxiliary spacetimes, which reminds the DSR paradigm \cite{Judes:2002bw}. As a result a non-local implementation of the Poincar\'e group is found to be a symmetry group of the theory in flat spacetime. However, the generators of the symmetries of this group do not form a closed algebra with the  generator of the symmetry of translations in time. When the theory is tried to be extended, it is unclear which of the symmetries are to be kept and which will be treated as accidental. Some works \cite{Giulini:2006uy} do not give much credit to non-locally implemented symmetries. On the other hand, in the absence of the Poincar\'e symmetry, the possibility of constructing a perturbatively renormalizable interacting theory of the noncanonical field is also unclear \cite{Carmona:2009ra}.

Two different paths have been followed in order to derive a QNCFT in Curved spacetimes depending on the symmetries that are required to be gauged: either the DSR-like symmetries of the free theory, though some of its elements are implemented non-locally, or the group of symmetries of the free theory that are locally implemented, i.e.~translations and rotations.

We find that curving the physical spacetime breaks the DSR-like symmetry. This maybe due to an obstruction in gauging symmetries generated by non-local operators. Consequently, if one wants to preserve the DSR-like implementation of the Poincar\'e group, one must take the auxiliary spacetime as the one to be curved. This leads to a relativistic QFT in Curved Auxiliary Spacetime, which is very similar to the standard one. The correspondence between the fields in the auxiliary frame and fields in the physical spacetime can be trivially done in regions in which the auxiliary spacetime is asymptotically flat. Strictly speaking, the two theories are not really equivalent due to the presence of $\bar{S}^\theta$ part of the physical action. The role of the $\bar{S}^\theta$ modes, which cannot be mapped to auxiliary spacetime, is to fix time ordering in the original frame but they do not seem to have any observable influence on the auxiliary frame, for example they do not affect the Bogoliubov transformations.

On the other hand, if only the symmetries that are implemented locally are required to characterize spacetime at short scales, the physical spacetime of the QNCFT can be made curve using the ADM prescription. The commutation relations must be adapted to fit in a general covariant frame. The action of the field coupled to the three-metric, lapse and shift functions can be written unambiguously. General coordinate invariance is broken by the commutation relations into invariance under general coordinate transformations on the slices of simultaneity, and reparameterizations of time. The resulting QNCFT in Curved Physical Spacetime seems to have the same structure as the ordinary one, but new phenomena related to the UV scale may appear. The main difference between this theory and the standard QFT in Curved spacetimes is that, due of the different energies of particles and antiparticles, the two terms appearing in the internal product (\ref{intprod}) become different not only in sign, but also in magnitude.

We conclude that the theories introduced offer clearly distinguishable outcomes for related phenomenology. This will be further studied in future work.

We would like to thank C.~Barcel\'o, J.~M.~Carmona, J.~L.~Cort\'es, D.~Maz\'on, J.~Rubio, L.~Sindoni and M.~Visser for fruitful discussions. J.I. would like to thank the Scuola Internazionale Superiore di Studi Avanzati (SISSA) for the hospitality during the development of this work. Financial support was provided by CICYT (project FPA2003-02948), DGIID-DGA (project 2008-E24/2). J.I. acknowledges an FPU grant from the Spanish Ministerio de Ciencia e Innovaci\'on (MICIIN).

%%%%%%%%%%%%%%%%%%%%%%%%%%%%%%%
\chapter{Summary and Conclusions}
%%%%%%%%%%%%%%%%%%%%%%%%%%%%%%%

The General Theory of Relativity (GR) and the Standard Model of particle physics (SM) have offered a very accurate description of all known phenomena with two remarkable exceptions, the neutrino oscillations and the accelerated expansion of the universe in the early universe and the present age. The neutrino oscillations and the present acceleration of the universe accept a quite conventional explanation in terms of neutrino masses and a cosmological constant/vacuum energy, respectively. However, these explanations are not by themselves very satisfactory because:

\begin{itemize}
\item
they introduce an enormous naturalness problem,

\item
they introduce  a new ingredient in the theory that is hard or impossible to measure aside from in the phenomena that caused their proposal,

\item
they are unable to explain the striking coincidence in scale between the effective energy density attached to the cosmological constant ($\rho_{vac}\sim (10^{-3} \, eV)^4 $) and the lowest neutrino mass difference ($\Delta m^2 \sim (10^{-3} \, ev)^2$),

\item
the present experimental and observational data are still unable to rule out some of the alternative proposals to these explanations.
\end{itemize}

Despite these questions that have been raised, the neutrino masses and the cosmological constant have been taken as a confirmed fact by many because they are the simplest explanation in agreement with the experimental and observational data, and because they do not compromise any of the basic principles underlying GR or the SM.

However nothing is sacred in science, aside from the proof of experiment. Simplicity is also important, but aesthetics cannot be used as a substitute of experimental verification of a prediction. The physical principles that are taken as basic must be rejected in case they conflict a single experiment or observation, and thus must always be subject of scrutiny.

Therefore we would like to build models that are alternatives to neutrino masses and the cosmological constant, which are consistent with all the known phenomena, and that could be contrasted with the more conventional explanations in the light of future experiments and surveys. We will not fear to violate the basic principles underlying GR and the SM, as long as the departures from them is small enough to be consistent with the available data.

For instance, neutrino masses are the only explanation to neutrino oscillations in vacuum within Special Relativity (SR), and thus we have considered a small Lorentz Invariance Violation (LIV) in neutrinos. On the other side, the too high quantum corrections to a vacuum energy are computed in the framework of Effective Field Theory (EFT), which is based in the Decoupling Theorem, which in turn is based in locality, so we have considered a small departure from locality.

Not always it is required to violate the underlying principles in a theory in order to go beyond it, as is the case of GR. We have considered more general actions than the Einstein-Hilbert action in the search for new physics that can explain the accelerated expansion.

In both cases, the departure from the standard framework has been characterized by energy scales which determine the onset of new physics. These scales can be either ultraviolet (UV) or infrared (IR), depending on the requirements of the phenomena we want to explore. Neutrino oscillations as well as the present accelerated expansion of the universe require the introduction of a new IR scale, whereas the explanation of the horizon and flatness problems requires the introduction of a UV scale. It is possible that the two IR scales are the same, as it is suggested by the coincidence mentioned above, but we have not been able to explain this coincidence within a certain consistent model.

Instead of resorting to a top-down approach like String Theories (ST), Loop Quantum Gravity (LQG) or Grand Unified Theories (GUT), which have grown far from the contact with experiment, we have prefer to adopt a bottom-up approach in which we have started from simple models that offer a description of the data and may grow in the future if they are not ruled out by future data first.

In order to explain the two epochs of accelerated expansion of the universe, we have introduced a phenomenological extension of the Cosmological Standard Model  (CSM) based on GR. The model is similar to other phenomenological descriptions that have been proposed, which relate the expansion of the universe with more ``anthropocentric'' parameters such as the redshift $z$. The Asymptotic Cosmological Model (ACM) is based (as GR) in a metric theory of spacetime in the homogeneous approximation, and it is defined through the relation between the Hubble rate $H$ and the energy density of the matter content of the universe defined by the SM particles and the Dark Matter. The effect of new physics is encoded in the dependence of these relation on an UV and an IR scales which act as bounds of the open interval to which the Hubble rate is constrained. For dimensional arguments, in the limit in which the Hubble rate is far from these bounds, the standard behavior predicted by GR is recovered, whereas the universe tends asymptotically to a deSitter universe in the infinite past and the infinite future when the bounds are approached. The CSM supplied with a cosmological constant ($\Lambda$CDM) is included as a particular case.

The ACM is built with four free parameters, two of them characterizing the early accelerated expansion and two of them characterizing the present accelerated expansion. We have constrained the two of them relevant for the present accelerated expansion, contrasting the results derived from simulations using Monte Carlo-Markov Chains with the observational data from Type Ia Supernovae (SNe), the distance to the recombination surface deduced from the Cosmic Microwave Background (CMB), direct measurements of the energy density and the distance parameter of Baryon Acoustic Oscillations (BAO). We have found that there are better fits to the expansion of the universe than $\Lambda$CDM, but that the quality of the fit depends too crucially on the dataset used. Thus an improvement on the datasets is important in order to favor $\Lambda$CDM or one of its competitors.

The homogeneous behavior of the ACM can be realized in the framework of modified gravity (for instance $f(R)$-theories), scalar-tensor theories or phenomenological Dark Energy models indistinguishably. Differences between these descriptions may be found in the behavior of perturbations.

The lack of an action for the ACM is a serious obstacle in building a theory of perturbations in the ACM. However, under an additional assumption based on phenomenological consistency (the differential equations being of order two), the homogeneous behavior of the ACM determines univocally the linearized behavior of scalar perturbations.

With the assumption of a Harrison-Zel'dovich scale invariant spectrum from the very early universe, this linear treatment suffices to derive the CMB spectrum deduced from the ACM. The comparison between this prediction and the observed spectrum imposes similar bounds on parameter space than the homogeneous analysis. This is due to the fact that the Integrated Sachs-Wolfe effect (ISW) of the ACM for the parameters lying within the confidence regions of our homogeneous analysis is masked by the cosmic variance, i.e.: the intrinsic statistical error associated to the lowest multipoles (cosmic censorship).

The treatment of linear perturbations is able to distinguish the ACM from the most na\"{\i}ve Dark Energy forms, such as scalar-tensor theories or f(R)-theories. These examples usually present fine-tuning problems such as the requirement of a chameleon mechanism for a fifth long-range interaction to disappear.

In the future we plan to explore further the consequences of this model, such as the nonlinear behavior of perturbations, the behavior of vector and tensor modes, or the weak gravity limit.

It is often stated that the neutrino oscillation in vacuum experiments prove that neutrinos are massive particles. We have shown that this statement is false. Neutrino oscillations can be also explained in terms of oscillations between particles with different dispersion relations, wether this difference relies on different masses or not.

Neutrinos with different constant velocities or dispersion relations modified by a UV scale are certainly ruled out, with the exception of a single very fine-tuned model. However, neutrino oscillation experiments only probe the high energy (few MeV to few GeV) regime of neutrinos, and therefore {\it any} modification of the dispersion relation in the IR, such that the high energy effect of the modification amounts to a term proportional to the square of the IR scale, can naturally explain neutrino oscillations in vacuum.

It is difficult to find such a modification of the dispersion relation different from a mass term in QFT. We have found a model in which this is possible also respecting the gauge symmetry and lepton number conservation of the standard model. However, we have had to sacrifice not only Lorentz invariance but also locality. We plan to study a modification of this model allowing for lepton number violation in the future.

In order to distinguish between massive neutrinos or other IR modification of the standard model, it will be necessary to probe the low energy regime (few meV to few eV) of the neutrinos. The Cosmic Neutrino Background (C$\nu$B) predicted by cosmology offers the possibility of measuring such neutrinos in the future.

However, in order to study the effect of a Lorentz non-invariant theory in cosmology it is important to build a quantum field theory in curved spacetime which preserves the symmetry properties of such a theory. Although this is a very complicated task, outside the scope of this thesis, as a first step we have explored the possibility of building the scalar Quantum Theory of Noncanonical Fields (QNCFT) in a curved spacetime.

Respecting either the non-locally implemented Poincar\'e group of symmetries or the locally implemented Euclidean group of symmetries of the theory in flat spacetime leads to two different ways of defining the theory in curved spacetime.

If one chooses to gauge the non-locally implemented Poincar\'e group of symmetries by the standard procedure, an obstruction is found. The solution is to introduce curvature in the auxiliary spacetime in which the Poincar\'e symmetry can be interpreted as a spacetime symmetry, instead of curving the physical spacetime. The resulting theory is nonlocal in time. In the limit in which the parameter $\theta$ tends to zero both spacetimes coincide and one recovers the standard QFT in curved spacetime. The consequences of this theory will be similar to the ones of standard QFT in curved spacetime but the particles will have a modified dispersion relation.

Alternatively, one may curve the physical spacetime, but then the non-locally implemented symmetry of the theory in flat spacetime is broken. In the resulting theory a preferred reference frame arises, and spacetime is foliated in spacelike simultaneity hypersurfaces. New phenomena related to the UV scale $\theta$ may arise, which were not present in the standard QFT in curved spacetime.

We plan to study how the Hawking radiation is affected by the noncanonical commutation relation in both theories. We would also like to go further in determining what are the symmetries of in the model of noncanonical neutrinos and how they can be coupled to gravity.

With a na\"{\i}ve assumption of about how the modified dispersion relation of the noncanonical neutrinos can be introduced in cosmology, we have computed the effect of a IR modified dispersion which mimics a cosmological constant in cosmology. Using the bounds on `squared neutrino mass differences' and on the `spatial curvature', the resulting ``effective cosmological constant'' turns out to be three orders of magnitude above the observed value.

If there exist a connection between the IR scale pointed out by neutrino oscillations and by the present acceleration of the universe, much work is still to be done to uncover this relation. We hope that this PhD thesis will serve as inspiration for future research in this direction.

We will now proceed to enunciate the results of this thesis

\begin{itemize}
\item
We have built a cosmological model (ACM) defined by the relation between the energy density and the Hubble parameter in the homogeneous approximation. The model contains the $\Lambda$CDM as a particular case. This model suffices to explain the horizon problem and the present accelerated expansion. Through a simulation, the late time behavior of the model has been constrained. Although a cosmological constant is permitted at $1\sigma$, the observations favor other regions of the parameter space. However, the parameter determination depends strongly on the dataset which is being used.
\item
Assuming order two differential equations for the perturbations, we have found that it is possible to find a unique set of equations for the scalar perturbations over a FRW universe whose homogeneous behavior departs from GR. In contrast to common belief, this is doable without resorting to higher order equations or new degrees of freedom, at least at the linearized level. The model is distinguishable from other proposals such as $f(R)$-gravity, scalar tensor theories or the most na\"{\i}ve dark fluid models.
\item
We have used the equations for the scalar perturbations associated to the ACM to derive the spectrum of the CMB in a universe following an expansion governed by the ACM. The fit to observations does not add new information with respect to the homogeneous fits, due to a masking of the late ISW effect by the cosmic variance at low multipoles.
\item
It is often stated that neutrino oscillation experiments imply that neutrinos have mass. We have found a counter example to this statement. We have built a model in which the effect of new physics governed by an IR scale is able to explain all the neutrino oscillation experiments. Thus neutrino oscillations imply that neutrinos are sensitive to new physics in the IR. The most promising observation that would be able to distinguish between massive and noncanonical neutrinos is the cosmological effect of neutrinos, and in particular the C$\nu$B.
\item
As a first step towards determining the cosmological effect of noncanonical neutrinos, we have built two possible quantum theories of scalar noncanonical fields in curved spacetimes, depending on which symmetry of the theory in flat spacetime is broken when curving spacetime. In one of them there is no preferred reference frame and particles with modified dispersion relations are allowed, but it has an unclear interpretation. The other involves a preferred time direction and new effects may arise, but the renormalizability of the QFT is not guaranteed.
\item
Our work has tried to explore two phenomena which seem to point to new physics in the IR: neutrino oscillations and the present cosmic acceleration. If these phenomena are related, there is still much work to do to uncover this relation.
\end{itemize}

\newpage

\appendix

\newpage

%%%%%%%%%%%%%%%%%%%%%%%%%%
\chapter{Implications of noncanonical neutrinos on cosmology}
%%%%%%%%%%%%%%%%%%%%%%%%%%

Violating Lorentz invariance in QFT through a deformation of the anticommutation rules of quantum fields yields the possibility of the neutrinos having a IR modified dispersion relation such that mimics the role of the cosmological constant in the present cosmic acceleration. This possibility is further studied.

\section{Motivation}

Our last work on LIV in neutrino physics caused by a deformation of the anticommutation rules of the fermion fields opens the possibility of the neutrinos having a dispersion relation very different from the relativistic one. Experimental data from high energy physics have shown that the neutrino dispersion relation differs from that of a relativistic massless particles at low energies. The scale of energy of that deviation is $\lambda < 1 eV$ and the differences between the dispersion relation of the different neutrino species can be parameterized up to first order corrections as

\begin{equation}
 E_i(p) = p + \frac{m_i^2}{2 p} + \mathcal{O}(\frac{\lambda^3}{p^2})\, ,
\end{equation}
where we suppose that $m_i \sim \lambda$, with $\vert\Delta m^2_{12}\vert = 7.9^{+0.27}_{-0.28} \times 10^{-5} (eV)^2$ and $\vert\Delta m^2_{13}\vert = 2.6 \pm 0.2  \times 10^{-3} (eV)^2$ as shown by neutrino oscillation experiments.

However, a modification in the dispersion relation of the neutrinos should have a visible impact on cosmology, due to the effective temperature of the cosmic neutrino background being at least one or two orders of magnitude below the scale $\lambda$ ($T_\nu = 1.95\, K = 1.68 \times 10^{-4}\, eV$). Moreover cosmological observations seem to imply the existence of a cosmological constant with energy density

\begin{equation}
 \rho_\Lambda = \rho_c \Omega_\Lambda = \frac {3 H_0 \Omega_\Lambda}{8 \pi G} = \frac {3 \Lambda}{8 \pi G} = 3.0 \times 10^{-11}\, (eV)^4 \, .
\end{equation}

The similarity between the energy scale of the cosmological constant and the energy scale of neutrino masses seem to suggest that both phenomena are connected. Might neutrinos be the Dark Energy?

Let us check the order of magnitude of a cosmological constant-like  contribution to the energy density of the universe. Neutrinos would mimic a cosmological constant if their dispersion relation satisfies $E(p_d \frac{a_d}a )n_\nu = const$,  where $n_\nu$ is the number density of neutrinos ($n_{\nu} \propto a^{-3}$, $n_{\nu, now} \sim n_{\gamma, now} \sim 10^{-12}\, (eV)^3 $), $p_d$ is the momentum of a neutrino at neutrino decoupling ($T_d \sim 1.5\, MeV$) and $a_d$ is the scale factor at neutrino decoupling. Therefore, the contribution of neutrinos to the energy density will be constant if they have a dispersion relation $E(p) \sim \frac {\lambda^4}{p^3}$ for $p \ll \lambda$. If we assume that the average momentum of the neutrinos at neutrino decoupling is similar to the decoupling temperature and that the number of photons and neutrinos is more or less conserved during the expansion of the universe after they decouple,  we get, for $\lambda^2\sim\Delta m_{12}^2$:

\begin{equation}
 \Omega_{\nu, now} \approx 0.68 \Omega_\gamma \frac {E_\nu (T_\nu)}{T_\nu} \sim 0.68\Omega_\gamma \frac {\lambda^4}{T^4_\nu} \sim 0.68 \times 10^{-5} \frac {(10^{-5}\, (eV)^2)^2}{(10^{-4}\, eV)^4} \sim 1\, .
\end{equation}

It is known that $\Omega_\Lambda = 0.73$ in $\Lambda$CDM and the contribution of this neutrinos seems to be of the same order, so maybe we can explain what is the Dark Energy by considering some specific neutrino dispersion relation.

We will now see how a polynomial dispersion relation in the ratio $\lambda/p$ is able to reproduce the effect of a cosmological constant. However, a detailed computation will show that the resulting effective cosmological constant is too high.

\section{A polynomial dispersion relation}

We will consider a universe composed by the following:
\begin{itemize}
 \item \textbf{Photons}: Their contribution to the total energy density today is $\Omega_\gamma= 4.92\times10^{-5}$. They form a Plankian spectrum with temperature $T_\gamma= 2.73\, K= 2.35\times10^{-4}\, eV$, which implies an energy denisity of $\rho_\gamma = 4.64 \times 10^{-34}\, g/cm^3 = 2.01\times 10^{-15} \, (eV)^4$.
\item \textbf{Baryons}: By baryons it is meant ordinary mater: nuclei and electrons (the contribution of electrons to the energy density is negligible compared to that of nulei). $\Omega_b = 0.0415 \pm 0.0016$.
\item \textbf{Dark Matter}: It doesn't interact with photons, it is neither baryonic. $\Omega_{DM} = 0.236^{+0.016}_{-0.024}$.
\item \textbf{Primordial Neutrinos}: Consider 3 families of neutrinos whose abundance would be $\rho_{\nu_i} \sim \frac {0.68}3 \rho_\gamma$ if they were  massless relativistic particles. They form a Plankian distribution with temperature $T_\nu = 1.95 \, K = 1.68 \times 10^{-4}\, eV$. In this section we will consider that their energy dispersion relation is
\begin{equation}
E_{\nu_i}(p)=p\left( 1+\frac{c_i}{p^2}\right) ^2\, .
\label{disp1}
\end{equation}
We can obtain more information about the coeficient $c_i$ from neutrino oscillation experiments: $ \vert c_2-c_1\vert\equiv\frac{\Delta m^2_{12}}4 = (1.98 \pm 0.07) \times 10^{-5} (eV)^2$, $ \vert c_3-c_1\vert\equiv\frac{\Delta m^2_{13}}4 = (6.5 \pm 0.5)\times10^{-4} (eV)^2$.
\end{itemize}

\section{Equal momenta approximation}

As a first approximation we will consider that today the momentum of every neutrino is equal to the temperature $T_\nu$. Taking into account the dispersion relation (\ref{disp1}), the contribution from each of the neutrino species to the total energy density will be:

\begin{equation}
\Omega_{\nu_i} = \frac{\rho_{\nu_i}}{\rho_c} \sim \frac{0.68}3\Omega_\gamma \left( 1+\frac{c_i}{T_\nu^2} \right)^2 \sim
1.12 \times10^{-5} +793 \frac{c_i}{(1 eV)^2}+1.40\times10^{10} \frac{c_i^2}{(1 eV)^4}\, .
\label{approx1}
\end{equation}

The first term equals the contribution of massless relativistic neutrinos to the energy density today. The second term plays the role of an effective spatial curvature contribution to the energy density $\equiv -\frac k{(aH_0)^2}$. The coefficients $c_i$ must be chosen to make this contribution compatible with the upper bounds on a spatial curvature term from WMAP. The last term mimics the contribution of a cosmological constant to the total energy density.

It is easy to see that the condition for $\vert c_2 -c_1 \vert$ is easy to satisfy consistently with all the experimental data. However, the condition for $\vert c_3-c_1\vert$ forces at least one of the $c_i$ to be of order $10^{-4} (eV)^2$, and this causes the last term in (\ref{approx1}) to be too large compared to the experimental data for $\Omega_\Lambda$.

A particular example is

$$
\begin{array}{l}
c_1=-2.23\times 10^{-4} (eV)^2 \, ,\\
c_2=-2.03\times 10^{-4} (eV)^2 \, ,\\
c_3=+4.27\times 10^{-4} (eV)^2 \, ,\\
\Rightarrow \Omega_\Lambda \sim 3800 \gg 0.73\, .
\end{array}
$$

\section{Taking into account the Planckian spectrum}

It might be possible that this disease is cured by taking into account that the momenta of the neutrinos are distributed forming a Planckian spectrum. We must have into account that there are two degrees of freedom ($\nu$ and $\bar{\nu}$) per neutrino species. The distribution in thermal equilibrium at temperature T is then

\begin{equation}
 \frac{d n_i (p)}{d p} = \frac 1{\pi^2} \frac {p^2}{e^{(E_i(p)-\mu_i)/T}+1}\, .
\end{equation}

The neutrinos were in thermal equilibrium before they decoupled. Then they keep their distribution with $p=p_d \frac {a_d}a$. The chemical potential can be neglected since

\begin{equation}
 \frac {\mu_i} T = \frac {n_{\nu_i}-n_{\bar{\nu_i}}}{n_\gamma}\equiv L_i<10^{-9}\, .
\end{equation}

At energies close to decoupling ($T_d \sim 1.5 \, MeV$) we can use the approximation $E_d \approx p_d$. Therefore, at neutrino decoupling we have:

\begin{equation}
 n_i \approx \frac 1{\pi^2} \int dp p^2 (e^{p/T} +1)^{-1} = \frac {3 T^3 \zeta(3)}{2 \pi^2} \approx 0.1827 \times T^3\, ,
\end{equation}

\begin{equation}
\left\langle  p\right\rangle_i \approx \frac 1{n_i}\frac 1{\pi^2} \int dp p^3 (e^{p/T} +1)^{-1} = \frac {7 \pi^2 T^4}{120 n_i} \approx 3.1514 \times T\, ,
\end{equation}

\begin{equation}
\left\langle  \frac 1p\right\rangle_i \approx \frac 1{n_i}\frac 1{\pi^2} \int dp p (e^{p/T} +1)^{-1} = \frac {T^2}{12 n_i} \approx \frac{0.4561}T\, .
\end{equation}
$\left\langle p^{-3} \right\rangle_i$ diverges due to the contribution of small momenta to the integral, unless we take into account that $E(p) = p\left( 1+\frac {c_i}{p^2}\right) ^2$. This integral cannot be solved analytically, so we will solve it numerically using the values for the $c_i$ in the previous example:

\be
\ba{l}
\left\langle p^{-3} \right\rangle _i = \frac 1{n_i}\frac 1{\pi^2} \int \frac {d p}p (e^{p\left( 1+\frac {c_i}{p^2}\right) ^2 /T_d} +1)^{-1}\\

\approx \left\lbrace

\begin{array}{r}
\frac {1.23609}{n_i} \approx 6.76 T_d^{-3} \;\text{for}  \; c_i = -2.23 \times 10^{-4}\, (eV)^2\, ,\\
\frac {1.23926}{n_i} \approx 6.78 T_d^{-3} \;\text{for}  \; c_i = -2.03 \times 10^{-4}\, (eV)^2\, ,\\
\frac {1.21415}{n_i} \approx 6.64 T_d^{-3} \;\text{for}  \; c_i = 4.27 \times 10^{-4}\, (eV)^2\, .
\end{array}
\right.

\ea
\ee

The coefficients coming out of this strict derivation are of order one, which means the dispersion relation (\ref{disp1}) gives a too large constant contribution to the total energy density. In fact,

\begin{equation}
 \Omega_\nu\equiv\sum_{i}\Omega_{\nu_i}^{(\gamma)}+\Omega_{\nu_i}^{(k)}+\Omega_{\nu_i}^{(\Lambda)}\, ,
\end{equation}
where $\sum_{i}\Omega_{\nu_i}^{(\gamma)}=3.37 \times 10^{-5}$, $\sum_{i}\Omega_{\nu_i}^{(k)}<0.02$, $\sum_{i}\Omega_{\nu_i}^{(\Lambda)}\sim 8100$. This means that the simplest energy-momentum dispersion relation which we have thought is ruled out by the cosmological data.

\section{Discussion}

Although it is possible to find {\it ad hoc} dispersion relations for the neutrinos that mimic the effect of the Dark Energy and are consistent with the cosmological data, the simplest examples are ruled out. This indicates that maybe our assumption that neutrinos were driving the present accelerated expansion was too ambitious.

Thus cosmology imposes great constraints in the low energy behavior of the dispersion relation of neutrinos. Any nontrivial behavior should be non-polynomial in $\lambda/p$ in order not to be ruled out by cosmological observations. Otherwise it would enter in conflict with the cosmological bounds on Hot Dark Matter (HDM)\footnote{If the neutrino dispersion relation mimics that of a massive particle in the IR, then one should obviously choose the simplest explanation and work with massive neutrinos.}, spatial curvature, or a cosmological constant, and we have proven that this bounds are tight.

The effect of neutrinos with a modified dispersion relation in the IR will be further studied in future work. In particular, the effect of a modified dispersion relation in the distance to neutrino decoupling surface or the matter power spectrum will be addressed elsewhere.

\end{document}